\documentclass[11pt,letterpaper]{article}
\usepackage{amsfonts,mathrsfs,natbib,hyperref}

\usepackage{pdflscape}

\usepackage[section]{placeins}
\usepackage{amssymb}
\usepackage{amsmath}
\usepackage{float}
\usepackage[margin=1.0in]{geometry}
\usepackage{color}
\usepackage[dvips]{graphicx}
\usepackage{Here,multirow,lscape}
\usepackage[most]{tcolorbox}
\usetikzlibrary{patterns,shadows}
\usepackage{amsthm}

\newtheorem{asu}{Assumption}
\newcounter{subassumption}[asu]
\renewcommand{\thesubassumption}{(\textbf{\alph{subassumption}})}
\makeatletter
\renewcommand{\p@subassumption}{\theasu}
\makeatother
\newcommand{\subasu}{
	\refstepcounter{subassumption}%
	\thesubassumption~\ignorespaces}
\newtheorem{theorem}{Theorem}[section]

\newtheorem{prop}{Proposition}

\newtheorem{algorithm}[theorem]{Algorithm}

\date{\today}
\newtcolorbox{boxone}{opacityback=0,enhanced jigsaw}

\usepackage{authblk}

\begin{document}
\title{Estimation of Heterogeneous Treatment Effects Using a Conditional Moment Based Approach}
\author[1]{Xiaolin Sun}
\affil[1]{Department of Economics, Simon Fraser University}
\date{\today}
\maketitle

\begin{abstract}
We propose a new estimator for heterogeneous treatment effects in a partially linear model (PLM) with multiple exogenous covariates and a potentially endogenous treatment variable. Our approach integrates a Robinson transformation to handle the nonparametric component, the Smooth Minimum Distance (SMD) method to leverage conditional mean independence restrictions, and a Neyman-Orthogonalized first-order condition (FOC). By employing regularized model selection techniques like the Lasso method, our estimator accommodates numerous covariates while exhibiting reduced bias, consistency, and asymptotic normality. Simulations demonstrate its robust performance with diverse instrument sets compared to traditional GMM-type estimators. Applying this method to estimate Medicaid's heterogeneous treatment effects from the Oregon Health Insurance Experiment reveals more robust and reliable results than conventional GMM approaches.

\vspace{.5cm}

\vspace{.5cm}
\end{abstract}

\newpage

\section{Introduction}
\label{sec:intro}
Heterogeneous treatment effects have been a central focus in the literature since \cite{Imbens1994}. These models explore how the effects of policies, programs, or other interventions on a specific outcome may vary across individuals with distinct characteristics. To quantify heterogeneous treatment effects, researchers often augment standard linear models by including interaction terms between the treatment variable and relevant covariates (as discussed in Section 21.4, Chapter 21 of \citet{WooldridgeJeffreyM.2010Eaoc} and Section 7.6, Chapter 7 of \citet{imbens_rubin_2015}, Page 125). This extended model becomes linear in the treatment variable, the interaction term, and the covariates, with the parameters of interest typically encompassing the effects of the treatment and its interaction with some covariates. Furthermore, significant attention has been devoted to addressing unobserved heterogeneity (\cite{Wooldridge1997b}, \cite{Wooldridge1999c}). In this paper, we adopt a standard and widely used semi-parametric model to demonstrate the advantages of integrating a conditional moment-based approach into the estimation of the observed heterogeneous treatment effects, that is, the parameters in front of the treatment and the interaction terms.

Our work introduces a novel estimator, D-RSMD, extending the R-SMD estimator proposed by \cite{AS2021} to accommodate big data settings. Specifically, the D-RSMD estimator combines the debiased method with R-SMD, integrating these approaches for improved estimation procedures. This new estimator offers two key properties essential for applied researchers. While versatile and applicable across various scenarios, we specifically focus on estimating heterogeneous treatment effects to illustrate those properties. We look into the case where the outcome variable adopts a classic partially linear structure encompassing linear and non-parametric components. Within the linear part, the treatment variable is endogenous, and the interaction terms are the treatment multiplied by exogenous covariates.

The primary advantage of the D-RSMD estimator is its reliance on a single valid instrument, bypassing the need to explicitly model the relationship between treatment and the instrument. This eliminates the requirement to generate additional exogenous variables, assume specific functional forms for the first stage (as in 2SLS), or select moments for GMM estimation. The choice of generated variables, model specifications in the first stage, or moments significantly impacts estimation results and the asymptotic properties of treatment effects. This vital advantage is shared by Smooth Minimum Distance (SMD) estimators like the SMD estimator proposed and examined in \cite{LavergnePatilea2013} and the R-SMD estimator introduced by \cite{AS2021}. Such an approach is crucial for avoiding model specification issues and ensuring robust estimation of treatment effects.

For example, consider a scenario where the treatment variable is endogenous and interacts with a covariate. The traditional GMM method typically requires two moments: a valid instrument and an additional generated variable, for instance, the interaction between the instrument and covariate, to estimate both parameters. If we employ 2SLS, we need a model specification assumption. However, if the covariate in the interaction term exhibits limited variation or lacks information, GMM estimates can become inconsistent with high variance, resembling a weak identification problem. In contrast, our D-RSMD method relies solely on a single valid instrument, eliminating the need for additional generated variables, first-stage model specification, or moment selection.

Another approach to addressing this issue brought about by interacting with the covariate is splitting the data set, which is effective when dealing with binary covariates and a known outcome model with a substantial number of observations. However, the D-RSMD method offers distinct advantages by leveraging all information from a single conditional moment restriction across all observations, producing competitive results with smaller standard errors (SEs), sometimes even smaller SEs than those from traditional GMM with the correct model specification. We illustrate this idea using simulations in the appendix.

The second advantage of the D-RSMD estimator is its applicability in big data settings. The R-SMD estimator introduced by \cite{AS2021} combines Robinson transformation with SMD estimation to estimate parameters in the linear portion of a partially linear model\footnote{see \cite{Robinsontrans} and \cite{LiRacine2006} for combining Robinson transformation and 2SLS estimators.}. In their work, \cite{AS2021} employ the Nadaraya-Watson estimator to estimate conditional means, restricting the number of covariates to fewer than four (refer to Chapter 7 of \cite{LiRacine2006}). However, when dealing with models involving a large number of covariates, as in their empirical application, they resort to dimension reduction techniques such as Principal Component Analysis (PCA).

In contrast, the D-RSMD method is tailored for big data scenarios, thus relaxing restrictions on the number of covariates. This is achieved by applying a method that works in big data settings, then incorporating the debiased method proposed by \cite{ChernozhukovVictor2018Dmlf}. In their work, \cite{ChernozhukovVictor2018Dmlf} employ various techniques such as Lasso, random forests, neural networks, boosted regression trees, ridge regression, and so on. Then they employ the debiased method to reduce the effect of bias introduced by those techniques, which involves Neyman-Orthogonalization. This method yields an orthogonalized first-order condition (FOC) akin to the Robinson transformation proposed in \cite{Robinsontrans} for partially linear models.

Our work employs the Neyman-Orthogonalized method following Robinson transformation. This adaptation is necessary due to the additional layer introduced by the SMD estimation procedure within the objective function. Orthogonalizing the FOC with this layer helps mitigate bias arising from employing the method in big data settings. Consequently, our D-RSMD estimator extends the debiased method proposed by \cite{ChernozhukovVictor2018Dmlf} into the settings of SMD and U-statistics. Recent work by \cite{escanciano2023debiased} provides general results for the Neyman orthogonalization of U-statistic type moments, offering a general construction of orthogonal quadratic moment functions for obtaining debiased estimators and valid inference in settings with U-statistics. While our objective function resembles a special case of the function in \cite{escanciano2023debiased}, it incorporates a special weighting component based on the Fourier Transform using two observations from SMD, adding another layer of complexity to the construction of the Neyman-orthogonal FOC. With this extra weighting component, our identifying moment cannot be linearized. Hence, the methods and the inference results of \cite{escanciano2023debiased} assuming the existence of a linearized component in Equation (2.8) of \cite{escanciano2023debiased} are not applicable in our work.

Our D-RSMD focuses on using the Lasso method. The Lasso method is not the only option we have; as discussed in \cite{ChernozhukovVictor2018Dmlf}, there are lots of available estimation methods. Here, we use Lasso as an illustration. Some researchers use sieves for the non-parametric part of the model, and they need to specify how to select the number of sieves in their estimation procedure in practice. In comparison, our new method does not need to choose the number of polynomials for the sieve method. When we use the Lasso method, the tuning parameter is inside the penalty term for the estimation of the nuisance parameter. In practice, there are several ways to choose Lasso tuning parameters (see \citet{vandeGeerSaraEaTU}), and we use cross-validation (\cite{ChernozhukovVictor2018Dmlf}). When we know there are fewer than four covariates, the Nadaraya-Watson estimator is an alternative method. This estimation procedure also needs one tuning parameter, namely the bandwidth. The bandwidth can be chosen by rule of thumb or cross-validation in practice. We show that our D-RSMD estimator is consistent and $\sqrt{n}$ asymptotically normal under mild regularity conditions. Monte-Carlo simulations show that the D-RSMD estimator outperforms the R-SMD and GMM-type estimators in terms of much smaller bias and lower standard errors.

To demonstrate the benefits of our method, we examine the impact of Medicaid using data from the Oregon Health Insurance Experiment.\footnote{I would like to acknowledge A. Colin Cameron for drawing my attention to the Oregon Health Insurance Experiment and providing access to the dataset. The dataset is sourced from his and Pravin K. Trivedi's book in 2022 and is also available on the website \url{https://www.stata-press.com/data/mus2.html}.} Our estimator, using a single valid instrument in the Oregon Health Insurance Experiment, yields statistically significant results for heterogeneous treatment effects, unlike the GMM estimator. Moreover, our approach produces more reliable results without requiring a generated instrument variable. This section presents an empirical analysis of the new D-RSMD estimator, emphasizing its advantages in estimating treatment effects and associated standard errors. Furthermore, it contributes to the literature on the Oregon Health Insurance Experiment by uncovering heterogeneous treatment effects that traditional methods fail to detect.

The paper is organized as follows: Section \ref{section:framework} introduces our framework, motivation, and the estimator. Section \ref{section:large sample theory} states the large sample properties for our estimator. Section \ref{section:simulation} presents the simulation results for finite samples with a large number of covariates. Section \ref{section:application} uses the D-RSMD estimator to estimate the effects of enrollment in Medicaid on outcomes of interest, based on the Oregon Medicaid health experiment. Additional results and proofs are in the Appendix.

\section{Framework and Motivation} \label{section:framework}

Our framework is based on a partially linear model featuring a binary treatment variable $W_i$. Our focus lies in understanding the treatment's impact on the outcome variable. However, it is important to note that the treatment is not always exogenous. Take, for instance, the Oregon Medicaid health experiment, where treatment corresponds to enrollment in Medicaid—a decision influenced by the choice of individual lottery winner, rendering enrollment endogenous. Moreover, the treatment effect may not be homogeneous; it can vary depending on certain covariate variables in $X_i$, a vector of $q_X$ covariates ($X_i = [X_{i1}', X_{i2}']'$). Thus, we seek to examine the treatment effect conditional on specific groups determined by the values of the covariates $X_{i1}$. The remaining part of the model consists of an unknown function of exogenous $X_i$. While we maintain a fixed dimension for $X_{i1}$ for illustrative purposes, the dimension of $X_i$ is unrestricted and can be either high or low-dimensional. Additionally, the outcome variable $y_i$ is not constrained to be binary. For studies involving binary outcome variables, a large number of covariates, and similar restricted dimension settings for estimating heterogeneous treatment effects, see \cite{NekipelovDenis2018ROML}. Generally, we focus solely on scenarios where we have knowledge of the functional form of the treatment and its heterogeneity, representing an initial step toward utilizing conditional moment restriction to estimate unobserved heterogeneity.
\begin{equation} \label{eq:model01}
y_i =\theta_{w0} W_i + W_i \cdot X_{i1}' \theta_{wx0} + f_{0,1}(X_i)+ \epsilon_i
\end{equation}
The formal definition for $W_i \cdot X_{i1}' \theta_{wx0}$ is $\sum_{k=1}^{q_{X_1}} W_i X_{i1,k}\times \theta_{wx0,k}$ where $X_{i1,k}$ is a $k$-th element inside the vector $X_{i1}$. The treatment part of interest is $W_i + W_i \cdot X_{i1}$. The parameters that measure heterogeneous treatment effects, i.e. $\theta_{w0}$ and $\theta_{wx0}$, are our key parameters.

As in the previous discussion, $W_i$ and the interaction term between $W_i$ and $X_{i1}$ are endogenous. To estimate the key parameters, a vector\footnote{$Z_i$ denotes the general vector of instruments. If we only use one instrument, $Z_i$ will be the random assignment. If we use a vector of instruments, $Z_i$ is the vector including the random assignment and other instruments.} of instruments $Z_i$ is introduced. For instance, in the Oregon Medicaid health experiment, the instrument is the lottery outcome, that is, the random assignment\footnote{In the experiment, the lottery winners are allowed to enroll in the Medicaid program. Not every lottery winner enrolled. The treatment variable is the actual enrollment status.}. We maintain the conditional mean independence assumption for the instrument. With valid instrument and exogenous control variables, the conditional moment restriction is
\begin{equation}
E(\epsilon_i|X_i, Z_i)=0 \ a.s.
\end{equation}

We consider the single treatment case, where the covariates may be correlated with the instruments. Traditional estimators, such as GMM or 2SLS, typically require $1 + q_{X_1}$ moments or exogenous variables to identify (and estimate) $1 + q_{X_1}$ parameters within $\theta_{w0}$ and $\theta_{wx0}$. These moments or exogenous variables include functions of $Z_i$ and $X_{i}$. For instance, the interaction terms $Z_i \cdot X_{i1}$ are often employed as generated instruments or in constructing moments.

However, as discussed in Section \ref{sec:intro}, if there is little variation in $X_{i1}$, the $Z_i \cdot X_{i1}$ interaction terms become less informative. In such cases, traditional GMM methods will fail to provide reliable estimates. For example, if $X_{i1}$ is binary, limited variation means that the majority of observations in $X_{i1}$ are either 0 or 1. With only one valid instrument, GMM encounters the problem of generating new moments when we need to estimate more than one parameter, leading to under-identification. While splitting the sample in this case can solve the problem, it decreases the number of observations, thus resulting in larger standard errors, as discussed in the appendix.

This dilemma is resolved by directly employing the conditional moment restriction. This restriction encapsulates all the information of each instrument, regardless of the number of instruments used. Consequently, this approach allows us to utilize just one instrument, such as the random assignment $Z_i$, to identify and estimate parameters associated with both $W_i$ and $W_i \cdot X_{i1}$. Similar approaches have been employed in the construction of other Bierens-type estimators, such as the Integrated Conditional Moment (ICM) estimators. Notable references include \cite{Bierens1982}, \cite{Antoine2014}, \cite{AS2021}, and \cite{Tsyawo2022}. We will demonstrate that our estimation strategy, relying solely on a single valid instrument (e.g., the random assignment $Z_i$), yields reliable inference for both parameters.

The function $f_{0,1}(.)$ represents the nuisance parameter, which is not of interest in our analysis. The unknown form of $f_{0,1}(.)$ allows $X_i$ to enter the model flexibly. The initial step in constructing our estimator involves employing a Robinson transformation, as described in \cite{Robinsontrans}. This transformation involves subtracting the conditional expectation of $y_i$ with respect to the controls $X_i$. As a result, the term $f_{0,1}(X_i)$ vanishes, eliminating the need to assume any specific form for $f_{0,1}(X_i)$. With the Robinson transformation and the exogeneity assumption on $X_i$, Equation \ref{eq:model01} simplifies to:
$$y_i - E(y_i|X_i) = (P_i-E(P_i|X_i))'\theta_0+ \epsilon_i$$ where $P_i = [W_i, W_i \cdot X_{i1} ']'$ and $\theta_0 = [\theta_{w0},\theta_{wx0}']'$.
Denote
\begin{equation} \label{eqmodel02}
\epsilon_j(\theta,g_0) = \tilde y_j - \tilde P_j' \theta
\end{equation}
with $\tilde y_j \equiv y_j - E(y_j|X_j)$ and $\tilde P_j \equiv P_j-E(P_j|X_j)$. 

This equation also illustrates that $P_j$ may encompass any known function form of the treatment, interactions, covariates, and so forth. The symbol $g_0$ represents all the unknown real-valued functions, i.e., nuisance parameters. At this stage, $g_0$ comprises a vector of $g_{0,y}(X_j)$ (or $E(y_j|X_j)$) and $g_{0,P}(X_j)$ (or $E(P_j|X_j)$). Through this transformation, we exclude $f_{0,1}(X_i)$ from the model and introduce $g_0$. This is also the first step in \cite{AS2021}.

Even though we only have a conditional moment restriction, we can exploit all the information in it by using an infinite number of unconditional moments. This equivalence between a conditional moment restriction and an infinite number of unconditional moment restrictions is from \cite{Bierens1982} and used in constructing SMD-type or ICM-type estimators. Based on Equation (\ref{eqmodel02}) and the equivalence, we have the following results:
\begin{eqnarray}
E[\epsilon_j(\theta_0, g_0) e^{it'Z_j}] = 0 \; \forall t \in \mathbb R^{q_z} \qquad \Longleftrightarrow \qquad E(\epsilon_j(\theta_0, g_0)|Z_j)=0 \ a.s.
\end{eqnarray}
$t$ is a vector of any real number.\footnote{Using higher order Fourier terms, for instance, $E[\epsilon_j(\theta_0, g_0) e^{it^2 Z_j}]$ will lead to implicit choice of $\mu(t)$ and assumptions about $k(.)$ introduced in the later discussion and in practice. We choose to work with $e^{it'Z_j}$ because of its explicit and easy-to-use practice results. This is also the choice of \cite{AS2021}.} Next, we define the population objective function in Equation (\ref{eqboj1}). Inside Equation (\ref{eqboj1}), $\mu(t)$ is a strictly positive measure on the vector $t$.
\begin{eqnarray} \label{eqboj1}
M_{\infty}(\theta, g_0) = \int_{\mathbb R^{q_z}} |E[\epsilon_j(\theta,g_0) e^{it'Z_j}]|^2 d\mu(t)
\end{eqnarray}
The objective function involves the norm of a complex function. To estimate $\theta_0$, we need to find the derivative of the objective function, which is difficult to compute. Under the independent assumption for the population, the objective function has an alternative expression. In the Appendix, we show the equivalence between these two objective functions.
\begin{eqnarray} \label{eqboj2}
\forall j \neq l, \;  \; M_{\infty}(\theta,g_0) = E[\epsilon_j(\theta,g_0) \epsilon_l(\theta,g_0)\kappa_{j,l}] \;  \; \text{with} \;  \;  \kappa_{j,l}= \int_{\mathbb R^{q_z}} e^{it'(Z_j-Z_l)}d\mu(t)
\end{eqnarray}

The objective function defined in Equation (\ref{eqboj2}) depends on the parameters $\theta$ and the nuisance parameters $g_0$, as specified in Equation (\ref{eqmodel02}) after the Robinson transformation. Under the regularity assumptions provided subsequently, $\theta_0$ serves as the unique minimizer of the objective function $M_{\infty}(\theta,g_0)$, where $g_0$ represents a vector of $g_{0,y}(X_j)$ and $g_{0,P}(X_j)$ at this stage. \cite{AS2021} similarly define an objective function incorporating $\kappa_{j,l}$ as the special weighting component from the SMD method, summarizing all the information from the conditional moment. While this may resemble a special case of the general moment considered in \cite{escanciano2023debiased}, we find that with $\kappa_{j,l}$, it is not possible to identify a function and constants satisfying Equation (2.8) of \cite{escanciano2023debiased}. Consequently, the methodology and inference results of \cite{escanciano2023debiased} do not apply. Therefore, we make assumptions and derive theorems tailored to our specific setting.

The FOC of $M_{\infty}(\theta,g_0)$ with respect to $\theta$ is
\begin{eqnarray} \label{eqfoc1}
E[ \tilde P_j (\tilde y_l- \tilde P_l' \theta)\kappa_{j,l}]=0
\end{eqnarray}
When we do not know $g_0$, the FOC of $M_{\infty}(\theta,g)$ is as follows:
\begin{eqnarray} \label{eqfoc0}
E(\left( P_{j} - g_{P}(X_j) \right) [y_l-g_y(X_l) - (P_{l} - g_{P}(X_l))' \theta] \kappa_{j,l})=0
\end{eqnarray}

The FOC defined in Equation (\ref{eqfoc1}) provides an explicit form for $\theta_0$. This FOC extends the framework established by \cite{AS2021} to incorporate interaction terms within the parametric component of the model. Thus, the estimator, derived as the sample analog of $\theta_0$, is a direct extension of \cite{AS2021}. It also represents a specialized instance of the methodology proposed by \cite{LavergnePatilea2013} when the bandwidth within $\kappa_{j,l}$ is held constant.

\begin{asu} (Regularity assumptions) \label{ass:regul 1} \\
\subasu \label{ass:exo} $E(\epsilon_i|X_i, Z_i) = 0$.\\
\subasu \label{ass:inc4} $E(\tilde P_i|Z_i) \neq 0$ a.s. (with probability 1) with $\tilde P_i= P_i- E(P_i|X_i)$. \\
\subasu \label{ass:inc1} $E(\tilde P_i \tilde P_i')$ is nonsingular. \\ 
\subasu \label{ass:m1} Let $f_{Z}(.)$ denote the density function of $Z_j$. We assume that $E(\tilde{P}_{j}|Z_j=.)f_Z(.)$ is $L_q$ for some $1\leq q\leq2$. \\
\subasu \label{ass:indept} $(y_l, W_l, X_l, Z_l)$ is an independent and identical copy of $(y_j, W_j, X_j, Z_j)$ with $l, j \in \{1,..., n\}$. \\
\subasu \label{ass:biereK} Let $\mu$ be a given strictly positive measure on $\mathbb{R}^{q_z}$. Let $k(.)$ be the Fourier inversion integral induced by $\mu$, $k(Z_j-Z_l)=\int_{\mathbb{R}^{q_z}} e^{it'(Z_j-Z_l)} d \mu(t)$. We assume that $k(.)$ is a symmetric bounded density function on $\mathbb{R}^{q_z}$ and that its Fourier transform is strictly positive.
\end{asu}
Assumption \ref{ass:exo} is the exogeneity assumption for the controls and instruments. In the potential outcomes setting, Assumption \ref{ass:exo} is also the Random Assignment Assumption. The model form implies the Exclusion Restriction Assumption as in \citet{imbens_rubin_2015}, Page 558, that is, the value of the instrument does not affect the potential outcomes directly. Assumption \ref{ass:inc4} is the relevant instrument assumption. Assumption \ref{ass:inc1} and the corresponding Assumption in \cite{Robinsontrans} are not the same. This assumption is made in a high-dimensional setting. For not invertible $E(\tilde P_i \tilde P_i')$, see the solution provided in \cite{ChernozhukovVictor2018Dmlf}. 

Assumption \ref{ass:biereK} is for the measure $\mu(.)$. These conditions in Assumption \ref{ass:biereK} are not very restrictive. $\mu$ has a $\mathbb{R}^{q_z}$ support and is symmetric. This leads us to have a symmetric $k(Z_j-Z_l)$. Also, the fourier transform of $k(Z_j-Z_l)$ is $\mu$. Thus, the fourier transform of $k(Z_j-Z_l)$ is strictly positive. There are many different available measures. Measures after Fourier inversion have the respective results. For the simplicity of computation, we focus on those $k(.)$ with explicit formulas. Specifically, we use the CDF of the Gaussian distribution in simulations and applications.

Under Assumption \ref{ass:regul 1}, $\theta_0$ is the unique minimizer of $M_{\infty}(\theta,g)$ when $g=g_0$. Specifically, assumptions \ref{ass:inc1} and \ref{ass:m1} guarantee the identification of the parameters that we are interested in under the FOC defined in Equation (\ref{eqfoc1}). This is established in \cite{AS2021} and is easily extended to our setting.

However, given that the nuisance parameters in $g_0$ are unknown, \cite{AS2021} consider the sample analog of $\theta_0$ defined in Equation (\ref{eqfoc1}) as an infeasible estimator. To obtain a feasible estimator, \cite{AS2021} estimate the nuisance parameters as the first step. They achieve this by estimating their nuisance parameters using Nadaraya-Watson estimators. They demonstrate that, under appropriate assumptions on the bandwidth, the bias introduced by the Nadaraya-Watson estimators on the estimation of the key parameters is sufficiently small, ensuring that the feasible and infeasible estimators of the key parameter share the same asymptotic properties. However, the Nadaraya-Watson estimator imposes a constraint on the number of covariates to satisfy these assumptions. This constraint may not be suitable when there are a large number of available covariates. Moreover, even with a small number of covariates, when $g \neq g_0$, the bias resulting from estimating the nuisance parameters affects the estimation of the key parameters, particularly when the sample size is small. This is because the FOC defined in Equation (\ref{eqfoc0}) is not orthogonal to the nuisance parameter, leading to a more pronounced effect of the bias on the estimate of the key parameter. This is illustrated in the proof of Appendix.

To address this bias issue, we propose a new FOC that extends the Neyman-orthogonal estimator from \cite{ChernozhukovVictor2018Dmlf} to the SMD and U-statistic settings. The Neyman-Orthogonal method constructs an FOC that is orthogonal to the nuisance parameters. This new FOC ensures that all partial derivatives with respect to the nuisance parameters are zero, effectively making it orthogonal to the bias introduced by estimators. The concept of partial derivatives with respect to a function is elucidated in \cite{ChernozhukovVictor2018Dmlf}.

Following Neyman-orthogonalization, the new FOC for $\theta_0$ becomes less sensitive to the bias of the estimator for the nuisance parameters. It is, however, sensitive to the square of the bias. As long as the bias is of the order $o_p(n^{-1/4})$, we still obtain a $\sqrt{n}$-asymptotically normally distributed estimator for the key parameter. It is important to note that under this order, there are numerous estimation methods to choose from, including Lasso, Sieves, and Random Forest, among others.

Our new FOC for $\theta_0$ corresponding to Equation (\ref{eqfoc1}) is
\begin{equation}\label{eq:ofoc01}
E[\Psi(D; \theta, g_0)] \equiv E\left[ \left( \tilde P_j  - \frac{g_{0,\tilde P_{m}}(X_l)}{g_{0,\kappa_{m,l}}(X_l)} \right) [\tilde y_l - \tilde P_l' \theta] \kappa_{j,l} \right]  = 0
\end{equation}
with $g_{0,\tilde P_{m}}(X_l) \equiv E[(P_{m} - g_{0,P}(X_m)) \kappa_{m,l} |X_l]$ and $g_{0,\kappa_{m,l}}(X_l) \equiv E[\kappa_{m,l} |X_l]$. $g_{0,\tilde P_{m}}(X_l)$ has the same dimension as $\tilde P_j$, and $g_{0,\kappa_{m,l}}(X_l)$ is a function of $X_l$: $ \mathbb{R}^{q_X} \to \mathbb{R}$.

$g_{0,\tilde P_{m}}(X_l)$ and $g_{0,\kappa_{m,l}}(X_l)$ are two additional parameters inside the nuisance parameter vector. The FOC defined in Equation (\ref{eq:ofoc01}) has a partial derivative with respect to all nuisance parameters equal to zero. This is shown in the Appendix.

\begin{asu} \label{ass:identification}
The singular values of the matrix $E\left[ \kappa_{j,l} \left( \tilde P_j  - g_{0,\tilde P_{m}}(X_l)g^{-1}_{0,\kappa_{m,l}}(X_l) \right) \tilde P_l' \right]$ are between $c_0$ and $c_1$ with $0<c_0<c_1$. 
\end{asu}

Assumption \ref{ass:identification} is the identification assumption. This is the similar assumption as Assumption 3.1 (e) in \cite{ChernozhukovVictor2018Dmlf}. As a contrary, \cite{Chernozhukov2022} and \cite{escanciano2023debiased} do not impose the identification assumption, because the structure of the $\Psi(.; \theta, g)$ of their works guarantee the identification. Here, due to the fact that our $\Psi(D; \theta, g)$ is constructed under the SMD method with the special weight $\kappa_{j,l}$, we need to assume additional identification assumption as in \cite{ChernozhukovVictor2018Dmlf}. Additionally, this assumption can be examined. We show this detailed information in the identification section of the Appendix, Section \ref{section:ident}\footnote{In the Appendix, we illustrate the identification issue for the group of SMD estimators, such as SMD, RSMD, and D-RSMD. We consider cases involving various types of instruments and models.}. In this section, we show that with the information of the true Data Generating Process (DGP), the determinant of the matrix $E\left[ \kappa_{j,l} \left( \tilde P_j  - g_{0,\tilde P_{m}}(X_l)g^{-1}_{0,\kappa_{m,l}}(X_l) \right) \tilde P_l' \right]$ is deducible. Also, in the applied setting, if the matrix is singular, the resulting standard error of the estimator will be huge and abnormal. This provides an intuitive way to determine whether the matrix is invertible.

Equation (\ref{eq:ofoc01}) gives us the identification of the true key parameter and the explicit forms of the infeasible and feasible estimators.

\begin{prop}\label{prop:iden} (Identification of $\theta_0$ using the orthogonalized FOC) \\
Under Assumptions \ref{ass:regul 1}-\ref{ass:identification} and FOC defined in Equation (\ref{eq:ofoc01})
$$\theta^*_0 = E\left[\kappa_{j,l}\left( \tilde P_j - \frac{g_{0,\tilde P_{m}}(X_l)}{g_{0,\kappa_{m,l}}(X_l)} \right)\tilde P_l' \right]^{-1} E\left[\kappa_{j,l}\left( \tilde P_j - \frac{g_{0,\tilde P_{m}}(X_l)}{g_{0,\kappa_{m,l}}(X_l)} \right) \tilde y_l \right]$$
\end{prop}

If we plug Equation (\ref{eqmodel02}) into the formula for $\theta^*_0$ with error term $\epsilon_l$, we will have $\theta^*_0= \theta_0$. Thus, the sample analog of $\theta^*_0$ delivers an estimator for $\theta_0$. Because we have two distinct individuals inside the expectation (e.g., j and l), we need to replace the expectation by the average of a double summation to obtain the infeasible estimator under Equation (\ref{eq:ofoc01}). The closed-form expression for the infeasible estimator, $\tilde \theta_{n,o}$, is:
\begin{equation*}
\tilde \theta_{n,o}= \left[\sum_{j = 1}^n \sum_{l \neq j}^n \kappa_{j,l}\left( \tilde P_j - \frac{g_{0,\tilde P_{m}}(X_l)}{g_{0,\kappa_{m,l}}(X_l)} \right)\tilde P_l' \right]^{-1} \left[\sum_{j = 1}^n \sum_{l \neq j}^n \kappa_{j,l}\left( \tilde P_j - \frac{g_{0,\tilde P_{m}}(X_l)}{g_{0,\kappa_{m,l}}(X_l)} \right) \tilde y_l \right]
\end{equation*}

The infeasible estimator $\tilde \theta_{n,o}$ depends on unknown nuisance parameters: $g_{0,\tilde P_{m}}(X_l)$, $g_{0,\kappa_{m,l}}(X_l)$, $E(y_l|X_l)$, and $E(P_l|X_l)$. All of the nuisance parameters are conditional expectation functions on covariates. Hence, in practice, we need to find estimators for these conditional expectation functions to obtain a feasible estimator. Replacing every nuisance parameter $g_0$ with estimators $\hat g$ such that $\hat g$ converges to $g_0$ at a rate of $o_p(n^{-1/4})$ will deliver the feasible estimator on $\theta_0$ with $\sqrt{n}$ asymptotic normality.

The nuisance parameters are functions of $X$ involving $q_X$ variables. If $q_X$ is very large, such as $q_X \geq n$, there will be overfitting. To mitigate this, we need to reduce the number of variables included in the regression. Using Lasso will be one of the solutions. In order to apply Lasso, we assume sparsity for the nuisance parameters; that is, the conditional means can be described with only a few non-zero parameters in front of $X$. The number of non-zero parameters, $s_0$, is allowed to grow at the rate of $o_p(n^{1/2}/log(q_X))$. Under the assumed rate for $s_0$, the Lasso estimation has the following property (as Equation (2.1) of Chapter 2 in \citet{vandeGeerSaraEaTU}) on the order of mean square error: $$||X(\hat \beta - \beta_0)||_2^2/n = \mathcal{O}_p \left(\frac{s_0 log(q_X)}{n} \right) $$ where $\hat \beta$ is the Lasso estimator (\cite{ChernozhukovVictor2018Dmlf} and \citet{vandeGeerSaraEaTU}), and $||.||_2$ is the $\mathcal{L}_2$ norm. When $3<q_X <n$, we can still use the Lasso method with a higher assumed rate for $s_0$ to select variables. If $q_X \leq 3$, we can still use the Nadaraya-Watson estimator to estimate the nuisance parameter with a second-degree kernel. It is worth noting that $q_X \leq 3$ is a constraint that is seldom realistic in practice. Nevertheless, our estimation procedure allows us to accommodate larger values of $q_X$. Upon replacing every nuisance parameter with its estimate, we obtain the feasible estimator $\hat \theta_{n,o}$, referred to as the D-RSMD estimator.
\begin{equation}\label{eq:fdrsmd}
\hat \theta_{n,o}= \left[\sum_{j = 1}^n \sum_{l \neq j}^n \kappa_{j,l} \left(\widehat{\tilde{P}}_j - \frac{\widehat{g_{0,\tilde P_{m}}}(X_l)}{\widehat{g_{0,\kappa_{m,l}}}(X_l)} \right) \widehat{\tilde{P_l}}'\right]^{-1}  \left[\sum_{j = 1}^n \sum_{l \neq j}^n \kappa_{j,l} \left(\widehat{\tilde{P}}_j - \frac{\widehat{g_{0,\tilde P_{m}}}(X_l)}{\widehat{g_{0,\kappa_{m,l}}}(X_l)} \right) \widehat{\tilde{y}}_l\right]
\end{equation}
since $E[\kappa_{j,l} \left(\widehat{\tilde{P}}_j - \frac{\widehat{g_{0,\tilde P_{m}}}(X_l)}{\widehat{g_{0,\kappa_{m,l}}}(X_l)} \right) \widehat{\tilde{P_l}}']$ is invertible.

This leads to the algorithm of our D-RSMD estimation procedure.
\begin{algorithm}(Implementation of the D-RSMD estimator)
\begin{enumerate}
\item Conduct the Robinson transformation. This step involves estimating the conditional means inside $\widehat{\tilde{P}}_j$ and $\widehat{\tilde{y}}_l$. Any estimator $\hat g$ that converges to $g_0$ at a rate of $o_p(n^{-1/4})$ can be applied.
\item Calculate $\kappa_{j,l}$. Within $\kappa_{j,l}$, $\mu(.)$ represents the CDF of the Gaussian distribution. Given that the Fourier transform of the Gaussian distribution is also Gaussian, computing $\kappa_{j,l}$ is straightforward.
\item Estimate the nuisance parameters $\widehat{g_{0,\tilde P_{m}}}(X_l)$ and $\widehat{g_{0,\kappa_{m,l}}}(X_l)$ within the orthogonal FOC.
\item Calculate the estimate based on Equation (\ref{eq:fdrsmd}) for $\hat \theta_{n,o}$.
\end{enumerate}
\end{algorithm}

All of the nuisance parameters for the D-RSMD estimator in the simulation and empirical application are estimated by the Lasso method with cross validation. The maximum degree of the polynomial for the nuisance parameters is 5, which guarantees that the nuisance parameters can be approximated by 5 degree polynomials in all controls. Lasso with cross validation helps us select the controls and their polynomials.

\section{Large Sample Theory} \label{section:large sample theory}

References such as \cite{LavergnePatilea2013} provide a general framework for analyzing the asymptotic properties of SMD estimators, even in cases where explicit forms of these estimators are not available. Building upon this work, \cite{AS2021} extend the methodology to accommodate semiparametric models and derive the asymptotic properties specific to the RSMD estimator. In this section, we establish the asymptotic properties of the D-RSMD estimator, demonstrating that the feasible estimator $\hat \theta_{n,o}$ and the infeasible estimator $\tilde \theta_{n,o}$ exhibit equivalent asymptotic behavior. The properties of the infeasible estimator $\tilde \theta_{n,o}$ are outlined  as follows:

\begin{prop} (Consistency and Asymptotic normality of $\tilde \theta_{n,o}$) \label{thm:infeas consistency normality} \\
Under Assumption \ref{ass:regul 1} and iid assumption for the sample, $\tilde \theta_{n,o}$ is consistent for $\theta_0$, that is $ \tilde \theta_{n,o} \stackrel{p}{\rightarrow} \theta_0 $, and asymptotically normally distributed,
$$\sqrt{n}(\tilde \theta_{n,o} - \theta_0) \xrightarrow{d} N \left(0, A^{-1}\Sigma \left(A^{-1} \right)' \right)$$
where $A = E\left[\kappa_{j,l}\left( \tilde P_j - \frac{g_{0,\tilde P_{m}}(X_l)}{g_{0,\kappa_{m,l}}(X_l)} \right)\tilde P_l'\right]$, and $\Sigma$ stands for the variance-covariance matrix involving terms of $(\tilde P_i,\epsilon_i,Z_i,X_i)$, which is defined in the proof section of the Appendix.
\end{prop}
The asymptotic properties of $\tilde \theta_{n,o}$ are based on the corresponding properties of U-statistics. This is thoroughly discussed in \cite{Hoeffding1948}, particularly in Theorem 7.1 on Page 320 of the book. 
\begin{asu} \label{ass:ad1}
$\hat g$ converges to $g_0$ at a rate of $o_p(n^{-1/4})$. For the Lasso method, the number of non-zero parameters $s_0$ grows at the rate of $o_p(n^{1/2}/log(q_X))$. For the Nadaraya-Watson estimator, $\sqrt n \left( \sum_{s=1}^{q_z} h_s^4+ \left[\frac{1}{n h_1...h_{q_z}} \right] \right) = o(1)$ where $h$ is the bandwidth.
\end{asu}
\begin{theorem} (Consistency and Asymptotic normality of the D-RSMD estimator: $\hat \theta_{n,o}$) \label{thm:consis asymp norm Debias RSMD}
Under Assumptions \ref{ass:regul 1} - \ref{ass:ad1}, $\hat \theta_{n,o}$ is consistent, and has an asymptotically normal distribution, that is,
$$\sqrt{n}(\hat \theta_{n,o} - \theta_0) \xrightarrow{d} N \left(0, A^{-1} \Sigma \left(A^{-1} \right)' \right),$$ 
\end{theorem}
Notably, the asymptotic variances of the D-RSMD and R-SMD estimators proposed by \cite{AS2021} differ, primarily due to additional terms introduced through the debiasing process in $A$ and $\Sigma$. Analytical comparison of these asymptotic variances is challenging, given the complexity introduced by these extra terms. Furthermore, we provide the explicit form of the estimator for variance under heteroskedasticity in the following:
\footnotesize
\begin{eqnarray} \label{eq: HAC estim}
[n(n-1) C_n]^{-1} \sum^n_{j=1}((\sum^n_{l=1}\kappa_{j,l} \left(\widehat{\tilde{P}}_l - \frac{\widehat{g_{0,\tilde P_{l}}}(X_j)}{\widehat{g_{0,\kappa_{m,j}}}(X_j)} \right))(\sum^n_{l=1}\kappa_{j,l} \left(\widehat{\tilde{P}}_l - \frac{\widehat{g_{0,\tilde P_{l}}}(X_j)}{\widehat{g_{0,\kappa_{m,j}}}(X_j)} \right)')\hat \epsilon_j^2)[n(n-1) C_n']^{-1}
\end{eqnarray}
\normalsize
with $C_n =\frac{1}{n(n-1)} \sum_{j = 1}^n \sum_{l \neq j}^n \kappa_{j,l} \left(\widehat{\tilde{P}}_j - \frac{\widehat{g_{0,\tilde P_{m}}}(X_l)}{\widehat{g_{0,\kappa_{m,l}}}(X_l)} \right) \widehat{\tilde{P_l}}'$.

Based on Theorem \ref{thm:consis asymp norm Debias RSMD}, the feasible and infeasible estimators exhibit the same asymptotic distribution and demonstrate consistency. This is from Assumption \ref{ass:ad1} on the convergence rate, which ensures that the bias introduced during nuisance parameter estimation has negligible impact on subsequent steps. For example, when the number of covariates ($q_X$) and the sample size are small, the D-RSMD estimator exhibits less bias compared to the R-SMD estimator, especially when both use the same Nadaraya-Watson estimator for nuisance parameter estimation. In \cite{ChernozhukovVictor2018Dmlf}, simulations illustrate a comparison between Neyman orthogonal estimators and non-orthogonal estimators. Further details on bias analysis are presented in the subsequent section, specifically in the simulation section.

\section{Simulation Study} \label{section:simulation}

We consider a partially linear model with one endogenous variable and its interaction in the linear part.
\begin{eqnarray}
y_i &=& \theta_{w0} W_i + W_i \cdot X_{i1} \theta_{wx0} + \sum_{q=1}^{q_X}  \beta_{1q} X_{qi} +  \sum_{q=1}^{q_X} \beta_{2q} X_{qi}^2 +\epsilon_i \label{eq:jmpsim design}\\
W_i &=& I(\theta_{z0} Z_i + \theta_{z3}  Z_i ^3 + \sum_{q=1}^{q_X} \alpha_{1q} X_{qi} + \sum_{q=1}^{q_X} \alpha_{3q}  X_{qi}^3> -v_i)\nonumber
\end{eqnarray}
$I(.)$ is the indicator function. It will take the value one if the statement inside is true and zero otherwise. In the simulations, the key parameters $\theta_{w0}$ and $\theta_{wx0}$ are 2 and 3 respectively. The other parameters are in the following. $\theta_{z0} = 3$, $\theta_{z3} = 4$, $\alpha_{3q} =2$, $\beta_{2q} =-3$, $\alpha_{1q}=\beta_{1q} = 1$ for $q \leq S$ with $S$ the number of non-zero parameters, and $q_X$ ($q_X = 30$) the number of covariates. There are nonlinear terms in the model, so $S$ is chosen to be a half of $s_0$, the sparsity level. $\alpha_{1q} = \alpha_{3q}  = \beta_{1q} = \beta_{2q} = 0$ when $q > S$. Here, $S = 5$. If $\beta_{2q}$ is 0 for all $q$, $y_i$ is linear in $X_i$. Otherwise, the model is partially linear.
 
We use a binary instrument to create a treatment variable and explore simulation outcomes. The results when we use a categorical instrument in the DGP are in the appendix. $Z_i$ is the instrument with $E(Z_i) = 0.318$. The covariate $X_i$ is modeled as $X_i^* + 0.4Z_i$, where $X_i^*$ follows a multivariate standard normal distribution. This implies a correlation between $X_i$ and $Z_i$. The errors $(\epsilon, v)$ are bivariate normally distributed with mean 0, variance 1, and covariance $\frac{4}{9}$. Consequently, the treatment variable $W_i$ is endogenous.

To assess the performance of our D-RSMD estimators, we conduct 5,000 Monte Carlo replications for each scenario, where $q_X \in \{3, 30\}$ and the sample size $n \in \{3000, 5000\}$. Our benchmark scenario involves 3000 observations and 30 control variables, which mirrors our empirical example. For instance, when examining the Oregon Health Insurance Experiment, we segment the dataset into three age-based groups, each containing around 6,000 observations and 21 covariates.

We report and compare simulation results for several estimators in our study. These include the D-RSMD estimator proposed in this paper (referred to as DRSMD-Lasso in Section \ref{section:application}), where we use the Lasso method for estimating nuisance parameters. We also consider the R-SMD estimator by \citet{AS2021} using Lasso, the R-GMM estimator combining Robinson Transformation with GMM (GMM-Lasso in Section \ref{section:application}), a GMM estimator treating $f_{0,1}(X_i)$ as linear in $X_i$, and a GMM (Oracle) estimator utilizing the true $f_{0,1}(X_i)$. We present results for the D-RSMD estimator with one and two instruments, along with results for the R-SMD estimator under both conditions. GMM-type estimators require at least two instruments to estimate two parameters. We list the available instrument sets used for estimation in the third column of each table.

All nuisance parameters for the D-RSMD estimator in both the simulation and empirical application sections are estimated using the Lasso method with cross-validation. We employ a maximum polynomial degree of 5 for the nuisance parameters. This involves generating 5-degree polynomials for each covariate and using the Lasso method with cross-validation to select relevant controls and their respective polynomials, resulting in predicted conditional expectations. This approach provides a non-parametric estimation similar to the sieve method by selecting polynomial degrees. Cross-validation is utilized to select the penalty level and mitigate the risk of overfitting. For RSMD and RGMM estimators with $q_X = 30$, we also adopt the same Lasso method for polynomial selection to ensure comparability across estimators.

\begin{table}[!ht]
\begin{center}
\scalebox{.75}{
\begin{tabular}{ l l c | c c c c | c c c c}
\hline \hline
\multicolumn{3}{c|}{\textbf{}} & \multicolumn{4}{c|}{\textbf{$\theta_{w0}$}} & \multicolumn{4}{c}{\textbf{$\theta_{wx0}$}} \\ \hline
 &\textit{Estimator} & \textit{Instrument} & Med.Bias  & MAD & Med.SE & RR & Med.Bias  & MAD & Med.SE & RR \\ \hline
    $n=3,000$ & D-RSMD  & $Z_1$      & -0.006  & 0.146 & 0.215 & 0.048 & -0.005  & 0.039 & 0.057 & 0.052 \\
    $q_X=3$ & D-RSMD  & $Z_2$         & -0.004  & 0.212 & 0.314 & 0.043 & -0.004  & 0.110 & 0.182 & 0.050 \\
    & D-RSMD & $(Z_1, Z_1 X_{1})$   & -0.026  & 0.164 & 0.247 & 0.048 & -0.004  & 0.124 & 0.186 & 0.047 \\
    & D-RSMD & $(Z_2, Z_2 X_{1})$   & -0.035  & 0.230 & 0.340 & 0.049 & -0.001  & 0.200 & 0.301 & 0.040 \\
& GMM (Oracle) & $(Z_1, Z_1 X_{1})$ & -0.004  & 0.164 & 0.250 & 0.048 & -0.001  & 0.126 & 0.191 & 0.047 \\ \hline
    $n=3,000$ & D-RSMD  & $Z_1$      &  0.028  & 0.288 & 0.427 & 0.045 & -0.012  & 0.045 & 0.059 &  0.064\\
    $q_X=30$   & D-RSMD  & $Z_2$      &  0.033  & 0.524 & 0.773 & 0.030 & -0.012  & 0.126 & 0.206 &  0.037 \\
    &  D-RSMD & $(Z_1, Z_1 X_{1})$  & -0.259  & 0.291 & 0.431 & 0.082 &  0.005  & 0.111 & 0.163 &  0.049 \\
    & D-RSMD  & $(Z_1, Z_2)$        &  0.041  & 0.333 & 0.493 & 0.041 & -0.013  & 0.181 & 0.287 &  0.039 \\
    & RSMD  & $Z_1$                 &  0.128  & 0.408 & 0.563 & 0.072 & -0.020  & 0.053 & 0.065 &  0.105 \\
    & RSMD  & $Z_2$                 &  0.154  & 0.555 & 0.790 & 0.039 & -0.024  & 0.147 & 0.302 &  0.018 \\
    & RSMD & $(Z_1, Z_1X_{1})$      &  0.230  & 0.463 & 0.572 & 0.110 & -0.337  & 0.137 & 0.273 &  0.144  \\
    & RSMD  & $(Z_1, Z_2)$          &  0.140  & 0.438 & 0.605 & 0.060 & -0.020  & 0.195 & 0.328 &  0.033 \\
    & GMM & $(Z_1, Z_1 X_{1})$      &  7.471  & 2.968 & 4.368 & 0.356 & -6.735  & 0.953 & 1.430 &  0.999\\
    & GMM & $(Z_1, Z_2)$            &  6.857  & 5.691 & 11.948& 0.043 & -4.444  & 21.525& 47.007&  0.000 \\
& GMM (Oracle) & $(Z_1, Z_1 X_{1})$ &  0.007  & 0.266 & 0.392 & 0.048 & -0.004  &  0.107& 0.157 &  0.049  \\
& GMM (Oracle) & $(Z_1, Z_2)$       & -0.011  & 0.635 & 1.424 & 0.003 &  0.114  & 2.502 & 5.548 &  0.000  \\ \hline
  $n=5,000$ & D-RSMD  & $Z_1$        &  0.003  & 0.209 & 0.314 & 0.052 & -0.005  & 0.029 & 0.040 &  0.062\\
$q_X=30$ & D-RSMD & $(Z_1, Z_1 X_{1})$& -0.158  & 0.210 & 0.308 & 0.083 &  0.006  & 0.078 & 0.120 &  0.045 \\
& GMM (Oracle) & $(Z_1, Z_1 X_{1})$ &  0.001  & 0.203 & 0.304 & 0.051 &  0.002  & 0.080 & 0.121 &  0.045 \\ \hline
\hline
\end{tabular}
}
\caption{Performance of Estimators with $Z_1$ as the Instrument in the DGP}
\label{sim:iv2table1}
\end{center}
\par \footnotesize Note: Simulation Results for $\theta_{w0}$ and $\theta_{wx0}$ in the benchmark model using D-RSMD estimator 5,000 replications. We report the Monte-Carlo Median Bias (Med.Bias), Median Absolute Deviation (MAD), median of asymptotic standard error under heteroskedasticity (Med.SE), and Rejection Rate (RR) using a 5\% t-test. 
$Z_1$: binary instrument.
 $Z_2$: variable representing the sum of $Z_1$ and another binary variable, resulting in three possible values for $Z_2$.
$(Z_1, Z_1X_{1})$: pair consisting of $Z_1$ and the interaction term $Z_1X_{1}$ (where $X_{1}$ is a covariate).
$(Z_2, Z_2 X_{1})$: pair comprising $Z_2$ and the interaction term $Z_2X_{1}$.
$(Z_1, Z_2)$: pair including both $Z_1$ and $Z_2$. The correlation between $Z_1$ and $Z_2$ is around 0.7.
\end{table}

We present the results in Table \ref{sim:iv2table1}. Firstly, focusing on the first panel with $q_X = 3$, we observe that D-RSMD using $Z_1$ achieves the lowest MAD and Med.SE compared to all estimators. The RRs are close to 5\% for $\theta_{w0}$ and slightly oversized for $\theta_{wx0}$. Using the instrument $Z_2$ (not in the DGP) for D-RSMD results in higher MAD and Med.SE compared to utilizing the correct instrument. The D-RSMD estimator also performs well when employing $(Z_1, Z_1 X_{1})$ and $(Z_2, Z_2 X_{1})$. Comparing D-RSMD estimators, those using $Z_1$ or the pair of instruments including the correct instrument tend to yield lower MADs and Med.SEs, which is reasonable because working with the correct variable increases estimation precision. The Med.Bias column indicates that for $\theta_{w0}$, D-RSMD estimators using $(Z_1, Z_1 X_{1})$ and $(Z_2, Z_2 X_{1})$ exhibit higher bias than other D-RSMD estimators, suggesting that using $Z_1$ leads to lower bias in this DGP. This property of D-RSMD also holds when there are 30 controls in the model. However, using one instrument may not always bring a lower bias. This is shown in the appendix when $Z_2$ is the instrument used in the DGP.

The second panel of Table \ref{sim:iv2table1} presents results for $q_X = 30$. Similar to the scenario with $q_X = 3$, among D-RSMD estimators, using $Z_1$ as the instrument yields better results, as evidenced by lower Med.Bias, MAD, and Med.SE values. Comparing the results obtained with $Z_1$ versus $Z_2$, we notice a slight distortion in the RR when $Z_1$ is employed. However, this size distortion diminishes when using a sample size of 5,000 observations per replication, as shown in the last panel of the table. Comparing D-RSMD with RSMD underscores the advantage of D-RSMD in producing less biased estimates, particularly when employing the Lasso method for nuisance parameter estimation.

Furthermore, comparing D-RSMD results with the GMM (Oracle) estimator reveals that D-RSMD estimates using $Z_1$ closely approximate the GMM (Oracle) estimates for $\theta_{w0}$ in terms of MAD, Med.SE, and RR. For $\theta_{wx0}$, D-RSMD results with $Z_1$ outperform the GMM (Oracle) estimator using $(Z_1, Z_1 X_{1})$, exhibiting lower MAD and Med.SE values. These findings suggest that D-RSMD with $Z_1$ achieves satisfactory performance due to its utilization of an infinite number of unconditional moments, equivalent to the conditional moment, unlike the GMM (Oracle) estimator, which relies solely on two unconditional moments.

Analyzing outcomes from instrument sets $(Z_1, Z_1 X_{1})$ and $(Z_1, Z_2)$ reveals notable differences between D-RSMD and GMM (Oracle) estimators. Specifically, for GMM (Oracle) estimators, $(Z_1, Z_1 X_{1})$ emerges as the preferred choice, exhibiting the lowest Med.Bias for both $\theta_{w0}$ and $\theta_{wx0}$. Conversely, GMM estimators using $(Z_1, Z_2)$ demonstrate scattered results and numerous extreme cases, with the RR approaching 0 for $\theta_{wx0}$, highlighting the challenges posed by instrument correlation within $(Z_1, Z_2)$ for GMM type estimators.

In contrast, the choice of instrument sets has a smaller impact on D-RSMD results compared to its effect on GMM (Oracle) estimators. This observation suggests that D-RSMD exhibits more stable performance when utilizing at least one valid instrument or an instrument closely related to the valid instrument in the instrument set, compared with GMM-type estimators.

\section{Empirical Application}  \label{section:application}

In this section, we investigate the heterogeneous treatment effects of Medicaid enrollment from a Randomized Controlled Trial (RCT) conducted in Oregon. In early 2008, Oregon expanded Medicaid enrollment through a lottery system for low-income, uninsured adults. This lottery provided random selection, enabling researchers to analyze the effects of health insurance on various outcomes, including medical, financial, and labour market impacts. This experiment has been extensively studied in various articles, such as \citet{Baicker2014} and \citet{Finkelstein2012}. Detailed information and datasets related to this RCT are available on the public website.\footnote{See \url{https://www.nber.org/research/data/oregon-health-insurance-experiment-data}} In this paper, we utilize a dataset derived from the original datasets.\footnote{Special thanks to A. Colin Cameron for providing the dataset and bringing the Oregon Health Insurance Experiment to my attention through his and Pravin K. Trivedi's book in 2022. The dataset is also accessible at \url{https://www.stata-press.com/data/mus2.html}} The dataset comprises 18,572 observations.

Many studies have focused on the homogeneous treatment effects of Medicaid. \citet{Finkelstein2012} assumed homogeneity in their analysis but also explored the potential for heterogeneous treatment effects through regression analysis. Estimates from 2SLS rule out the possibility of heterogeneous treatment effects. However, this may be due to the misspecification of the first stage or problems with the generated instrument variables.

To address this problem, we employ the D-RSMD estimator to estimate heterogeneous treatment effects. We compare the results between the new and traditional estimation methods, both with and without the generated instrument variable, to obtain reliable inference.

Here, the heterogeneity arises from interaction terms involving indicators ($X_{1}$) for household income above 50\% of the federal poverty line in 2008 (income), household income (hhincome), TANF (cash welfare assistance to low-income families), or cigarette smoking level (smoke). The lottery serves as the instrumental variable. We adopt similar controls to those used by \citet{Baicker2014} and \citet{Finkelstein2012}, including household controls, lottery and survey wave indicators, and individual characteristics. The dependent variables include the current employment indicator, constructed from three indicators: hours of employment (employment), total out-of-pocket spending on medical care (out-of-pocket cost), and whether the individual currently owes money to a healthcare provider (debt for health). All dependent variables are derived from a mail survey conducted between July 2009 and March 2010, approximately one year after treatment.

Given that health conditions, employment status, and other outcomes are closely linked to individuals' ages, we identify age as a key source of heterogeneity. We derive an age group variable from the year of birth data. First, we calculate the age of each individual, resulting in ages ranging from 21 to 64 years. Next, we partition the original dataset into three or five subsets based on age to create the age-group variable. Each subsample contains approximately 5900 to 6900 observations. Individuals within the same age group share more similarities. In Section \ref{app2:results}, we present results categorized into three age groups (e.g., 21 to 35, 35 to 50, and 50 to 64). Here, the `agegroup` variable represents a categorical variable with three values, mirroring the age groups.

To provide a clearer comparison, we contrast the D-RSMD estimator with traditional estimators, such as the GMM estimator. The D-RSMD estimator is designed to cope with a nonparametric first stage and a partially linear second stage. The framework for the D-RSMD estimator is outlined as follows:
\begin{align*}
Debt_i &= \theta_{w0} Medicaid_i  +  \theta_{wx0} Medicaid_i \times income_{i} + \theta_{x10} income_{i}\\
&+ \theta_{x20} agegroup_i + \theta_{x30}  Medicaid_i \times agegroup_i  + f_{0,1}(X_i)+ \epsilon_i \\
Medicaid_i &= I(f_{0,2}(X_i, Lottery_i) > v_i )
\end{align*}

The framework for GMM estimators includes a linear first stage within the indicator function as well as a linear second stage for the dependent variable.
\begin{align*}
Debt_i &= \theta_{w0} Medicaid_i  +  \theta_{wx0} Medicaid_i \times income_{i} + \theta_{x10} income_{i}\\
 &+ \theta_{x20} agegroup_i + \theta_{x30}  Medicaid_i \times agegroup_i  +  X_i' \beta_{x}+ \epsilon_i \\
Medicaid_i &= I(\alpha_{z} Lottery_i + X_i' \alpha_{x} > v_i )
\end{align*}

The model for GMM-Lasso estimators includes a linear first stage for the indicator function and a partially linear second stage for the dependent variable.
\begin{align*}
Debt_i &= \theta_{w0} Medicaid_i  +  \theta_{wx0} Medicaid_i \times income_{i} + \theta_{x10} income_{i}\\
 &+ \theta_{x20} agegroup_i + \theta_{x30}  Medicaid_i \times agegroup_i  +  f_{0,1}(X_i)+ \epsilon_i \\
Medicaid_i &= I(\alpha_{z} Lottery_i + X_i' \alpha_{x} > v_i )
\end{align*}

In Table \ref{app2:table1}, we present preliminary results for three estimators of $\theta$. D-RSMD stands out as the sole estimator utilizing the lottery for estimating multiple parameters. We report D-RSMD results in two columns: using only the lottery ($Z_1$) and incorporating the lottery and its interaction term $(Z_1, Z_1 X_{1})$. This setup allows for a direct comparison of results across the two instrument sets and three estimators.

Using the DRSMD-Lasso estimator with the lottery as the sole instrument ($Z_1$), we obtain statistically significant estimates for Medicaid and the interaction between Medicaid and income at a 5\% significance level. Similarly, we observe statistically significant estimates for income and the interaction term between Medicaid and age at a 15\% significance level with the lottery. These findings corroborate the conclusions drawn in Section \ref{app2:results}, suggesting heterogeneous treatment effects in Medicaid related to age and income when employing the valid instrument alone. Conversely, analyzing outcomes using the GMM estimator or GMM-Lasso with the instrument set $(Z_1, Z_1 X_{1})$ would erroneously lead to the conclusion of no heterogeneity in treatment effects, underscoring the problems associated with using generated instruments or moments.

\begin{table}[!ht]
\begin{center}
\scalebox{.8}{
\begin{tabular}{l | c c | c c | c c  }
\hline \hline
\multicolumn{7}{l}{\textbf{Heterogeneous treatment effects}} \\ \hline
\textit{Debt for Health} & \multicolumn{2}{c|}{GMM}  & \multicolumn{2}{c|}{GMM-Lasso} &  \multicolumn{2}{c}{DRSMD-Lasso}\\ \hline
\textit{Variables}& $Z_1$  & $(Z_1, Z_1 X_{1})$ & $Z_1$  & $(Z_1, Z_1 X_{1})$ & $Z_1$  & $(Z_1, Z_1 X_{1})$ \\ \hline
\multirow{2}{*}{Medicaid}         &   & -0.148***  &   & -0.206*  & -0.187***  & -0.178*** \\
                                  &   & (0.044) &   & (0.115) & (0.037) & (0.046) \\ 
\multirow{2}{*}{Medicaid*income}  &   &  0.003 &   &  -0.045 &  0.074*** &  -0.018\\
                                  &   & (0.053) &   & (0.068) & (0.020) & (0.056) \\ 
\multirow{2}{*}{income}           &   &  0.003 &   &  0.014 &  -0.025. &  0.005\\
                                  &   & (0.020) &   & (0.062) & (0.016) & (0.025) \\
\multirow{2}{*}{agegroup}         &   &  -0.007 &   &  -0.010 &  -0.006 &  -0.007\\
                                  &   & (0.010) &   & (0.021) & (0.005) & (0.009) \\ 
\multirow{2}{*}{Medicaid*agegroup}&   &  -0.019 &   &  -0.016 &  -0.018. &  -0.016\\
                                  &   & (0.032) &   & (0.054) & (0.011) & (0.028) \\ 
\hline \hline
\end{tabular}
}
\caption{Heterogeneous Treatment Effects for Robustness Check (5 parameters)}
\label{app2:table1}
\end{center}
\par \footnotesize Note: *** Significant at 1\%, ** at 5\%, * at 10\%, . at 15\%.
\end{table}

\subsection{Results} \label{app2:results}

In this section, we examine the potential heterogeneous treatment effects of Medicaid on health-related debt, with a focus on household income as a source of heterogeneity. We partition the dataset into three subsets based on age to facilitate intuitive interpretation. The complete table presenting all treatment effects is provided at the end of this section. Additional results for robustness checks are in Appendix Section \ref{app2:robust}.

We consider two sets of instruments: the lottery (used exclusively for D-RSMD) and the pair of the lottery along with its interaction with $X_1$ (indicating income greater than 50\% of the poverty line). To illustrate, the model for debt is specified as:
\begin{eqnarray*}
Debt_i &=& \theta_{w0} Medicaid_i +  \theta_{wx0} Medicaid_i \times X_{i1} + f_{0,1}(X_i) + \epsilon_i
\end{eqnarray*}

Table \ref{app2:table16} presents results for three age groups: 21–35, 35–50, and 50–64 years old. For the age group 35–60, the total number of observations is 6693. Heterogeneous treatment effects are observed among individuals. The estimates obtained by D-RSMD (using the lottery as an instrument or the lottery and its interaction with $X_1$ as instruments) are very similar to each other and to the GMM estimator (using the lottery and its interaction with $X_1$ as instruments). This indicates a strong and reliable interaction between the lottery and $X_1$. Notably, only the D-RSMD estimation procedure reveals statistically significant heterogeneous effects. The age group exhibits significant results for both $\theta_{w0}$ and $\theta_{wx0}$ when employing our new estimator with the valid lottery instrument, suggesting that the effect of Medicaid on debt for health varies depending on income level, thus indicating heterogeneity.

For the age group 35–60, the interpretation of the D-RSMD estimate (using $Z_1$ or $(Z_1, Z_1 X_{1})$) for $\theta_{w0}$ is that for people with an income 50\% below the federal poverty line, Medicaid enrollment decreases their probability of owing money to health providers by 23.2 log points on average, ceteris paribus. For people with an income above 50\% of the federal poverty line, Medicaid enrollment reduces their probability of owing money, but not by that much. This is reasonable because enrollment in Medicaid is not that critical to reducing debt for individuals with higher incomes compared with those with lower incomes. 

Now focus on the $\theta_{wx0}$. Comparing the results of D-RSMD between the two instrument sets ($Z_1$ or $(Z_1, Z_1 X_{1})$), we find that the estimates are close, but the standard errors using $Z_1$ are substantially lower, suggesting that using one valid instrument provides a different result for statistical significance.

For the age group 21–35, all estimators generate similar results using $(Z_1, Z_1 X_{1})$. The coefficients for $\theta_{w0}$ are not statistically significant, and the ones for $\theta_{wx0}$ are statistically significant. It suggests that when households' incomes are above the 50\% federal poverty line, individuals with Medicaid will be less likely to owe money to their health providers. It also suggests that Medicaid helps people with higher income levels more than it helps people with lower incomes. Using only the lottery as the instrument, our new procedure generates the opposite results. The interpretation is that for people with lower incomes, Medicaid enrollment decreases their probability of owing money to health providers by 19.9 log points on average, holding other variables constant. For people with higher incomes, the effect of Medicaid decreases.

Based on the D-RSMD results using only the lottery, we do not find support in the data to say that there are heterogeneous treatment effects for individuals between 21 and 35 years old. Furthermore, we discover that the results for individuals aged 21–35 differ significantly between the two instrument sets. It suggests that the generated interaction $Lottery \times X_{1}$ is invalid.

Next, we re-estimate a homogeneous model. In a homogeneous model, the interaction term is not included in the regression model. The traditional estimation method only requires one instrument to estimate one parameter. When we examine the homogeneous treatment effects using both instrument sets in Table \ref{app2:table16}, all estimators yield similar results to the DRSMD-Lasso method using only the lottery as the instrument in the heterogeneous treatment effects panel. This suggests that using a valid instrument to estimate both parameters produces more reliable outcomes. In practical applications where the homogeneity of the model is uncertain, the DRSMD-Lasso method proves to be highly valuable. Its ability to provide reliable estimates under both homogeneous and heterogeneous conditions, coupled with its smaller standard errors compared to traditional methods, makes it a preferred choice. In particular, the smaller standard errors enhance the significance of results obtained through t-tests. 

The estimates and standard errors of average treatment effects in Table \ref{app2:table16ATE} are calculated based on Table \ref{app2:table16}. We conduct the robustness check in the Appendix.

\begin{table}[!ht]
\begin{center}
\scalebox{.75}{
\begin{tabular}{l | c c | c c | c c   }
\hline \hline
\multicolumn{7}{l}{\textbf{Panel A: Heterogeneous treatment effects}} \\ \hline
\textit{Debt for Health} & \multicolumn{2}{c|}{Age: 21 - 35}  & \multicolumn{2}{c|}{Age: 36 - 50} &  \multicolumn{2}{c}{Age: 51 - 64}  \\ \hline
\textit{Estimator for $\theta_{w0}$}& $Z_1$  & $(Z_1, Z_1 X_{1})$ & $Z_1$  & $(Z_1, Z_1 X_{1})$ & $Z_1$  & $(Z_1, Z_1 X_{1})$ \\ \hline
\multirow{2}{*}{GMM}&   &  -0.069 & & -0.213***  &  &  -0.193*** \\
& & (0.055) &   & (0.041)  &  & (0.046) \\ 
\multirow{2}{*}{GMM-Lasso}&   & -0.091  &  & -0.259*** &  &  -0.245***  \\
& & (0.093) &  & (0.078) & & (0.090) \\ 
\multirow{2}{*}{DRSMD-Lasso}& -0.199*** & -0.053  & -0.232*** & -0.232*** & -0.182** &  -0.190*** \\
& (0.071) & (0.065) & (0.062) & (0.050) & (0.072) & (0.056) \\
\hline
\textit{Estimator for $\theta_{wx0}$}& $Z_1$  & $(Z_1, Z_1 X_{1})$ & $Z_1$  & $(Z_1, Z_1 X_{1})$ & $Z_1$  & $(Z_1, Z_1 X_{1})$  \\ \hline
\multirow{2}{*}{GMM}&   & -0.204** &  & 0.108 &  &  0.069\\
&   & (0.096) &  & (0.085)  &  & (0.094) \\ 
\multirow{2}{*}{GMM-Lasso}&   & -0.184  &  & 0.008 &  &   -0.081 \\
&   & (0.131) &   & (0.128) &    & (0.153) \\ 
\multirow{2}{*}{DRSMD-Lasso}&  0.058 & -0.252**  & 0.091*** & 0.088 & 0.076* &  0.096 \\
& (0.036) & (0.105) & (0.031) & (0.096) & (0.040) & (0.103) \\
\hline
\multicolumn{7}{l}{\textbf{Panel B: Homogeneous treatment effects}} \\ \hline
\multirow{2}{*}{GMM}& -0.161***  & -0.130***  & -0.170***  & -0.189***  & -0.164*** &  -0.177*** \\
&  (0.049) & (0.046) &   (0.040)  & (0.037)  &  (0.044)  & (0.041) \\ 
\multirow{2}{*}{GMM-Lasso}& -0.180*  & -0.113  & -0.256*** & -0.260*** & -0.279** & -0.244*** \\
& (0.104)  & (0.092) &  (0.097)  & (0.078) & (0.112)  & (0.089) \\ 
\multirow{2}{*}{DRSMD-Lasso}& -0.172*** & -0.187***  & -0.198*** & -0.198*** & -0.150** & -0.145** \\
& (0.063) & (0.066) & (0.056) & (0.057) & (0.063) & (0.064) \\ \hline
\textit{N} & \multicolumn{2}{c|}{5962}  & \multicolumn{2}{c|}{6693} &  \multicolumn{2}{c}{5917} \\
\hline \hline
\end{tabular}
}
\caption{Heterogeneous Treatment Effects of Medicaid on Debt for Health}
\label{app2:table16}
\end{center}
\par \footnotesize Note: *** Significant at 1\%, ** at 5\%, * at 10\%. Each row shows the estimates and robust standard errors for the same type of estimator. In the columns, we present the instruments these estimators used and the age groups. The interaction term is Medicaid $\times$ Above 50\% Federal Poverty Line.
\end{table}

\begin{table}[!ht]
\begin{center}
\scalebox{.8}{
\begin{tabular}{l | c c | c c | c c}
\hline \hline
\multicolumn{7}{l}{\textbf{Average Treatment Effects}} \\ \hline
\textit{Debt for Health} & \multicolumn{2}{c|}{Age: 21 - 35}  & \multicolumn{2}{c|}{Age: 36 - 50} &  \multicolumn{2}{c}{Age: 51 - 64}  \\ \hline
\textit{Estimator for LATE} & $Z_1$  & $(Z_1, Z_1 X_{1})$ & $Z_1$  & $(Z_1, Z_1 X_{1})$ & $Z_1$  & $(Z_1, Z_1 X_{1})$   \\ \hline
\multirow{2}{*}{GMM}&   & -0.190*** &  & -0.150*** &  &  -0.149*** \\
                    &   & (0.055) &  & (0.048)  &  & (0.055) \\ 
\multirow{2}{*}{GMM-Lasso} &   & -0.201* &  & -0.255** &  &  -0.296** \\
                    &   & (0.112) &  & (0.110)  &  & (0.133) \\
\multirow{2}{*}{DRSMD-Lasso} & -0.164***  & -0.202***  & -0.179*** & -0.181*** & -0.135** & -0.129*  \\
                             & (0.061) & (0.069) & (0.054) & (0.067) & (0.060) & (0.075) \\\hline
\textit{N} & \multicolumn{2}{c|}{5962}  & \multicolumn{2}{c|}{6693} &  \multicolumn{2}{c}{5917} \\
\hline \hline
\end{tabular}
}
\caption{Heterogeneous Treatment Effects of Medicaid on Debt for Health}
\label{app2:table16ATE}
\end{center}
\par \footnotesize Note: *** Significant at 1\%, ** at 5\%, * at 10\%. Each row shows the estimates and robust standard errors for the same type of estimator. In the columns, we present the instruments these estimators used and the age groups. The interaction term is Medicaid $\times$ Above 50\% Federal Poverty Line. The expression for LATE is $\theta_{w0} + \theta_{wx0} E(X)$.
\end{table}

\clearpage

\bibliographystyle{chicago}
\bibliography{refs.bib}

\newpage
\appendix

\section{Appendix: Additional Results of Monte Carlo Simulation }

\subsection{The Advantage of the D-RSMD Estimator vs. Splitting the Sample}

Table \ref{sim:iv2intro} presents simulation results using a binary covariate in the interaction term. This simulation follows a similar DGP as the one in the simulation section, except that $X_1$ is a binary variable. From the table, the D-RSMD estimator provides much better estimation in terms of MAD and Med.SE than the GMM with the correct model specification (GMM (Oracle)). However, without the correct model specification, RGMM (which incorporates Lasso to approximate the unknown nonparametric part and then GMM) does not perform as effectively. The situation worsens with fewer observations and when the mean of $X_1$ is closer to 0 or 1. Here, $E(X_1) = 0.2$. This is shown when we use the subsets of the dataset, for instance, the rows when $X_1 = 1$.

\begin{table}[!ht]
\begin{center}
\scalebox{.7}{
\begin{tabular}{ l l c | c c c c | c c c c}
\hline \hline
\multicolumn{11}{l}{\textbf{n = 3,000 and there are 30 covariates}}\\
\hline
\multicolumn{3}{c|}{\textbf{}} & \multicolumn{4}{c|}{\textbf{$\theta_{w0}$}} & \multicolumn{4}{c}{\textbf{$\theta_{wx0}$}} \\ \hline
 \textit{Dataset} &\textit{Estimator} & \textit{Instrument} & Med.Bias  & MAD & Med.SE & RR & Med.Bias  & MAD & Med.SE & RR \\ \hline
Full dataset          & D-RSMD       & $Z_1$            &  0.056  & 0.219 & 0.327 & 0.049 & -0.053  & 0.120 & 0.180 &  0.058\\
Rows when $X_{1} = 1$ & RGMM         & $Z_1$            & -0.293  & 0.413 & 0.962 & 0.003 &         &       &       &   \\
Rows when $X_{1} = 0$ & RGMM         & $Z_1$            &  0.676  & 0.295 & 0.885 & 0.003 &         &       &       &   \\
Full dataset          & RGMM & $(Z_1, Z_1 X_{1})$       &  0.592  & 0.326 & 0.817 & 0.003 & -0.822  & 0.180 & 0.412 &  0.527\\
Rows when $X_{1} = 1$ & GMM (Oracle) & $Z_1$            & -0.006  & 0.409 & 0.575 & 0.044 &         &       &       &    \\
Rows when $X_{1} = 0$ & GMM (Oracle) & $Z_1$            &  0.004  & 0.265 & 0.396 & 0.046 &         &       &       &    \\
Full dataset          & GMM (Oracle) &$(Z_1, Z_1 X_{1})$&  0.005  & 0.234 & 0.347 & 0.048 & -0.003  & 0.155 & 0.228 &  0.053\\
\hline \hline
\end{tabular}
}
\caption{Performance of Estimators with and without Splitting the Sample}
\label{sim:iv2intro}
\end{center}
\par \footnotesize Note: Simulation Results for $\theta_{w0}$ and $\theta_{wx0}$ in the benchmark model with $Z_1$ as the Instrument inside DGP. There are 5,000 replications. We report the Monte-Carlo Median Bias (Med.Bias), Median Absolute Deviation (MAD), median of asymptotic standard error under heteroskedasticity (Med.SE), and Rejection Rate (RR) using a 5\% t-test. $Z_1$ is the binary instrument. $X_1$ is the binary covariate with $P_r(X_{1} = 1) = 0.2$. The dataset is splitted for different $X_1$ values when employing GMM and RGMM to handle the heterogeneity of the treatment. 
\end{table}

\subsection{The Advantage of the D-RSMD Estimator vs. Using Another Instrument Set}

Next, we demonstrate the performance of the GMM (Oracle) estimator using a pair of the instrument and the interaction term between a predicted endogenous variable and a covariate, while avoiding the use of the interaction term between the instrument and the covariate. This approach yields favourable results; however, the D-RSMD estimator still achieves competitive performance. In scenarios where the model specification is unknown, the RGMM estimator using $(Z_1, \hat{E}(W|X) X_{1})$ exhibits higher bias, MAD, Med.SE, and a larger rejection rate (RR), as illustrated in Table \ref{sim:iv2introcomp1}.

\begin{table}[!ht]
\begin{center}
\scalebox{.7}{
\begin{tabular}{ l l c | c c c c | c c c c}
\hline \hline
\multicolumn{3}{c|}{\textbf{}} & \multicolumn{4}{c|}{\textbf{$\theta_{w0}$}} & \multicolumn{4}{c}{\textbf{$\theta_{wx0}$}} \\ \hline
 &\textit{Estimator} & \textit{Instrument} & Med.Bias  & MAD & Med.SE & RR & Med.Bias  & MAD & Med.SE & RR \\ \hline
$n=3,000$ & D-RSMD  & $Z_1$                      &  0.028  & 0.288 & 0.427 & 0.045 & -0.012  & 0.045 & 0.059 &  0.064\\
$P=30$   & RGMM & $(Z_1, \hat{E}(W|X) X_{1})$   &  0.196  & 0.358 & 0.812 & 0.002 & -0.166  & 0.092 & 0.178 &  0.130 \\
& GMM (Oracle) & $(Z_1, \hat{E}(W|X) X_{1})$    &  0.010  & 0.252 & 0.372 & 0.050 & -0.001  & 0.049 & 0.073 &  0.051  \\
& GMM (Oracle) & $(Z_1, Z_2)$                   & -0.011  & 0.635 & 1.424 & 0.003 &  0.114  & 2.502 & 5.548 &  0.000  \\ \hline
\hline
\end{tabular}
}
\caption{Performance of Estimators  with and without using an additional instrument variable}
\label{sim:iv2introcomp1}
\end{center}
\par \footnotesize Note: Simulation Results for $\theta_{w0}$ and $\theta_{wx0}$ in the benchmark model  with $Z_1$ as the Instrument inside DGP. There are 5,000 replications. We report the Monte-Carlo Median Bias (Med.Bias), Median Absolute Deviation (MAD), median of asymptotic standard error under heteroskedasticity (Med.SE), and Rejection Rate (RR) using a 5\% t-test.  $Z_2$: the sum of $Z_1$ and another binary variable. There are three values in $Z_2$. $(Z_1, \hat{E}(W|X) X_{1})$: pair consisting of $Z_1$ and the interaction term $\hat{E}(W|X) X_{1}$ (where $\hat{E}(W|X)$ is the estimated conditional expectation and $X_{1}$ is a covariate).
$(Z_1, Z_2)$: pair including both $Z_1$ and $Z_2$. The correlation between $Z_1$ and $Z_2$ is around 0.7.
\end{table}

\subsection{Results for the Model with a Categorical Instrument}

In this section, we introduce a categorical instrument $Z$ with three values. Specifically, for the benchmark model, we generate the endogenous treatment using $Z_2$.

When the number of covariates is small, for example, $q_X = 1$, we apply the Nadaraya-Watson estimator with a rule-of-thumb bandwidth $h = \sigma_x n^{-0.2}$ to estimate the nuisance parameters for the R-SMD and R-GMM estimators. In our setting with $n = 3,000$ and $q_X = 1$, the Nadaraya-Watson estimators for nuisance parameters use a bandwidth of $0.202\hat{\sigma}_x$. For the D-RSMD, we use the Lasso method to estimate the nuisance parameters. Both D-RSMD and R-SMD estimators utilize the CDF of a standard Gaussian distribution as $\mu(\cdot)$ inside $\kappa(Z_j - Z_l) = \int_{\mathbb{R}^{q_z}} e^{it'(Z_j - Z_l)} d\mu(t)$.

Table \ref{sim:iv3table1} presents the simulation results of D-RSMD, R-SMD, and GMM-type estimators for $\theta_{w0}$ and $\theta_{wx0}$. The sample size is 3,000 with 5,000 replications and only one covariate ($q_X = 1$). The table displays results for D-RSMD and R-SMD estimators using one instrument, either $Z_1$ or $Z_2$, and two instruments, namely $(Z_1, Z_1 X_{1})$ or $(Z_2, Z_2 X_{1})$. GMM-type estimators utilize pairs of instruments: $(Z_1, Z_1 X_{1})$ or $(Z_2, Z_2 X_{1})$.

\begin{table}[!ht]
\begin{center}
\scalebox{.75}{
\begin{tabular}{ l c | c c c c | c c c c}
\hline \hline
\multicolumn{2}{c|}{\textbf{}} & \multicolumn{4}{c|}{\textbf{$\theta_{w0}$}} & \multicolumn{4}{c}{\textbf{$\theta_{wx0}$}} \\ \hline
\textit{Estimator} & \textit{Instrument} & Med.Bias  & MAD & Med.SE & RR & Med.Bias  & MAD & Med.SE & RR \\ \hline
D-RSMD  & $Z_1$ & -0.007  &  0.069 & 0.103  & 0.047    & 0.019  & 0.036 & 0.052 & 0.067\\
D-RSMD  & $Z_2$ & -0.017 & 0.129 & 0.211  & 0.035        & 0.034  & 0.313 & 0.523 & 0.035\\
D-RSMD  & $(Z_1, Z_1X_{1})$ & 0.002 & 0.077         & 0.112        & 0.050         & -0.008  & 0.089 & 0.130 & 0.044 \\
D-RSMD & $(Z_2, Z_2X_{1})$ & 0.001  &  0.045 & 0.065 & 0.048      & -0.006  & 0.055 & 0.077 & 0.054\\
\hline
R-SMD  & $Z_1$ & -0.207  &  0.072 & 0.236  & 0.007   & 0.424  & 0.065 & 0.449 & 0.000\\
R-SMD & $Z_2$ &  0.027   & 0.172  & 0.280  & 0.022  & -0.060 & 0.416 & 0.693 & 0.018  \\
R-SMD & $(Z_1, Z_1X_{1})$  & 0.064  & 0.077 & 0.127 & 0.052 & -0.112  & 0.083 & 0.220 & 0.003 \\
R-SMD & $(Z_2, Z_2X_{1})$  & 0.028  & 0.039 & 0.072 & 0.031  & -0.053  & 0.048 & 0.076 & 0.079\\
\hline
R-GMM & $(Z_1, Z_1X_{1})$ & -0.044  & 0.076 & 0.182 & 0.003  & -0.239  & 0.089 &  0.245 & 0.026\\
R-GMM & $(Z_2, Z_2X_{1})$ & -0.015 & 0.042 & 0.105 & 0.001 & -0.172  & 0.053 &  0.156 & 0.043\\
\hline
GMM & $(Z_1, Z_1X_{1})$ & 2.571  & 0.318 & 0.472 & 1.000  & -6.003  & 0.415 & 0.618 & 1.000 \\
GMM & $(Z_2, Z_2X_{1})$ & 2.176  & 0.231 & 0.322 & 1.000  & -5.233  & 0.319 & 0.454 & 1.000\\
\hline
GMM (Oracle) & $(Z_1, Z_1X_{1})$ & 0.000  & 0.076 & 0.113 & 0.048 & -0.003  & 0.093 & 0.137 & 0.045\\
GMM (Oracle) & $(Z_2, Z_2X_{1})$ & 0.000  & 0.042 & 0.061 & 0.049   & 0.000  & 0.053 & 0.078 & 0.057 \\ \hline \hline
\end{tabular}
}
\caption{Performance of Estimators with a Categorical Instrument in the DGP when $q_X=1$}
\label{sim:iv3table1}
\end{center}
\par \footnotesize Note: Simulation Results for $\theta_{w0}$ and $\theta_{wx0}$ in the benchmark model with $Z_2$ in the DGP. There are 5,000 replications. The sample size is 3,000. We report the Monte-Carlo Median Bias (Med.Bias), Median Absolute Deviation (MAD), the median of asymptotic standard error under heteroskedasticity (Med.SE), and Rejection Rate (RR) using a 5\% t-test. 
$Z_1$: binary instrument.
 $Z_2$: variable representing the sum of $Z_1$ and another binary variable, resulting in three possible values for $Z_2$.
$(Z_1, Z_1X_{1})$: pair consisting of $Z_1$ and the interaction term $Z_1X_{1}$ (where $X_{1}$ is a covariate).
$(Z_2, Z_2 X_{1})$: pair comprising $Z_2$ and the interaction term $Z_2X_{1}$.
$(Z_1, Z_2)$: pair including both $Z_1$ and $Z_2$. The correlation between $Z_1$ and $Z_2$ is around 0.7.
\end{table}

Recall that $Z_2$ is the instrument used to generate the treatment variable. Intuitively, estimators that employ $Z_2$ or $(Z_2, Z_2 X_{1})$ are expected to yield better results compared to those using $Z_1$ or $(Z_1, Z_1 X_{1})$. This is indeed evident from Table \ref{sim:iv3table1}, particularly for R-GMM, GMM, and GMM (Oracle) estimators. However, for D-RSMD and RSMD, this is not consistently observed. For instance, the MAD and Med.SE for D-RSMD with $Z_1$ are smaller than those with $Z_2$ for both parameters, although this property may be case-specific and does not persist in the subsequent case with $q_X = 30$. SMD-type estimators generally use all available information from the instrument. These estimators remain effective when using a closely related instrument.

Comparing D-RSMD and GMM (Oracle), when both employ $(Z_2, Z_2 X_{1})$, we find that their results are similar. This suggests that D-RSMD with $(Z_2, Z_2 X_{1})$ performs comparably to GMM (Oracle). Notably, D-RSMD exhibits a lower median bias than R-SMD with $(Z_2, Z_2 X_{1})$, indicating the effectiveness of bias reduction in D-RSMD. The Med.SE for D-RSMD is also smaller than that for R-SMD, and the rejection rate (RR) for D-RSMD is closer to 5\% compared to R-SMD.

Next, we consider the scenario with $q_X = 30$. We set the sparsity level to 10, where $\alpha_{3q} = 2$, $\beta_{2q} = -3$, and $\alpha_{1q} = \beta_{1q} = 1$ for $q \leq 5$. We continue to use the benchmark model in our simulation study.

\begin{table}[!ht]
\begin{center}
\scalebox{.7}{
\begin{tabular}{ l l c | c c c c | c c c c}
\hline \hline
\multicolumn{3}{c|}{\textbf{}} & \multicolumn{4}{c|}{\textbf{$\theta_{w0}$}} & \multicolumn{4}{c}{\textbf{$\theta_{wx0}$}} \\ \hline
 &\textit{Estimator} & \textit{Instrument} & Med.Bias  & MAD & Med.SE & RR & Med.Bias  & MAD & Med.SE & RR \\ \hline
  $n=2,000$ & D-RSMD  & $Z_2$        & -0.041  & 0.115 & 0.146 & 0.064 & -0.019  & 0.106 & 0.158 & 0.064 \\
 & D-RSMD & $(Z_2, Z_2 X_{1})$      & -0.059  & 0.119 & 0.144 & 0.072 & -0.033  & 0.079 & 0.101 & 0.064 \\
 & GMM (Oracle)& $(Z_2, Z_2 X_{1})$ & -0.006  & 0.110 & 0.160 & 0.050 &  0.001  & 0.059 & 0.087 & 0.053 \\ \hline
  $n=3,000$ & D-RSMD  & $Z_1$        &  0.139  & 0.765 & 1.140 & 0.012 & -0.013  & 0.091 & 0.158 & 0.020 \\
& D-RSMD  & $Z_2$                   & -0.018  & 0.119 & 0.158 & 0.044 & -0.019  & 0.145 & 0.236 & 0.048 \\
& D-RSMD  & $(Z_1, Z_1 X_{1})$      & -0.115  & 0.557 & 0.843 & 0.025 & -0.015  & 0.105 & 0.154 & 0.037 \\
& D-RSMD  & $(Z_2, Z_2 X_{1})$      & -0.028  & 0.100 & 0.118 & 0.055 & -0.023  & 0.068 & 0.083 & 0.059 \\
& D-RSMD  & $(Z_1, Z_2)$            & -0.019  & 0.189 & 0.267 & 0.036 & -0.018  & 0.274 & 0.435 & 0.034 \\
& RSMD  & $Z_1$                     & -0.476  & 0.773 & 1.173 & 0.024 & -0.049  & 0.101 & 0.161 & 0.052 \\
& RSMD  & $Z_2$                     & -0.134  & 0.178 & 0.313 & 0.083 & -0.052  & 0.231 & 0.421 & 0.058 \\
& RSMD  & $(Z_1, Z_1X_{1})$         &  0.082  & 0.790 & 1.158 & 0.015 & -0.279  & 0.135 & 0.260 & 0.071 \\
& RSMD & $(Z_2, Z_2X_{1})$          & -0.202  & 0.111 & 0.190 & 0.157 & -0.074  & 0.054 & 0.085 & 0.129 \\
& RSMD  & $(Z_1, Z_2)$              & -0.128  & 0.267 & 0.411 & 0.073 & -0.058  & 0.374 & 0.602 & 0.045 \\
& GMM & $(Z_2, Z_2 X_{1})$          &  3.379  & 0.897 & 1.354 & 0.715 & -5.957  & 0.509 & 0.726 & 1.000 \\
& GMM & $(Z_1, Z_2)$                & 12.082  & 4.798 & 7.355 & 0.230 & -26.653 & 9.961 & 15.090& 0.312 \\
& GMM (Oracle) & $(Z_2, Z_2 X_{1})$ &  0.000  & 0.089 & 0.130 & 0.056 &  0.000  & 0.048 & 0.071 & 0.052 \\
& GMM (Oracle) & $(Z_1, Z_2)$       &  0.003  & 0.506 & 0.802 & 0.004 &  0.015  & 1.065 & 1.712 & 0.002 \\ \hline
$n=5,000$ & D-RSMD  & $Z_2$          & -0.005  & 0.397 & 0.694 & 0.021 & -0.028  & 0.714 & 1.267 & 0.022 \\
& D-RSMD & $(Z_2, Z_2 X_{1})$       & -0.016  & 0.082 & 0.096 & 0.053 & -0.013  & 0.054 & 0.066 & 0.056 \\
& GMM (Oracle) & $(Z_2, Z_2 X_{1})$ & -0.001  & 0.068 & 0.101 & 0.050 & -0.001  & 0.037 & 0.055 & 0.049 \\ \hline
\hline
\end{tabular}
}
\caption{Performance of Estimators with a Categorical Instrument in the DGP when $q_X=30$}
\label{sim:iv3table2}
\end{center}
\par \footnotesize Note: Simulation Results for $\theta_{w0}$ and $\theta_{wx0}$ in the benchmark model with $Z_2$ in the DGP. There are 5,000 replications with the sample size $n \in \{2000, 3000, 5000 \}$. We report the Monte-Carlo Median Bias (Med.Bias), Median Absolute Deviation (MAD), median of asymptotic standard error under heteroskedasticity (Med.SE), and Rejection Rate (RR) using a 5\% t-test. 
$Z_1$: binary instrument.
 $Z_2$: variable representing the sum of $Z_1$ and another binary variable, resulting in three possible values for $Z_2$.
$(Z_1, Z_1X_{1})$: pair consisting of $Z_1$ and the interaction term $Z_1X_{1}$ (where $X_{1}$ is a covariate).
$(Z_2, Z_2 X_{1})$: pair comprising $Z_2$ and the interaction term $Z_2X_{1}$.
$(Z_1, Z_2)$: pair including both $Z_1$ and $Z_2$. The correlation between $Z_1$ and $Z_2$ is around 0.7.
\end{table}

Table \ref{sim:iv3table2} presents simulation results for D-RSMD, RSMD, and GMM type estimators with sample sizes $n \in \{2000, 3000, 5000 \}$, each based on 5000 replications.

In the second panel of the table, D-RSMD estimators using $Z_2$ generally exhibit better performance than those using $Z_1$, showing smaller Med.Bias, MAD, and Med.SE when estimating $\theta_{w0}$. However, for $\theta_{wx0}$, D-RSMD with $Z_1$ yields competitive estimates.

Comparing D-RSMD and GMM type estimators in the 3000 observation panel of Table \ref{sim:iv3table2}, GMM (Oracle) demonstrates relatively better performance with the instrument set $(Z_2, Z_2 X_{1})$, characterized by lower Med.Bias and MAD. Nonetheless, D-RSMD using the same instrument set still produces promising results.

It's worth noting that the GMM (Oracle) estimator encounters challenges with the instrument set $(Z_1, Z_2)$, which is less problematic for the D-RSMD estimator. This is the same observation as that for the previous case with $q_X = 1$. 

With a categorical instrument in the DGP, the D-RSMD estimator generates a smaller bias compared to RSMD estimators. D-RSMD's flexibility allows for estimation with various instrument sets. Notably, D-RSMD remains competitive compared to GMM (Oracle) estimators.

With a categorical instrument in the DGP, the D-RSMD estimators have the proper size and a smaller bias than the RSMD estimators. As the sample size grows, the size distortions decrease for $(Z_2, Z_2 X_{1})$. D-RSMD allows us to use any possible instrument set to estimate heterogeneous treatment effects. Compared to GMM (Oracle), the D-RSMD estimator still generates competitive results.

\section{Appendix: Additional Results of Empirical Application}
 
\subsection{Robustness Check} \label{app2:robust}

In Section \ref{app2:results}, the interaction term includes a covariate indicating whether household income exceeds 50\% of the Federal Poverty Line. To offer further insights into these regressions, we conduct various robustness checks. The first subsection examines estimation results when the covariate is 1 for household income exceeding 100\% or 150\% of the Federal Poverty Line. The second subsection provides a brief overview of other robustness checks.

\begin{itemize}

\item \textbf{Estimation Results with Other Covariates}

The results are included in Table \ref{app2:table6} when $X_1$ is an indicator for "Income Above 100\% Federal Poverty Line". Comparing Table \ref{app2:table6} with Table \ref{app2:table16}, we find that the results are quite similar. For individuals between 36 and 50 years old, there are heterogeneous treatment effects. The estimates show that they benefit more from Medicaid health coverage when they have lower incomes. We do not find support in the data to say that there are heterogeneous treatment effects for individuals between 21 and 35 years old, and the DRSMD-Lasso esimator using only one valid instrument variable generates more reliable results.

\begin{table}[!ht]
\begin{center}
\scalebox{.8}{
\begin{tabular}{l | c c | c c | c c}
\hline \hline
\multicolumn{7}{l}{\textbf{Heterogeneous Treatment Effects of Medicaid on Debt for Health}} \\ \hline
\textit{Debt for Health} & \multicolumn{2}{c|}{Age: 21 - 35}  & \multicolumn{2}{c|}{Age: 36 - 50} &  \multicolumn{2}{c}{Age: 51 - 64}  \\ \hline
\textit{Estimator for $\theta_{w0}$} & $Z_1$  & $(Z_1, Z_1 X_{1})$ & $Z_1$  & $(Z_1, Z_1 X_{1})$ & $Z_1$  & $(Z_1, Z_1 X_{1})$   \\ \hline
\multirow{2}{*}{GMM}&   &  -0.123** & & -0.179*** & & -0.142*** \\
& & (0.048) &   & (0.038) &   & (0.041)  \\ 
\multirow{2}{*}{DRSMD-Lasso}&-0.178*** & -0.123**  & -0.202*** & -0.200*** & -0.164** & -0.131\\
& (0.063) & (0.061) & (0.057) & (0.049) & (0.065) & (0.053)\\
\hline
\textit{Estimator for $\theta_{wx0}$}& $Z_1$  & $(Z_1, Z_1 X_{1})$ & $Z_1$  & $(Z_1, Z_1 X_{1})$ & $Z_1$  & $(Z_1, Z_1 X_{1})$   \\ \hline
\multirow{2}{*}{GMM}&   & -0.200 &  & 0.072 & & -0.156 \\
&   & (0.148) &  & (0.159) &  & (0.185) \\ 
\multirow{2}{*}{DRSMD-Lasso}&  0.023 & -0.212  & 0.102*** & 0.032 & 0.096 &-0.154 \\
& (0.033) & (0.154) & (0.038) & (0.192) & (0.044) & (0.217)\\
\hline
\textit{N} & \multicolumn{2}{c|}{5962}  & \multicolumn{2}{c|}{6693} & \multicolumn{2}{c}{5917}  \\
\hline \hline
\end{tabular}
}
\caption{Income above 100\% Federal Poverty Line}
\label{app2:table6}
\end{center}
\par  Note: *** Significant at 1\%, ** at 5\%, * at 10\%.
\end{table}

Table \ref{app2:table7} reports the outcomes when $X_{1}$ is an indicator for ``Income Above 150\% Federal Poverty Line''. Results in Tables \ref{app2:table7} and \ref{app2:table6} are similar in estimates but different in standard errors. In the table, estimation results from using a dummy for income above 150\% Federal Poverty Line show that the treatment effects of Medicaid on debt for people between 36 and 50 are not supported by the data to be heterogeneous because the estimate for the parameter in front of the interaction term is not statistically significant at the 5\% significance level.

The effects of Medicaid on debt for people between 21 and 35 are heterogeneous because the estimate is statistically significant. The difference in the results using two instrument sets for the DRSMD-Lasso estimators suggests that using only the lottery variable as an instrument generates more reliable results. The difference between Tables \ref{app2:table7} and \ref{app2:table6} shows that the covariate inside the interaction is important for the estimation results. It can be explained by the fact that there are only 808 individuals with incomes above 150\% Federal Poverty Line for people between 36 and 50 and 794 individuals for people between 21 and 35 in this data set.

\begin{table}[!ht]
\begin{center}
\scalebox{.80}{
\begin{tabular}{l | c c | c c | c c }
\hline \hline
\multicolumn{7}{l}{\textbf{Heterogeneous Treatment Effects of Medicaid on Debt for Health}} \\ \hline
\textit{Debt for Health} & \multicolumn{2}{c|}{Age: 21 - 35}  & \multicolumn{2}{c|}{Age: 36 - 50} &  \multicolumn{2}{c}{Age: 51 - 64}  \\ \hline
\textit{Estimator for $\theta_{w0}$} & $Z_1$  & $(Z_1, Z_1 X_{1})$ & $Z_1$  & $(Z_1, Z_1 X_{1})$ & $Z_1$  & $(Z_1, Z_1 X_{1})$   \\ \hline
\multirow{2}{*}{GMM}&   &  -0.145*** & & -0.179***  & & -0.145***\\
& & (0.048) &   & (0.038)  & & (0.042)\\ 
\multirow{2}{*}{DRSMD-Lasso}&-0.181*** & -0.149**  & -0.207*** & -0.205*** & -0.154** & -0.121** \\
& (0.062) & (0.062) & (0.056) & (0.053) & (0.063) & (0.057)\\
\hline
\textit{Estimator for $\theta_{wx0}$} & $Z_1$  & $(Z_1, Z_1 X_{1})$ & $Z_1$  & $(Z_1, Z_1 X_{1})$ & $Z_1$  & $(Z_1, Z_1 X_{1})$   \\ \hline
\multirow{2}{*}{GMM}&   & -0.204 &  & 0.195 & & -0.521\\
&   & (0.245) &  & (0.340) & & (0.426)\\ 
\multirow{2}{*}{DRSMD-Lasso}&  0.095* & -0.254  & 0.067 & 0.030 & 0.078 & -0.620\\
& (0.055) & (0.292) & (0.061) & (0.453) & (0.067) & (0.559)\\
\hline
\textit{N} & \multicolumn{2}{c|}{5962}  & \multicolumn{2}{c|}{6693} & \multicolumn{2}{c}{5917} \\
\hline \hline
\end{tabular}
}
\caption{Income above 150\% Federal Poverty Line}
\label{app2:table7}
\end{center}
\par  Note: *** Significant at 1\%, ** at 5\%, * at 10\%.
\end{table}

\item \textbf{Other Robustness Checks}

We present additional robustness checks. One check focuses on the variable TANF, which exhibits limited variation with only 2\% of individuals receiving it. We examine whether TANF contributes to heterogeneity. Table \ref{app2:table3} displays the results for the Debt for Health variable across five age groups. The DRSMD-Lasso estimator with $Z_1$ yields statistically significant estimates of -0.215 for $\theta_{w0}$ and 0.307 for $\theta_{wx0}$ at a 5\% significance level in the first column. This suggests that, for individuals aged 21 to 29, TANF mitigates the negative effect of Medicaid on the likelihood of having debt. The new method, utilizing the valid instrument $Z_1$ continues to produce reliable results with the lowest standard error.

Tables \ref{app2:table10} and \ref{app2:table11} display the effects of treatment on employment and out-of-pocket costs, respectively. Both tables indicate heterogeneous treatment effects among young individuals (aged 21 to 29) when $X_{1}$ is a binary variable representing "Income Above 50\% of the Federal Poverty Line."

\end{itemize}

\begin{landscape}

\begin{table}[!ht]
\begin{center}
\scalebox{.7}{
\begin{tabular}{l | c c | c c | c c | c c | c c  }
\hline \hline
\multicolumn{11}{l}{\textbf{Panel A: Heterogeneous treatment effects}} \\ \hline
\textit{Debt for Health} & \multicolumn{2}{c|}{Age: 21 - 29}  & \multicolumn{2}{c|}{Age: 30 - 38} &  \multicolumn{2}{c|}{Age: 39 - 47} & \multicolumn{2}{c|}{Age: 48 - 56}  &  \multicolumn{2}{c}{Age: 57 - 64}  \\ \hline
\textit{Estimator for $\theta_{w0}$}& $Z_1$  & $(Z_1, Z_1 X_{1})$ & $Z_1$  & $(Z_1, Z_1 X_{1})$ & $Z_1$  & $(Z_1, Z_1 X_{1})$& $Z_1$  & $(Z_1, Z_1 X_{1})$ & $Z_1$  & $(Z_1, Z_1 X_{1})$    \\ \hline
\multirow{2}{*}{GMM}&   &  -0.193*** & &  -0.199*** &  & -0.113** &   & -0.196*** &  &  -0.157**\\
& & (0.065) &   & (0.061)  &  & (0.049) &  & (0.046) &   & (0.068)\\ 
\multirow{2}{*}{GMM-Lasso}&   &  -0.270* &  & -0.216 &  & -0.211*  &   & -0.351*** &  & -0.295 \\
& & (0.147) &  & (0.136) & & (0.110) &  & (0.112) &   & (0.182)\\ 
\multirow{2}{*}{DRSMD-Lasso}&-0.215**  & -0.217**  & -0.137* & -0.010 & -0.148** &  -0.135** &  -0.199*** & -0.192*** & -0.139 &  -0.141\\
& (0.087) & (0.103) & (0.082) & (0.337) & (0.068) & (0.068) & (0.062) & (0.062) & (0.107)  & (0.107)\\
\hline
\textit{Estimator for $\theta_{wx0}$}& $Z_1$  & $(Z_1, Z_1 X_{1})$ & $Z_1$  & $(Z_1, Z_1 X_{1})$ & $Z_1$  & $(Z_1, Z_1 X_{1})$& $Z_1$  & $(Z_1, Z_1 X_{1})$ & $Z_1$  & $(Z_1, Z_1 X_{1})$    \\ \hline
\multirow{2}{*}{GMM}&   & 0.212 &  & 1.959 &  & -0.791  &   & 2.059 &  & 3.074 \\
&   & (0.900) &  & (1.353)  &  & (0.652) &   & (2.040) &   & (3.827)\\ 
\multirow{2}{*}{GMM-Lasso}&   & 0.983  &  & 1.904 &  &  -0.785 &   & 1.083 &  & 0.818 \\
&   & (1.099) &   & (1.348) &    & (0.956) &  & (1.844) &   & (1.164)\\ 
\multirow{2}{*}{DRSMD-Lasso}& 0.307**  &  0.777 & 0.042 & -9.902 & -0.320*** &  -1.448 & 0.470  & -1.725 & -0.048 & -0.281 \\
& (0.142) & (4.239) & (0.153) & (19.627) & (0.085) & (1.554) & (0.417) & (1.821) & (0.522)  & (0.590)\\
\hline
\textit{N} & \multicolumn{2}{c|}{3504}  & \multicolumn{2}{c|}{3596} &  \multicolumn{2}{c|}{3941} & \multicolumn{2}{c|}{4599}  &  \multicolumn{2}{c}{2932}  \\
\hline \hline
\end{tabular}
}
\caption[]{Heterogeneous Treatment Effects of Medicaid on Debt for Health}
\label{app2:table3}
\end{center}
\par Note: *** Significant at 1\%, ** at 5\%, * at 10\%. Each row shows the estimates and robust standard errors for the same type of estimator. In the columns, we present the instruments these estimators used and the age groups. The interaction term is Medicaid$\times$ TANF.
\end{table}

\begin{table}[!ht]
\begin{center}
\scalebox{.7}{
\begin{tabular}{l | c c | c c | c c | c c | c c  }
\hline \hline
\multicolumn{11}{l}{\textbf{Panel A: Heterogeneous treatment effects}} \\ \hline
\textit{Employ} & \multicolumn{2}{c|}{Age: 21 - 29}  & \multicolumn{2}{c|}{Age: 30 - 38} &  \multicolumn{2}{c|}{Age: 39 - 47} & \multicolumn{2}{c|}{Age: 48 - 56}  &  \multicolumn{2}{c}{Age: 57 - 64}  \\ \hline
\textit{Estimator for $\theta_{w0}$}& $Z_1$  & $(Z_1, Z_1 X_{1})$ & $Z_1$  & $(Z_1, Z_1 X_{1})$ & $Z_1$  & $(Z_1, Z_1 X_{1})$& $Z_1$  & $(Z_1, Z_1 X_{1})$ & $Z_1$  & $(Z_1, Z_1 X_{1})$    \\ \hline
\multirow{2}{*}{GMM}&   & 0.134*  & &  -0.042 &  & 0.007  &   & -0.063 &  &  0.022\\
& & (0.071) &   & (0.067)  &  & (0.048) &  & (0.040) &   & (0.064)\\ 
\multirow{2}{*}{GMM-Lasso}&   &  0.143 &  & -0.125 &  & -0.069  &   & -0.105 &  &  -0.063\\
& & (0.112) &  & (0.118) & & (0.089) &  & (0.077) &   & (0.123)\\ 
\multirow{2}{*}{DRSMD-Lasso}& 0.255*** & 0.159**  & 0.014 & -0.067 & 0.010 & -0.001  & 0.033  & -0.058 & 0.086 & 0.072 \\
& (0.093) & (0.081) & (0.087) & (0.085) & (0.071) & (0.060) & (0.066) & (0.049) & (0.103)  & (0.086)\\
\hline
\textit{Estimator for $\theta_{wx0}$}& $Z_1$  & $(Z_1, Z_1 X_{1})$ & $Z_1$  & $(Z_1, Z_1 X_{1})$ & $Z_1$  & $(Z_1, Z_1 X_{1})$& $Z_1$  & $(Z_1, Z_1 X_{1})$ & $Z_1$  & $(Z_1, Z_1 X_{1})$    \\ \hline
\multirow{2}{*}{GMM}&   & -0.070 &  & 0.066 &  & -0.025  &   & 0.159 &  &  -0.046\\
&   & (0.128) &  & (0.120)  &  & (0.100) &   & (0.104) &   & (0.127)\\ 
\multirow{2}{*}{GMM-Lasso}&   & -0.014  &  &0.198 &  &  0.120 &   & 0.304* &  & 0.157 \\
&   & (0.179) &   & (0.145) &    & (0.150) &  & (0.162) &   & (0.193)\\ 
\multirow{2}{*}{DRSMD-Lasso}& -0.233*** &  -0.003 & -0.092** & 0.073 & 0.006 & 0.040  & -0.083**  &0.169  & -0.018 &  0.012\\
& (0.043) & (0.145) & (0.039) & (0.135) & (0.037) & (0.113) & (0.036) & (0.113) & (0.041)  & (0.153)\\
\hline
\textit{N} & \multicolumn{2}{c|}{3504}  & \multicolumn{2}{c|}{3596} &  \multicolumn{2}{c|}{3941} & \multicolumn{2}{c|}{4599}  &  \multicolumn{2}{c}{2932}  \\
\hline \hline
\end{tabular}
}
\caption[]{Heterogeneous Treatment Effects of Medicaid on Employ}
\label{app2:table10}
\end{center}
\par Note: *** Significant at 1\%, ** at 5\%, * at 10\%. Each row shows the estimates and robust standard errors for the same type of estimator. In the columns, we present the instruments these estimators used and the age groups. The interaction term is Medicaid $\times$ Above 50\% Federal Poverty Line.
\end{table}

\begin{table}[!ht]
\begin{center}
\scalebox{.7}{
\begin{tabular}{l | c c | c c | c c | c c | c c  }
\hline \hline
\multicolumn{11}{l}{\textbf{Panel A: Heterogeneous treatment effects}} \\ \hline
\textit{Out of Pocket Cost} & \multicolumn{2}{c|}{Age: 21 - 29}  & \multicolumn{2}{c|}{Age: 30 - 38} &  \multicolumn{2}{c|}{Age: 39 - 47} & \multicolumn{2}{c|}{Age: 48 - 56}  &  \multicolumn{2}{c}{Age: 57 - 64}  \\ \hline
\textit{Estimator for $\theta_{w0}$}& $Z_1$  & $(Z_1, Z_1 X_{1})$ & $Z_1$  & $(Z_1, Z_1 X_{1})$ & $Z_1$  & $(Z_1, Z_1 X_{1})$& $Z_1$  & $(Z_1, Z_1 X_{1})$ & $Z_1$  & $(Z_1, Z_1 X_{1})$    \\ \hline
\multirow{2}{*}{GMM}&   & -238.166**  & & -294.692***  &  & -32.148  &   &-149.770*** &  &  -153.110*\\
& & (94.947) &   & (92.333)  &  & (69.362) &  & (53.074) &   & (87.341)\\ 
\multirow{2}{*}{GMM-Lasso}&   &  -545.537*** &  & -569.684*** &  & -276.342**  &   & -321.387*** &  & -465.763** \\
& & (148.534) &  & (141.359) & & (131.711) &  & (91.886) &   & (189.849)\\ 
\multirow{2}{*}{DRSMD-Lasso}& -373.155*** & -250.372**  & -153.656 & -376.780*** & -55.406 &  18.780 & -162.227  & -185.345*** & 262.158 & -49.106 \\
& (136.072) & (119.078) & (142.101) & (126.053) & (99.185) & (84.691) & (101.883) & (68.418) & (197.082)  & (114.506)\\
\hline
\textit{Estimator for $\theta_{wx0}$}& $Z_1$  & $(Z_1, Z_1 X_{1})$ & $Z_1$  & $(Z_1, Z_1 X_{1})$ & $Z_1$  & $(Z_1, Z_1 X_{1})$& $Z_1$  & $(Z_1, Z_1 X_{1})$ & $Z_1$  & $(Z_1, Z_1 X_{1})$    \\ \hline
\multirow{2}{*}{GMM}&   &-101.939 &  & 384.290** &  &  -269.238* &   & 155.502  &  &  173.800\\
&   & (193.546) &  & (181.372)  &  & (156.752) &   & (165.046) &   & (211.235)\\ 
\multirow{2}{*}{GMM-Lasso}&   & 226.006  &  & 558.392** &  & -77.050  &   & 307.058 &  & 477.385 \\
&   & (245.546) &   & (243.326) &    & (209.610) &  & (276.655) &   & (370.165)\\ 
\multirow{2}{*}{DRSMD-Lasso}& 104.666*  & -179.667  & 46.142 &496.881** & -72.394* & -242.417  & 6.897  & 72.042 & -111.454* &  502.058*\\
& (62.096) & (214.131) & (53.675) & (214.081) & (43.318) & (167.287) & (45.935) & (181.569) & (57.602)  & (295.269)\\
\hline
\textit{N} & \multicolumn{2}{c|}{3504}  & \multicolumn{2}{c|}{3596} &  \multicolumn{2}{c|}{3941} & \multicolumn{2}{c|}{4599}  &  \multicolumn{2}{c}{2932}  \\
\hline \hline
\end{tabular}
}
\caption[]{Heterogeneous Treatment Effects of Medicaid on Out of Pocket Cost}
\label{app2:table11}
\end{center}
\par Note: *** Significant at 1\%, ** at 5\%, * at 10\%. Each row shows the estimates and robust standard errors for the same type of estimator. In the columns, we present the instruments these estimators used and the age groups. The interaction term is Medicaid $\times$ Above 50\% Federal Poverty Line.
\end{table}

\end{landscape}

\section{Appendix: Proofs of the Theoretical Results}

\subsection{Equivalence between the Objective Functions (\ref{eqboj1}) and (\ref{eqboj2})}
\begin{proof}
The objective function (\ref{eqboj1}) can be written as
$$M_{\infty}(\theta, g) = \int_{\mathbb R^{q_z}} E[\epsilon_j(\theta,g) e^{it'Z_j}] E[\epsilon_l(\theta,g) e^{-it'Z_l}] d\mu(t)$$
From Assumption \ref{ass:indept}, for all $j \neq l$, $Cov(\epsilon_j(\theta,g) e^{it'Z_j}, \epsilon_l(\theta,g) e^{-it'Z_l}) = 0$. Thus, for all $j \neq l$, we have:
\begin{eqnarray*}
M_{\infty}(\beta) &=& \int_{\mathbb{R}^{q_z}} E(\epsilon_j(\theta,g) e^{it'Z_j}\epsilon_l(\theta,g) e^{-it'Z_l}) d \mu(t) \\
&=& \int_{\mathbb{R}^{q_z}} E( \epsilon_j(\theta,g) \epsilon_l(\theta,g)   e^{it'(Z_j-Z_l)}) d \mu(t)  \\
&=& E(\int_{\mathbb{R}^{q_z}}  \epsilon_j(\theta,g) \epsilon_l(\theta,g)  e^{it'(Z_j-Z_l)}d \mu(t))
\end{eqnarray*}
Thus, the objective function becomes
$$M_{\infty}(\beta) =E( \epsilon_j(\theta,g) \epsilon_l(\theta,g)  \kappa_{j,l})$$
where $\kappa_{j,l} = k(Z_j-Z_l)=\int_{\mathbb{R}^{q_z}} e^{it'(Z_j-Z_l)} d \mu(t)$. And $k(u)$ is the inverse Fourier transform of $d \mu(t)$ with $u=Z_j-Z_l$.
\end{proof}

\subsection{Proof of Proposition \ref{prop:iden}}

The orthogonal FOC in Equation (\ref{eq:ofoc01}) implies that
\begin{equation*}
E\left[ \left( \tilde P_j -  \frac{g_{0,\tilde P_{m}}(X_l)}{g_{0,\kappa_{m,l}}(X_l)} \right) (\tilde y_l - \tilde P_l' \theta) \kappa_{j,l} \right] = 0
\end{equation*}
\begin{equation*}
E\left[\kappa_{j,l} \left( \tilde P_j -  \frac{g_{0,\tilde P_{m}}(X_l)}{g_{0,\kappa_{m,l}}(X_l)} \right) \tilde y_l  - \kappa_{j,l} \left( \tilde P_j -  \frac{g_{0,\tilde P_{m}}(X_l)}{g_{0,\kappa_{m,l}}(X_l)} \right)\tilde P_l' \theta \right] = 0
\end{equation*}
\begin{equation*}
\theta_0 =  E\left[ \kappa_{j,l}\left( \tilde P_j -  \frac{g_{0,\tilde P_{m}}(X_l)}{g_{0,\kappa_{m,l}}(X_l)} \right)\tilde P_l' \right] ^{-1}  E\left[ \kappa_{j,l}\left( \tilde P_j -  \frac{g_{0,\tilde P_{m}}(X_l)}{g_{0,\kappa_{m,l}}(X_l)} \right) \tilde y_l \right]
\end{equation*}

The proof for invertibility of $E\left[ \kappa_{j,l}\left( \tilde P_j -  \frac{g_{0,\tilde P_{m}}(X_l)}{g_{0,\kappa_{m,l}}(X_l)} \right)\tilde P_l' \right]$ follows the same steps of proofs for identification in \cite{AS2021}.

\subsection{Proof of Orthogonal Properties of Equation (\ref{eqfoc0}) and (\ref{eq:ofoc01})}

In this section, we check the orthogonal properties of the two FOCs.

$$M_{\infty}(\theta,g) = - \frac{1}{2} E[\epsilon_j(\theta) \epsilon_l(\theta)\kappa_{j,l}] = - \frac{1}{2} E[(\tilde y_j - \tilde P_j' \theta) (\tilde y_l - \tilde P_l' \theta)\kappa_{j,l}]$$ where $g$ stands for all nuisance parameters.

\;

\textbf{The FOC defined in Equation (\ref{eqfoc0}) is not orthogonal}

The key parameter is $\theta$, the true value is $\theta_0$. The nuisance parameters are $g_{P}(X_i)$ and $g_{y}(X_i)$. Their true values are $E(P_i|X_i) = g_{0,P}(X_i)$ and $E(y_i|X_i) = g_{0,y}(X_i)$.

The first order condition is written as follows:
\begin{equation*}
E[(P_j - g_{P}(X_j)) (y_l-g_y(X_l) - [P_l - g_{P}(X_l)]' \theta)\kappa_{j,l}]=0
\end{equation*}
If $P_i = [W_i, f(W_i, X_i)]'$ and there are only one variable in $X_i$, $\partial_{\theta} \varphi(D; \theta, g)$ becomes:
\begin{equation*}
\partial_{\theta} \varphi(D; \theta, g) = \begin{bmatrix}(P_{1j} - g_{P_1}(X_j))[y_l-g_y(X_l) - (P_{1l} - g_{P_1}(X_l)) \theta_1 - (P_{2l} - g_{P_2}(X_l)) \theta_2] \kappa_{j,l}\\ (P_{2j} - g_{P_2}(X_j))[y_l-g_y(X_l) - (P_{1l} - g_{P_1}(X_l)) \theta_1 - (P_{2l} - g_{P_2}(X_l)) \theta_2] \kappa_{j,l} \end{bmatrix}
\end{equation*}

Prove that $\partial_{g} E[\partial_{\theta} \varphi(D; \theta_0, g_0)] \neq 0$.

\begin{proof}

The first row of $\partial_{\theta} \varphi(D; \theta_0, g_0 + r(g-g_0))$ is defined as $I$ where
\begin{eqnarray*}
I =& [P_{1j} - g_{0,P_1}(X_j) - r(g_{P_1}(X_j) - g_{0,P_1}(X_j))]\\
 & [y_l-g_{0,y}(X_l) - r(g_{y}(X_l) - g_{0,y}(X_l))\\
  -& (P_{1l} - g_{0,P_1}(X_l)- r(g_{P_1}(X_l) - g_{0,P_1}(X_l))) \theta_1\\
  -& (P_{2l} - g_{0,P_2}(X_l)- r(g_{P_2}(X_l) - g_{0,P_2}(X_l))) \theta_2] \kappa_{j,l}
\end{eqnarray*}
According to the definition for $\partial_{g} E[\partial_{\theta} \varphi(D; \theta_0, g_0)] \neq 0$ in \cite{ChernozhukovVictor2018Dmlf}, to show that $\partial_{g} E[\partial_{\theta} \varphi(D; \theta_0, g_0)] \neq 0$ we need to show $\partial_r E[ I]|_{r=0} \neq 0$. Under a regularity condition, for instance, $\int_{\theta}^{}\int_{r}^{}|\partial_r E[ I]|_{r=0} | dr d\theta < \infty$, we have the following:
\begin{equation*}
\partial_r E[ I]|_{r=0}  = -I_1 - I_2 + I_3 + I_4
\end{equation*}
\begin{eqnarray*}
I_1 &=& E[(g_{P_1}(X_j) - g_{0,P_1}(X_j))(y_l-g_{0,y}(X_l) \\
    &-& (P_{1l} - g_{0,P_1}(X_l)) \theta_{0,1} - (P_{2l} - g_{0,P_2}(X_l)) \theta_{0,2}) \kappa_{j,l}] \\
 &=&E\left[(g_{P_1}(X_j) - g_{0,P_1}(X_j)) \epsilon_l \kappa_{j,l}\right]\\
&=&\int_{\mathbb{R}^{q_z}} E[(g_{P_1}(X_j) - g_{0,P_1}(X_j)) \epsilon_l  e^{it'(Z_j - Z_l)} ]d \mu(t)\\
&=& \int_{\mathbb{R}^{q_z}}  E[(g_{P_1}(X_j) - g_{0,P_1}(X_j)) e^{it' Z_j}] E[\epsilon_l e^{-it' Z_l} ] d \mu(t) = 0
\end{eqnarray*}
The first equation sign is a result of notation. The second one is derived based on the outcomes following the Robinson Transformation. The third equality arises from the definition of $\kappa_{j,l}$. The fourth one is inferred from Assumption \ref{ass:indept}. The final equal sign is a consequence of Assumption \ref{ass:exo}.
\begin{eqnarray*}
I_2 &=& E\left[(P_{1j} - g_{0,P_1}(X_j)) (g_{y}(X_l) - g_{0,y}(X_l))  \kappa_{j,l}\right] \neq 0 \\
I_3 &=& E\left[(P_{1j} - g_{0,P_1}(X_j)) (g_{P_1}(X_l) - g_{0,P_1}(X_l)) \theta_1 \kappa_{j,l}\right] \neq 0 \\
I_4 &=& E\left[(P_{1j} - g_{0,P_1}(X_j)) (g_{P_2}(X_l) - g_{0,P_2}(X_l)) \theta_2  \kappa_{j,l}\right] \neq 0
\end{eqnarray*}
Hence, $\partial_{g} E[ \partial_{\theta} \varphi(D; \theta_0, g_0)]$ is probably not 0. 

\end{proof}

\textbf{The FOC defined in Equation (\ref{eq:ofoc01}) is orthogonal}

With $E[(P_{1m} - g_{0,P_1}(X_m)) \kappa_{m,l} |X_l] = g_{0,\tilde P_{1,j}}(X_l)$, $E[\kappa_{m,l} |X_l]  = g_{0,\kappa_{m,l}}(X_l)$,

$\Psi(D; \theta, g)$
\begin{equation*}
= \begin{bmatrix}\left( P_{1j} - g_{P_1}(X_j) - \frac{g_{\tilde P_{1,j}}(X_l)}{g_{\kappa_{j,l}}(X_l)} \right) [y_l-g_y(X_l) - (P_{1l} - g_{P_1}(X_l)) \theta_1 - (P_{2l} - g_{P_2}(X_l)) \theta_2] \kappa_{j,l}\\ \left( P_{2j} - g_{P_2}(X_j) - \frac{g_{\tilde P_{2,j}}(X_l)}{g_{\kappa_{j,l}}(X_l)}  \right) [y_l-g_y(X_l) - (P_{1l} - g_{P_1}(X_l)) \theta_1 - (P_{2l} - g_{P_2}(X_l)) \theta_2] \kappa_{j,l} \end{bmatrix}
\end{equation*}

Prove that $\partial_{g} E[ \Psi(D; \theta_0, g_0)] = 0$.

\begin{proof}

The first row of $\Psi(D; \theta_0, g_0 + r(g-g_0))$ is $I'$.
\begin{eqnarray*}
I' =& \left(P_{1j} - g_{0,P_1}(X_j) - r(g_{P_1}(X_j) - g_{0,P_1}(X_j)) - \frac{g_{0,\tilde P_{1,j}}(X_l) + r[g_{\tilde P_{1,j}}(X_l)- g_{0,\tilde P_{1,j}}(X_l)]}{g_{0,\kappa_{m,l}}(X_l) + r[g_{\kappa_{j,l}}(X_l)-g_{0,\kappa_{m,l}}(X_l)]} \right) \\
 & [y_l-g_{0,y}(X_l) - r(g_{y}(X_l) - g_{0,y}(X_l)) \\
 -& (P_{1l} - g_{0,P_1}(X_l)- r(g_{P_1}(X_l) - g_{0,P_1}(X_l))) \theta_1 \\
-& (P_{2l} - g_{0,P_2}(X_l)- r(g_{P_2}(X_l) - g_{0,P_2}(X_l))) \theta_2] \kappa_{j,l} \\
\end{eqnarray*}
\begin{equation*}
\partial_r E[I']|_{r=0}  = -I_1 - I_2' + I_3' + I_4' - I_5
\end{equation*}
\begin{eqnarray*}
I_1 &=& E[(g_{P_1}(X_j) - g_{0,P_1}(X_j))(y_l-g_{0,y}(X_l) - (P_{1l} - g_{0,P_1}(X_l)) \theta_{0,1} \\
 &-& (P_{2l} - g_{0,P_2}(X_l)) \theta_{0,2}) \kappa_{j,l}] \\
&=& E[(g_{P_1}(X_j) - g_{0,P_1}(X_j)) \epsilon_l \kappa_{j,l}] \\
&=& 0
\end{eqnarray*}
The $I_1$ term is the same as the $I_1$ in the previous proof, which demonstrates that the FOC defined in Equation (\ref{eqfoc0}) is not orthogonal. Therefore, it remains 0.
\begin{eqnarray*}
I_2' &=& E[(P_{1j} - g_{0,P_1}(X_j) - \frac{E[(P_{1m} - g_{0,P_1}(X_m)) \kappa_{m,l} |X_l]}{E[\kappa_{m,l} |X_l]}) (g_{y}(X_l) - g_{0,y}(X_l))  \kappa_{j,l}] \\
&=& E( E\left[ \left(P_{1j} - g_{0,P_1}(X_j) - \frac{E[(P_{1m} - g_{0,P_1}(X_m)) \kappa_{m,l} |X_l]}{E[\kappa_{m,l} |X_l]} \right) \kappa_{j,l} | X_l\right] \\
 & & (g_{y}(X_l) - g_{0,y}(X_l) ) \\
&=& E( E\left[ \left(P_{1j} - g_{0,P_1}(X_j) \right) \kappa_{j,l} - \frac{E[(P_{1m} - g_{0,P_1}(X_m)) \kappa_{m,l} |X_l]}{E[\kappa_{m,l} |X_l]} \kappa_{j,l} | X_l\right] \\
& &  (g_{y}(X_l) - g_{0,y}(X_l) ) \\
&=& 0
\end{eqnarray*} because $E\left[ \left(P_{1j} - g_{0,P_1}(X_j) \right) \kappa_{j,l} - \frac{E[(P_{1m} - g_{0,P_1}(X_m)) \kappa_{m,l} |X_l]}{E[\kappa_{m,l} |X_l]} \kappa_{j,l} | X_l\right] = 0$. To see this, notice that after conditional on $X_l$, it becomes $E\left[ \tilde P_{1j} \kappa_{j,l}| X_l\right] - E\left[ \frac{E[\tilde P_{1m} \kappa_{m,l} |X_l]}{E[\kappa_{m,l} |X_l]} \kappa_{j,l} | X_l\right]$, which is 0 with $\tilde P_{1j}=  \left(P_{1j} - g_{0,P_1}(X_j) \right)$. Following the same idea, we have the next equations for $I_3'$ and $I_4'$:
\begin{eqnarray*}
I_3' &=& E\left[(P_{1j} - g_{0,P_1}(X_j) - \frac{E[(P_{1m} - g_{0,P_1}(X_m)) \kappa_{m,l} |X_l]}{E[\kappa_{m,l} |X_l]}) (g_{P_1}(X_l) - g_{0,P_1}(X_l)) \theta_1 \kappa_{j,l}\right]\\
 &=& 0 \\
I_4' &=& E\left[(P_{1j} - g_{0,P_1}(X_j) - \frac{E[(P_{1m} - g_{0,P_1}(X_m)) \kappa_{m,l} |X_l]}{E[\kappa_{m,l} |X_l]}) (g_{P_2}(X_l) - g_{0,P_2}(X_l)) \theta_2  \kappa_{j,l}\right] \\
&=& 0 \\
I_5 &=&  E\left[\frac{ \left(g_{\tilde P_{1,j}}(X_l)- g_{0,\tilde P_{1,j}}(X_l) \right)g_{0,\kappa_{m,l}}(X_l)-\left(g_{\kappa_{j,l}}(X_l)-g_{0,\kappa_{m,l}}(X_l)\right)g_{0,\tilde P_{1,j}}(X_l)}{(g_{0,\kappa_{m,l}}(X_l))^2} \epsilon_l \kappa_{j,l}\right] \\
&=&  E\left[g(X_l) \epsilon_l \kappa_{j,l}\right ] =  \int_{\mathbb{R}^{q_z}} E[g(X_l) \epsilon_l  e^{it'(Z_j - Z_l)} ]d \mu(t) =  \int_{\mathbb{R}^{q_z}}  E[ e^{it' Z_j}] E[\epsilon_l g(X_l) e^{-it' Z_l} ] d \mu(t) \\
&=& 0
\end{eqnarray*}
For $I_5$, the first equal sign is a result of notation. The second one serves a notation purpose, summarizing the fraction into a function of $X_l$. The third equality arises from the definition of $\kappa_{j,l}$. The fourth one is inferred from Assumption \ref{ass:indept}. The final equal sign is a consequence of Assumption \ref{ass:exo}.

Hence, $\partial_{g} E[ \Psi(D; \theta_0, g_0)] = 0$.

\end{proof}
\subsection{Proof of Proposition \ref{thm:infeas consistency normality}}

\begin{proof}

\begin{eqnarray*}
\tilde \theta_{n,o} =& \left[ \frac{1}{n(n-1)} \sum_{j = 1}^n \sum_{l \neq j}^n \kappa_{j,l}\left( \tilde P_j -  \frac{g_{0,\tilde P_{m}}(X_l)}{g_{0,\kappa_{m,l}}(X_l)} \right)\tilde P_l' \right]^{-1} \\
 & \left[\frac{1}{n(n-1)}\sum_{j = 1}^n \sum_{l \neq j}^n \kappa_{j,l}\left( \tilde P_j -  \frac{g_{0,\tilde P_{m}}(X_l)}{g_{0,\kappa_{m,l}}(X_l)} \right) (\tilde P_l \theta_0 + \epsilon_l) \right] \\
=& \theta_0+ \left[ \frac{1}{n(n-1)} \sum_{j = 1}^n \sum_{l \neq j}^n \kappa_{j,l}\left( \tilde P_j -  \frac{g_{0,\tilde P_{m}}(X_l)}{g_{0,\kappa_{m,l}}(X_l)} \right)\tilde P_l' \right]^{-1} \\
 & \left[\frac{1}{n(n-1)}\sum_{j = 1}^n \sum_{l \neq j}^n \kappa_{j,l}\left( \tilde P_j -  \frac{g_{0,\tilde P_{m}}(X_l)}{g_{0,\kappa_{m,l}}(X_l)} \right) \epsilon_l \right]
\end{eqnarray*}
Denote $A_n =  \frac{1}{n(n-1)} \sum_{j = 1}^n \sum_{l \neq j}^n \kappa_{j,l}\left( \tilde P_j -  \frac{g_{0,\tilde P_{m}}(X_l)}{g_{0,\kappa_{m,l}}(X_l)} \right)\tilde P_l'$ and

$B_n = \frac{1}{n(n-1)}\sum_{j = 1}^n \sum_{l \neq j}^n \kappa_{j,l}\left( \tilde P_j -  \frac{g_{0,\tilde P_{m}}(X_l)}{g_{0,\kappa_{m,l}}(X_l)} \right) \epsilon_l$. We first show that $A_n$ is a U-statistic and find its probability limit. Then we show that $B_n$ is also a U-statistic and find its probability limit. \\
To show that $A_n$ is a U-statistic, notice that
$$A_n = \frac{1}{n(n-1)} \sum_{j = 1}^n \sum_{l \neq j}^n \kappa_{j,l}\left( \tilde P_j -  \frac{g_{0,\tilde P_{m}}(X_l)}{g_{0,\kappa_{m,l}}(X_l)} \right)\tilde P_l'$$
$$= \frac{1}{2}\frac{2}{n(n-1)} \sum_{j < l}^n \left(\kappa_{j,l} \left( \tilde P_j -  \frac{g_{0,\tilde P_{m}}(X_l)}{g_{0,\kappa_{m,l}}(X_l)} \right)\tilde P_l' + \kappa_{l,j} \left( \tilde P_l -  \frac{g_{0,\tilde P_{m}}(X_j)}{g_{0,\kappa_{m,j}}(X_j)} \right)\tilde P_j'\right)$$
According to WLLN for U-statistics under Assumption \ref{ass:regul 1},
$$A_n \xrightarrow{p} A \qquad \text{with} \qquad A \equiv E\left[\kappa_{j,l}\left( \tilde P_j -  \frac{g_{0,\tilde P_{m}}(X_l)}{g_{0,\kappa_{m,l}}(X_l)} \right)\tilde P_l'\right]$$
and under Assumption \ref{ass:identification}, $A$ is nonsingular.\\
To show that $B_n$ is also a U-statistic, notice that:
$$B_n = \frac{1}{n(n-1)} \sum_{j<l}^n  \left(\kappa_{j,l} \left( \tilde P_j -  \frac{g_{0,\tilde P_{m}}(X_l)}{g_{0,\kappa_{m,l}}(X_l)} \right) \epsilon_l + \kappa_{l,j} \left( \tilde P_l -  \frac{g_{0,\tilde P_{m}}(X_j)}{g_{0,\kappa_{m,j}}(X_j)} \right) \epsilon_j \right)$$
Define $h(\tilde p_1,e_1,z_1,x_1; \tilde p_2,e_2,z_2,x_2)= \kappa_{1,2} \tilde p_1 e_2+\kappa_{2,1} \tilde p_2 e_1 - \kappa_{1,2} \frac{g_{0,\tilde P_{1}}(x_2)}{g_{0,\kappa_{1,2}}(x_2)} e_2 - \kappa_{2,1} \frac{g_{0,\tilde P_{2}}(x_1)}{g_{0,\kappa_{2,1}}(x_1)} e_1$. Since $h$ is a symmetric function of observations 1 and 2, a U-statistic with kernel $h$ is defined as
$$B_n' = \frac{2}{n(n-1)} \sum_{j<l}^n h(\tilde P_j,\epsilon_j,Z_j,X_j; \tilde P_l,\epsilon_l,Z_l,X_l)
\qquad \text{and} \qquad B_n = \frac{1}{2} B_n'$$
And we have:
\begin{eqnarray*}
E(B_n') &=& E(\kappa_{j,l} \tilde P_j \epsilon_l+\kappa_{l,j} \tilde P_l \epsilon_j)- \left( E\left(  \kappa_{j,l} \frac{g_{0,\tilde P_{m}}(X_l)}{g_{0,\kappa_{m,l}}(X_l)} \epsilon_l\right) + E\left(\kappa_{l,j} \frac{g_{0,\tilde P_{m}}(X_j)}{g_{0,\kappa_{m,j}}(X_j)} \epsilon_j \right)\right)\\
&=& 2E(\kappa_{j,l} \tilde P_j \epsilon_l) - 2 E\left(  \kappa_{j,l} \frac{g_{0,\tilde P_{m}}(X_l)}{g_{0,\kappa_{m,l}}(X_l)} \epsilon_l\right) \\
&=& 2 \int_{\mathbb{R}^{q_z}} E[\tilde P_j \epsilon_l  e^{it'(Z_j - Z_l)} ]d \mu(t) - 2 \int_{\mathbb{R}^{q_z}} E[\frac{g_{0,\tilde P_{m}}(X_l)}{g_{0,\kappa_{m,l}}(X_l)} \epsilon_l  e^{it'(Z_j - Z_l)} ]d \mu(t) \\
&=& 2\int_{\mathbb{R}^{q_z}}  E[\tilde P_j e^{it' Z_j} \epsilon_l e^{-it' Z_l} ] d \mu(t) - 2 \int_{\mathbb{R}^{q_z}} E[\frac{g_{0,\tilde P_{m}}(X_l)}{g_{0,\kappa_{m,l}}(X_l)} e^{it' Z_j} \epsilon_l e^{-it' Z_l} ]d \mu(t) \\
&=& 2\int_{\mathbb{R}^{q_z}}  E[\tilde P_j e^{it' Z_j}] E[\epsilon_l e^{-it' Z_l} ] d \mu(t) - 2 \int_{\mathbb{R}^{q_z}} E[ e^{it' Z_j} ]E[\frac{g_{0,\tilde P_{m}}(X_l)}{g_{0,\kappa_{m,l}}(X_l)}\epsilon_l e^{-it' Z_l} ]d \mu(t) \\
&=& 0 \qquad \text{since $E[\epsilon_l'  e^{-it' Z_l} ]=0$ and $E[\epsilon_l|X_l,Z_l]=0$. }
\end{eqnarray*}
Hence, $E(B_n)= 0$. According to WLLN for U statistics, we have $B_n \xrightarrow{p} 0$, and we conclude that $\tilde\beta_n$ is a consistent estimator of $\beta_0$.

To derive the asymptotic normality, we first need to compute the asymptotic variance for the U-statistic $B_n'$, which means that we need to find the variance for the following:

$E(h(\tilde P_1,\epsilon_1,Z_1,X_1; \tilde P_2,\epsilon_2,Z_2,X_2)|\tilde P_1=\tilde p_1,\epsilon_1=e_1,Z_1=z_1,X_1 = x_1)$. \\
Let $h_1(\tilde p_1,e_1,z_1,x_1) \equiv E(h(\tilde P_1,\epsilon_1,Z_1,X_1; \tilde P_2,\epsilon_2,Z_2,X_2)|\tilde P_1=\tilde p_1,\epsilon_1=e_1,Z_1=z_1,X_1 = x_1)$. We have:
\begin{equation*}
h(\tilde p_1,e_1,z_1,x_1; \tilde p_2,e_2,z_2,x_2)= \kappa_{1,2} \tilde p_1 e_2+\kappa_{2,1} \tilde p_2 e_1 - \kappa_{1,2} \frac{g_{0,\tilde P_{1}}(x_2)}{g_{0,\kappa_{1,2}}(x_2)} e_2 - \kappa_{2,1} \frac{g_{0,\tilde P_{2}}(x_1)}{g_{0,\kappa_{2,1}}(x_1)} e_1
\end{equation*}
and
\begin{eqnarray*}
h_1(\tilde p_1,e_1,z_1,x_1)
& =& E[\int_{\mathbb{R}^{q_z}} e^{it'(z_1 - Z_2)} d \mu(t)\tilde p_1 \epsilon_2+\int_{R^{q_z}} e^{it'(Z_2 - z_1)} d \mu(t) \tilde P_2 e_1 \\
& -& \int_{\mathbb{R}^{q_z}} e^{it'(z_1 - Z_2)} d \mu(t)\frac{g_{0,\tilde P_{1}}(X_2)}{g_{0,\kappa_{1,2}}(X_2)} \epsilon_2-\int_{R^{q_z}} e^{it'(Z_2 - z_1)} d \mu(t)  \frac{g_{0,\tilde P_{2}}(x_1)}{g_{0,\kappa_{2,1}}(x_1)} e_1 ]\\
& =& E[\int_{\mathbb{R}^{q_z}} e^{it'(z_1 - Z_2)} d \mu(t)\tilde p_1 \epsilon_2]+E[\int_{R^{q_z}} e^{it'(Z_2 - z_1)} d \mu(t) \tilde P_2 e_1] \\
& -& E[\int_{\mathbb{R}^{q_z}} e^{it'(z_1 - Z_2)} d \mu(t)\frac{g_{0,\tilde P_{1}}(X_2)}{g_{0,\kappa_{1,2}}(X_2)} \epsilon_2] - E[\int_{R^{q_z}} e^{it'(Z_2 - z_1)} d \mu(t)  \\
& & \frac{g_{0,\tilde P_{2}}(x_1)}{g_{0,\kappa_{2,1}}(x_1)} e_1 ]
\end{eqnarray*}
The first element of the right hand side is
\begin{eqnarray*}
E[\int_{\mathbb{R}^{q_z}} e^{it'(z_1 - Z_2)} d \mu(t)\tilde p_1 \epsilon_2]
& =& \int_{\mathbb{R}^{q_z}}E[ e^{it'z_1} e^{- it'Z_2} \tilde p_1 \epsilon_2]d \mu(t) \\
& =& \int_{\mathbb{R}^{q_z}}e^{it'z_1} \tilde p_1 E[  e^{- it'Z_2} \epsilon_2]d \mu(t) = 0
\end{eqnarray*}
The third term is
\begin{eqnarray*}
E[\int_{\mathbb{R}^{q_z}} e^{it'(z_1 - Z_2)} d \mu(t)\frac{g_{0,\tilde P_{1}}(X_2)}{g_{0,\kappa_{1,2}}(X_2)} \epsilon_2]
& =& \int_{\mathbb{R}^{q_z}}E[ e^{it'z_1} e^{- it'Z_2} \frac{g_{0,\tilde P_{1}}(X_2)}{g_{0,\kappa_{1,2}}(X_2)} \epsilon_2]d \mu(t) \\
& =& \int_{\mathbb{R}^{q_z}}e^{it'z_1} \tilde p_1 E[  e^{- it'Z_2} \frac{g_{0,\tilde P_{1}}(X_2)}{g_{0,\kappa_{1,2}}(X_2)} \epsilon_2]d \mu(t) = 0
\end{eqnarray*}
The second term of the right hand side is
\begin{eqnarray*}
E[\int_{R^{q_z}} e^{it'(Z_2 - z_1)} d \mu(t) \tilde P_2 e_1]
& =& \int_{\mathbb{R}^{q_z}} E[e^{it'Z_2} e^{-it' z_1}  \tilde P_2 e_1]d \mu(t) \\
& =& \int_{\mathbb{R}^{q_z}}e^{-it' z_1} e_1 E[e^{it'Z_2} \tilde P_2 ]d \mu(t)
\end{eqnarray*}
The fourth term is
\begin{eqnarray*}
E[\int_{R^{q_z}} e^{it'(Z_2 - z_1)} d \mu(t)  \frac{g_{0,\tilde P_{2}}(x_1)}{g_{0,\kappa_{2,1}}(x_1)} e_1 ]
& =& \int_{\mathbb{R}^{q_z}} E[e^{it'Z_2} e^{-it' z_1}  \frac{g_{0,\tilde P_{2}}(x_1)}{g_{0,\kappa_{2,1}}(x_1)}  e_1]d \mu(t) \\
& =& \int_{\mathbb{R}^{q_z}}e^{-it' z_1} e_1  \frac{g_{0,\tilde P_{2}}(x_1)}{g_{0,\kappa_{2,1}}(x_1)}  E[e^{it'Z_2} ]d \mu(t)
\end{eqnarray*}
Hence, $h_1(\tilde p_1,e_1,z_1,x_1) = \int_{\mathbb{R}^{q_z}}e^{-it' z_1} e_1 E[e^{it'Z_2} \tilde P_2 ]d \mu(t) - \int_{\mathbb{R}^{q_z}}e^{-it' z_1} e_1  \frac{g_{0,\tilde P_{2}}(x_1)}{g_{0,\kappa_{2,1}}(x_1)}  E[e^{it'Z_2} ]d \mu(t)$.\\
$h_1(\tilde p_1,e_1,z_1,x_1) = \int_{\mathbb{R}^{q_z}}e^{-it' z_1} e_1 \left(E[e^{it'Z_2} \tilde P_2 ]-\frac{g_{0,\tilde P_{2}}(x_1)}{g_{0,\kappa_{2,1}}(x_1)}  E[e^{it'Z_2} ] \right) d \mu(t)$\\
Since $E(h_1(\tilde P_1,\epsilon_1,Z_1,X_1))=0$, we have:
\begin{eqnarray*}
Var[h_1(\tilde P_1,\epsilon_1,Z_1,X_1)]
&=& Var \left[\int_{\mathbb{R}^{q_z}}e^{-it' Z_1} \epsilon_1 \left(E[e^{it'Z_2} \tilde P_2 ]-\frac{g_{0,\tilde P_{2}}(X_1)}{g_{0,\kappa_{2,1}}(X_1)}  E[e^{it'Z_2} ] \right) d \mu(t) \right] \\
&=& Var\left[\int_{\mathbb{R}^{q_z}}e^{-it' Z_j} \epsilon_j \left(E[e^{it'Z_l} \tilde P_l ]-\frac{g_{0,\tilde P_{m}}(X_j)}{g_{0,\kappa_{m,j}}(X_j)}  E[e^{it'Z_l} ] \right) d \mu(t) \right]
\end{eqnarray*}
Following \cite{Hoeffding1948}, the asymptotic distribution for U-statistics yields:
$$\sqrt{n}(B_n'-0) \stackrel{d}{\rightarrow} \mathcal{N}(0, 4Var[h_1(\tilde P_1,\epsilon_1,Z_1,X_1)] )$$
Thus,
$$\sqrt{n}(\tilde \theta_{n,o} - \theta_0) \xrightarrow{d} N \left(0, A^{-1} \Sigma \left(A^{-1} \right)' \right)$$
with $A = E\left[\kappa_{j,l}\left( \tilde P_j - \frac{g_{0,\tilde P_{m}}(X_l)}{g_{0,\kappa_{m,l}}(X_l)} \right)\tilde P_l'\right]$ and $\Sigma= Var[h_1(\tilde P_1,\epsilon_1,Z_1,X_1)]$.
\end{proof}

\subsection{Proof of Theorem \ref{thm:consis asymp norm Debias RSMD}}
\begin{proof}
Recall that
$$\widehat{\tilde{y}}_i = y_i - \hat g_y(X_i) = y_i -E(y_i|X_i)+E(y_j|X_i)- \hat g_y(X_i)$$
For the first two terms of the right hand side, we have
\begin{eqnarray*}
y_i -E(y_i|X_i)
& =& (P_i- E(P_i|X_i))' \theta_0 + \epsilon_i \\
& =& (P_i - \hat g_{P}(X_i))' \theta_0 +(\hat g_{P}(X_i)- E(P_i|X_i))' \theta_0 + \epsilon_i
\end{eqnarray*}
In matrix form, the feasible estimator writes:
\begin{eqnarray*}
\hat \theta_{n,o}& =& \left[\frac{1}{n(n-1)} \sum_{j = 1}^n \sum_{l \neq j}^n  \kappa_{j,l} \left(\widehat{\tilde{P}}_j -  \frac{\widehat{g_{0,\tilde P_{m}}}(X_l)}{\widehat{g_{0,\kappa_{m,l}}}(X_l)} \right) \widehat{\tilde{P_l}}'\right]^{-1} \\
& & \left[\frac{1}{n(n-1)} \sum_{j = 1}^n \sum_{l \neq j}^n \kappa_{j,l} \left(\widehat{\tilde{P}}_j -  \frac{\widehat{g_{0,\tilde P_{m}}}(X_l)}{\widehat{g_{0,\kappa_{m,l}}}(X_l)} \right) \widehat{\tilde{y}}_l\right]
\end{eqnarray*}
Define $C_n =\frac{1}{n(n-1)} \sum_{j = 1}^n \sum_{l \neq j}^n  \kappa_{j,l} \left(\widehat{\tilde{P}}_j -  \frac{\widehat{g_{0,\tilde P_{m}}}(X_l)}{\widehat{g_{0,\kappa_{m,l}}}(X_l)} \right) \widehat{\tilde{P_l}}'$. We get:
\begin{eqnarray*}
\hat \theta_{n,o}
& =& [C_n]^{-1} [\frac{1}{n(n-1)} \sum_{j = 1}^n \sum_{l \neq j}^n \kappa_{j,l} \left(\widehat{\tilde{P}}_j -  \frac{\widehat{g_{0,\tilde P_{m}}}(X_l)}{\widehat{g_{0,\kappa_{m,l}}}(X_l)} \right) \\
& & [y_l -E(y_l|X_l)+E(y_l|X_l)- \hat g_y(X_l)]] \\
& =& [C_n]^{-1} [\frac{1}{n(n-1)} \sum_{j = 1}^n \sum_{l \neq j}^n \kappa_{j,l} \left(\widehat{\tilde{P}}_j -  \frac{\widehat{g_{0,\tilde P_{m}}}(X_l)}{\widehat{g_{0,\kappa_{m,l}}}(X_l)} \right) \\
&&[(P_l - \hat g_{P}(X_l))' \theta_0 +(\hat g_{P}(X_l)- E(P_l|X_l))' \theta_0 + \epsilon_l +E(y_l|X_l)- \hat g_y(X_l)] ] \\
& =& \theta_0 + [C_n]^{-1} [\frac{1}{n(n-1)} \sum_{j = 1}^n \sum_{l \neq j}^n \kappa_{j,l} \left(\widehat{\tilde{P}}_j -  \frac{\widehat{g_{0,\tilde P_{m}}}(X_l)}{\widehat{g_{0,\kappa_{m,l}}}(X_l)} \right) \\
&& [(\hat g_{P}(X_l)- E(P_l|X_l))' \theta_0 + \epsilon_l +E(y_l|X_l)- \hat g_y(X_l)] ]
\end{eqnarray*}
Consider now,
\begin{eqnarray*}
A_n -C_n
& =& \frac{1}{n(n-1)} \sum_{j = 1}^n \sum_{l \neq j}^n \kappa_{j,l}\left( \tilde P_j -  \frac{g_{0,\tilde P_{m}}(X_l)}{g_{0,\kappa_{m,l}}(X_l)} \right)\tilde P_l' \\
&& - \frac{1}{n(n-1)} \sum_{j = 1}^n \sum_{l \neq j}^n  \kappa_{j,l} \left(\widehat{\tilde{P}}_j -  \frac{\widehat{g_{0,\tilde P_{m}}}(X_l)}{\widehat{g_{0,\kappa_{m,l}}}(X_l)} \right) \widehat{\tilde{P_l}}' \\
& =& \frac{1}{n(n-1)} \sum_{j = 1}^n \sum_{l \neq j}^n \kappa_{j,l}\left( \tilde P_j -  \frac{g_{0,\tilde P_{m}}(X_l)}{g_{0,\kappa_{m,l}}(X_l)} \right)\tilde P_l' \\
&&- \frac{1}{n(n-1)} \sum_{j = 1}^n \sum_{l \neq j}^n  \kappa_{j,l}\\
&& \left(\tilde P_j + \widehat{\tilde{P}}_j -\tilde P_j -\frac{g_{0,\tilde P_{m}}(X_l)}{g_{0,\kappa_{m,l}}(X_l)} + \frac{g_{0,\tilde P_{m}}(X_l)}{g_{0,\kappa_{m,l}}(X_l)} -  \frac{\widehat{g_{0,\tilde P_{m}}}(X_l)}{\widehat{g_{0,\kappa_{m,l}}}(X_l)} \right) \widehat{\tilde{P_l}}' \\
& =& \frac{1}{n(n-1)} \sum_{j = 1}^n \sum_{l \neq j}^n  \kappa_{j,l}  [\left( \widehat{\tilde{P}}_j -\tilde P_j + \frac{g_{0,\tilde P_{m}}(X_l)}{g_{0,\kappa_{m,l}}(X_l)} -  \frac{\widehat{g_{0,\tilde P_{m}}}(X_l)}{\widehat{g_{0,\kappa_{m,l}}}(X_l)} \right) \widehat{\tilde{P_l}}' \\
&&+ \left( \tilde P_j -  \frac{g_{0,\tilde P_{m}}(X_l)}{g_{0,\kappa_{m,l}}(X_l)} \right) (-\widehat{\tilde{P}}_l +\tilde P_l) ]\\
&\stackrel{P}{\rightarrow}& 0
\end{eqnarray*}
Hence, we have $\verb'Plim' C_n = A$, since we showed in the proof of Proposition \ref{thm:infeas consistency normality} that $\verb'Plim' A_n = A$.

\noindent Define now the following quantities:
\begin{eqnarray*}
D_n &=& \frac{1}{n(n-1)} \sum_{j = 1}^n \sum_{l \neq j}^n \kappa_{j,l} \left(\widehat{\tilde{P}}_j -  \frac{\widehat{g_{0,\tilde P_{m}}}(X_l)}{\widehat{g_{0,\kappa_{m,l}}}(X_l)} \right)\\
& &[(\hat g_{P}(X_l)- E(P_l|X_l))' \theta_0 + \epsilon_l +E(y_l|X_l)- \hat g_y(X_l)] \\
E_n &=& \frac{1}{n(n-1)} \sum_{j = 1}^n \sum_{l \neq j}^n \kappa_{j,l} \left(\widehat{\tilde{P}}_j -  \frac{\widehat{g_{0,\tilde P_{m}}}(X_l)}{\widehat{g_{0,\kappa_{m,l}}}(X_l)} \right)(\hat g_{P}(X_l)- E(P_l|X_l))' \theta_0 \\
F_n &=& \frac{1}{n(n-1)} \sum_{j = 1}^n \sum_{l \neq j}^n \kappa_{j,l} \left(\widehat{\tilde{P}}_j -  \frac{\widehat{g_{0,\tilde P_{m}}}(X_l)}{\widehat{g_{0,\kappa_{m,l}}}(X_l)} \right) \epsilon_l\\
G_n &=& \frac{1}{n(n-1)} \sum_{j = 1}^n \sum_{l \neq j}^n \kappa_{j,l} \left(\widehat{\tilde{P}}_j -  \frac{\widehat{g_{0,\tilde P_{m}}}(X_l)}{\widehat{g_{0,\kappa_{m,l}}}(X_l)} \right) [E(y_l|X_l)- \hat g_y(X_l)]
\end{eqnarray*}
We have, $D_n= E_n+F_n+G_n$. For consistency, we show that the probability limits for $E_n$, $F_n$, and $G_n$ are all zero.
\begin{eqnarray*}
E_n
&=& \frac{1}{n(n-1)} \sum_{j = 1}^n \sum_{l \neq j}^n \kappa_{j,l} \left(\widehat{\tilde{P}}_j -  \frac{\widehat{g_{0,\tilde P_{m}}}(X_l)}{\widehat{g_{0,\kappa_{m,l}}}(X_l)} \right)(\hat g_{P}(X_l)- E(P_l|X_l))' \theta_0 \\
&=& \frac{1}{n(n-1)} \sum_{j = 1}^n \sum_{l \neq j}^n \kappa_{j,l} \left(\tilde P_j + \widehat{\tilde{P}}_j -\tilde P_j -\frac{g_{0,\tilde P_{m}}(X_l)}{g_{0,\kappa_{m,l}}(X_l)} + \frac{g_{0,\tilde P_{m}}(X_l)}{g_{0,\kappa_{m,l}}(X_l)} -  \frac{\widehat{g_{0,\tilde P_{m}}}(X_l)}{\widehat{g_{0,\kappa_{m,l}}}(X_l)} \right) \\
&& (\hat g_{P}(X_l)- E(P_l|X_l))' \theta_0 \\
&=& \frac{1}{n(n-1)} \sum_{j = 1}^n \sum_{l \neq j}^n \kappa_{j,l} \left(\tilde P_j -\frac{g_{0,\tilde P_{m}}(X_l)}{g_{0,\kappa_{m,l}}(X_l)} + \widehat{\tilde{P}}_j -\tilde P_j + \frac{g_{0,\tilde P_{m}}(X_l)}{g_{0,\kappa_{m,l}}(X_l)} -  \frac{\widehat{g_{0,\tilde P_{m}}}(X_l)}{\widehat{g_{0,\kappa_{m,l}}}(X_l)} \right) \\
&& (\hat g_{P}(X_l)- E(P_l|X_l))' \theta_0 \\
&=& \frac{1}{n(n-1)} \sum_{j = 1}^n \sum_{l \neq j}^n \kappa_{j,l} \left(\tilde P_j -\frac{g_{0,\tilde P_{m}}(X_l)}{g_{0,\kappa_{m,l}}(X_l)} \right)(\hat g_{P}(X_l)- E(P_l|X_l))' \theta_0 \\
&+&  \frac{1}{n(n-1)} \sum_{j = 1}^n \sum_{l \neq j}^n \kappa_{j,l} \left( \widehat{\tilde{P}}_j -\tilde P_j + \frac{g_{0,\tilde P_{m}}(X_l)}{g_{0,\kappa_{m,l}}(X_l)} -  \frac{\widehat{g_{0,\tilde P_{m}}}(X_l)}{\widehat{g_{0,\kappa_{m,l}}}(X_l)} \right) \\
& &(\hat g_{P}(X_l)- E(P_l|X_l))' \theta_0 \\
&\stackrel{P}{\rightarrow} 0 
\end{eqnarray*}
\begin{eqnarray*}
F_n
&=& \frac{1}{n(n-1)} \sum_{j = 1}^n \sum_{l \neq j}^n \kappa_{j,l} \left(\tilde P_j -\frac{g_{0,\tilde P_{m}}(X_l)}{g_{0,\kappa_{m,l}}(X_l)} + \widehat{\tilde{P}}_j -\tilde P_j + \frac{g_{0,\tilde P_{m}}(X_l)}{g_{0,\kappa_{m,l}}(X_l)} -  \frac{\widehat{g_{0,\tilde P_{m}}}(X_l)}{\widehat{g_{0,\kappa_{m,l}}}(X_l)} \right) \epsilon_l \\
&=& \frac{1}{n(n-1)} \sum_{j = 1}^n \sum_{l \neq j}^n \kappa_{j,l} \left(\tilde P_j -\frac{g_{0,\tilde P_{m}}(X_l)}{g_{0,\kappa_{m,l}}(X_l)} \right)\epsilon_l \\
&+&  \frac{1}{n(n-1)} \sum_{j = 1}^n \sum_{l \neq j}^n \kappa_{j,l} \left( \widehat{\tilde{P}}_j -\tilde P_j + \frac{g_{0,\tilde P_{m}}(X_l)}{g_{0,\kappa_{m,l}}(X_l)} -  \frac{\widehat{g_{0,\tilde P_{m}}}(X_l)}{\widehat{g_{0,\kappa_{m,l}}}(X_l)} \right) \epsilon_l \\
&\stackrel{P}{\rightarrow} 0 \\
\end{eqnarray*}
\begin{eqnarray*}
G_n
&=& \frac{1}{n(n-1)} \sum_{j = 1}^n \sum_{l \neq j}^n \kappa_{j,l} \left(\tilde P_j -\frac{g_{0,\tilde P_{m}}(X_l)}{g_{0,\kappa_{m,l}}(X_l)} \right)[E(y_l|X_l)- \hat g_y(X_l)] \\
&+&  \frac{1}{n(n-1)} \sum_{j = 1}^n \sum_{l \neq j}^n \kappa_{j,l} \left( \widehat{\tilde{P}}_j -\tilde P_j + \frac{g_{0,\tilde P_{m}}(X_l)}{g_{0,\kappa_{m,l}}(X_l)} -  \frac{\widehat{g_{0,\tilde P_{m}}}(X_l)}{\widehat{g_{0,\kappa_{m,l}}}(X_l)} \right) [E(y_l|X_l)- \hat g_y(X_l)] \\
&\stackrel{P}{\rightarrow} 0
\end{eqnarray*}
All in all, we have $\verb'Plim' D_n= \verb'Plim' (E_n+F_n+G_n) = 0$, so $\hat \theta_{n,o} \xrightarrow{P} \theta_0$.

\noindent In addition, we have:
$$\sqrt{n} (\hat \theta_{n,o} - \theta_0) = [C_n]^{-1} \sqrt{n} [E_n+F_n+G_n]$$
And we study each term separately:
\begin{eqnarray*}
\sqrt n F_n
&=& \sqrt n B_n +  \sqrt n \frac{1}{n(n-1)} \sum_{j = 1}^n \sum_{l \neq j}^n \kappa_{j,l} \\
& & \left( \widehat{\tilde{P}}_j -\tilde P_j + \frac{g_{0,\tilde P_{m}}(X_l)}{g_{0,\kappa_{m,l}}(X_l)} -  \frac{\widehat{g_{0,\tilde P_{m}}}(X_l)}{\widehat{g_{0,\kappa_{m,l}}}(X_l)} \right) \epsilon_l 
\end{eqnarray*}
The second term of $\sqrt n F_n$ is $o_p(1)$ because $\left( \widehat{\tilde{P}}_j -\tilde P_j + \frac{g_{0,\tilde P_{m}}(X_l)}{g_{0,\kappa_{m,l}}(X_l)} -  \frac{\widehat{g_{0,\tilde P_{m}}}(X_l)}{\widehat{g_{0,\kappa_{m,l}}}(X_l)} \right)$ is $o_p(1)$ under Assumption \ref{ass:ad1}, and $\sqrt n \frac{1}{n(n-1)} \sum_{j = 1}^n \sum_{l \neq j}^n \kappa_{j,l} \epsilon_l $ is $O_p(1)$.
\begin{eqnarray*}
\sqrt n E_n
& =& \sqrt n \frac{1}{n(n-1)} \sum_{j = 1}^n \sum_{l \neq j}^n \kappa_{j,l} \left(\tilde P_j -\frac{g_{0,\tilde P_{m}}(X_l)}{g_{0,\kappa_{m,l}}(X_l)} \right)(\hat g_{P}(X_l)- E(P_l|X_l))' \theta_0 \\
&+& \sqrt n \frac{1}{n(n-1)} \sum_{j = 1}^n \sum_{l \neq j}^n \kappa_{j,l} \left( \widehat{\tilde{P}}_j -\tilde P_j + \frac{g_{0,\tilde P_{m}}(X_l)}{g_{0,\kappa_{m,l}}(X_l)} -  \frac{\widehat{g_{0,\tilde P_{m}}}(X_l)}{\widehat{g_{0,\kappa_{m,l}}}(X_l)} \right)\\
& & (\hat g_{P}(X_l)- E(P_l|X_l))' \theta_0
\end{eqnarray*}

\begin{eqnarray*}
\sqrt n G_n
&=& \sqrt n \frac{1}{n(n-1)} \sum_{j = 1}^n \sum_{l \neq j}^n \kappa_{j,l} \left(\tilde P_j -\frac{g_{0,\tilde P_{m}}(X_l)}{g_{0,\kappa_{m,l}}(X_l)} \right)[E(y_l|X_l)- \hat g_y(X_l)] \\
&+& \sqrt n \frac{1}{n(n-1)} \sum_{j = 1}^n \sum_{l \neq j}^n \kappa_{j,l} \left( \widehat{\tilde{P}}_j -\tilde P_j + \frac{g_{0,\tilde P_{m}}(X_l)}{g_{0,\kappa_{m,l}}(X_l)} -  \frac{\widehat{g_{0,\tilde P_{m}}}(X_l)}{\widehat{g_{0,\kappa_{m,l}}}(X_l)} \right) \\
& &[E(y_l|X_l)- \hat g_y(X_l)]
\end{eqnarray*}

The second terms of $\sqrt n E_n$ and $\sqrt n G_n$ are $o_p(1)$ under Assumption \ref{ass:ad1}, because $$\left( \widehat{\tilde{P}}_j -\tilde P_j + \frac{g_{0,\tilde P_{m}}(X_l)}{g_{0,\kappa_{m,l}}(X_l)} -  \frac{\widehat{g_{0,\tilde P_{m}}}(X_l)}{\widehat{g_{0,\kappa_{m,l}}}(X_l)} \right)$$ and $$(\hat g_{P}(X_l)- E(P_l|X_l)) \text{ or } [E(y_l|X_l)- \hat g_y(X_l)]$$ are both $o_p(n^{-1/4})$. $o_p(n^{-1/4})o_p(n^{-1/4})$ is $o_p(n^{-1/2})$. Thus, the second terms are $o_p(1)$.

Therefore, the task is to demonstrate that both of these first terms are $o_p(1)$, meaning their magnitudes do not grow too fast relative to the sample size $n$. Once we establish this, we can conclude that the first terms of $\sqrt n E_n$ and $\sqrt n G_n$ are $o_p(1)$, completing the proof. 

In the first term of $\sqrt n E_n$, we have $(\hat g_{P}(X_l)- E(P_l|X_l))' \theta_0$, which is $o_p(n^{-1/4})$. Recall that $E[\tilde P_{j} \kappa_{j,l} - \frac{E[\tilde P_{m} \kappa_{m,l} |X_l]}{E[\kappa_{m,l} |X_l]} \kappa_{j,l} | X_l] = 0$. That is, $E[\tilde P_{j} \kappa_{j,l} - \frac{g_{0,\tilde P_{m}}(X_l)}{g_{0,\kappa_{m,l}}(X_l)} \kappa_{j,l} | X_l] = 0$. This is discussed in the proof that shows the FOC defined in Equation (\ref{eq:ofoc01}) is orthogonal, specifically the proof for $I_2'$ being 0.

The first term of $\sqrt n E_n$ is:
\begin{align*}
  \sqrt n \frac{1}{n(n-1)} & \sum_{j = 1}^n \sum_{l \neq j}^n \kappa_{j,l} \left(\tilde P_j -\frac{g_{0,\tilde P_{m}}(X_l)}{g_{0,\kappa_{m,l}}(X_l)} \right)(\hat g_{P}(X_l)- E(P_l|X_l))' \theta_0   \\
   & =   \frac{1}{n} \sum_{l = 1}^n  (\hat g_{P}(X_l)- E(P_l|X_l))' \theta_0 \sqrt n \frac{1}{n-1}\sum_{j \neq l}^n \kappa_{j,l} \left(\tilde P_j -\frac{g_{0,\tilde P_{m}}(X_l)}{g_{0,\kappa_{m,l}}(X_l)} \right)\\
   & =  \frac{1}{n} \sum_{l = 1}^n o_p(n^{-1/4}) \frac{\sqrt n}{\sqrt{n-1}} \frac{\sqrt{n-1}}{n-1}\sum_{j \neq l}^n \kappa_{j,l} \left(\tilde P_j -\frac{g_{0,\tilde P_{m}}(X_l)}{g_{0,\kappa_{m,l}}(X_l)} \right)
\end{align*}
The first equality arises because we sum across $j$ and $l$ except for cases when $j = l$ and rearranging the order of summation does not affect the outcome. The second equality holds under Assumption \ref{ass:ad1}. Additionally, $\frac{\sqrt{n-1}}{n-1}\sum_{j \neq l}^n \kappa_{j,l} \left(\tilde P_j -\frac{g_{0,\tilde P_{m}}(X_l)}{g_{0,\kappa_{m,l}}(X_l)} \right)$ is $O_p(1)$
because $\frac{1}{n-1}\sum_{j \neq l}^n \kappa_{j,l} \left(\tilde P_j -\frac{g_{0,\tilde P_{m}}(X_l)}{g_{0,\kappa_{m,l}}(X_l)} \right)$ represents the sample average of $\kappa_{j,l} \left(\tilde P_j -\frac{g_{0,\tilde P_{m}}(X_l)}{g_{0,\kappa_{m,l}}(X_l)} \right)$ when $X_l$ is fixed. The conditional expectation $E[\tilde P_{j} \kappa_{j,l} - \frac{g_{0,\tilde P_{m}}(X_l)}{g_{0,\kappa_{m,l}}(X_l)} \kappa_{j,l} | X_l]$ is 0, and the corresponding variance is bounded.

Hence, the first term of $\sqrt n E_n$ becomes $\frac{1}{n} \sum_{l = 1}^n o_p(n^{-1/4}) O_P(1)$. That is, $o_p(1)$. Based on a similar analysis, the first term of $\sqrt n G_n$ is also $o_p(1)$.

From Assumption \ref{ass:ad1}, $\sqrt{n} [E_n+F_n+G_n] = \sqrt n B_n + o_p(1)$.

Hence, $$\sqrt{n} (\hat \theta_{n,o} - \theta_0) = [C_n]^{-1} \sqrt{n}B_n + o_p(1)$$

$\hat \theta_{n,o}$ and $\tilde \theta_{n,o}$ share the same asymptotic distribution.
\end{proof}

\subsection{Consistent Estimator of the Asymptotic Variance (\ref{eq: HAC estim})}

\begin{eqnarray*}
&& Var\left[\int_{\mathbb{R}^{q_z}}e^{-it' Z_j} \epsilon_j \left(E[e^{it'Z_l} \tilde P_l ]-\frac{g_{0,\tilde P_{m}}(X_j)}{g_{0,\kappa_{m,j}}(X_j)}  E[e^{it'Z_l} ] \right) d \mu(t) \right] \\
&=& Var\left[ E_l \left[\int_{\mathbb{R}^{q_z}} e^{it'(Z_l - Z_j)} \epsilon_j \left(\tilde P_l -\frac{g_{0,\tilde P_{m}}(X_j)}{g_{0,\kappa_{m,j}}(X_j)} \right)d \mu(t) \right] \right] \\
&=& Var \left[E_l \left[k(Z_l- Z_j)\left(\tilde P_l -\frac{g_{0,\tilde P_{m}}(X_j)}{g_{0,\kappa_{m,j}}(X_j)} \right) \epsilon_j  \right] \right] \\
&=& E[\left(E_l \left[k(Z_l- Z_j)\left(\tilde P_l -\frac{g_{0,\tilde P_{m}}(X_j)}{g_{0,\kappa_{m,j}}(X_j)} \right)  \right] \right) \\
& & \left(E_l \left[k(Z_l- Z_j)\left(\tilde P_l -\frac{g_{0,\tilde P_{m}}(X_j)}{g_{0,\kappa_{m,j}}(X_j)} \right)' \right] \right)[\epsilon_j]^2 ]
\end{eqnarray*}
Under heteroskedasticity, the estimator for variance of the feasible D-RSMD is
\begin{eqnarray*}
& &[n(n-1) C_n]^{-1} \sum^n_{j=1}((\sum^n_{l=1}\kappa_{j,l} \left(\widehat{\tilde{P}}_l -  \frac{\widehat{g_{0,\tilde P_{l}}}(X_j)}{\widehat{g_{0,\kappa_{m,j}}}(X_j)} \right))(\sum^n_{l=1}\kappa_{j,l} \left(\widehat{\tilde{P}}_l -  \frac{\widehat{g_{0,\tilde P_{l}}}(X_j)}{\widehat{g_{0,\kappa_{m,j}}}(X_j)} \right)')\hat \epsilon_j^2)\\
& &[n(n-1) C_n']^{-1}
\end{eqnarray*}

\section{Appendix: Identification} \label{section:ident}

\subsection{Summary}

In this section, we provide a summary to illustrate the identification issue for D-RSMD. Due to the complexity of the estimator introduced by the debiasing process, we cannot assume higher-order assumptions for the identification, as in \cite{AS2021}. Hence, in this section, we will discuss the issue step by step. If there are constants in the model, the conditions for identification will be slightly different, but the logic will remain the same.

We will start with the simplest model first. Then move on to the complicated models. Models are provided with coherent simulation results. Those results are included in the Table \ref{sim:idenfi}.

\begin{enumerate}
  \item Model (1\label{identi:model01}): $$y_i = \beta_1 W_{i,1}+ \beta_2 W_{i,2} + u_i$$
$$W_{i,1} = \pi_1 Z_i+ v_{i,1}$$
$$W_{i,2} = \pi_2 Z_i+ v_{i,2}$$

SMD cannot identify $\beta_1$ and $\beta_2$ with continuous or discrete $Z_i$. To identify $\beta_1$ and $\beta_2$ with SMD, the model need to have a nonlinear part of $Z_i$ (continuous) or another control variable $X_i$ inside one of the equations for $W_i$.

  \item $$y_i =  \beta_1 W_{i,1}+ \beta_2 W_{i,2} + u_i$$
$$W_{i,1} = \pi_{1,1} Z_{i,1} + \pi_{1,2} Z_{i,2} + v_{i,1}$$
$$W_{i,2} = \pi_{2,1} Z_{i,1}  + \pi_{2,2} Z_{i,2} + v_{i,2}$$

When (continuous or discrete) $Z_{i,1}$ and $Z_{i,2}$ are independent, we are unable to identify both parameters using SMD with only one instrument $Z_{i,1}$ under some specific conditions for the mean of $E(Z_{i,2})$, $\pi_{1,1}$, $\pi_{1,2}$, $\pi_{2,1}$, and $\pi_{2,2}$. We are able to identify both parameters $\beta_1$ and $\beta_2$ when these conditions are not met.

\item Model (3\label{identi:model02}): $$y_i = \beta_1 W_{i,1}+ \beta_2 W_{i,1} * X_{i} + u_i$$
$$W_{i,1} = \pi_{1}  Z_i+ v_{i,1}$$
$$W_{i,1} * X_{i} = \pi_{1} Z_i * X_{i} + v_{i,1} * X_{i}$$

If $Z_i$ and $X_{i}$ are dependent, when $Z_i$ is continuous, both parameters are identified, and when $Z_i$ is binary, using SMD we can identify 0 or 1 parameter based on the value of $E(X_i|Z_i)$. If $Z_i$ and $X_{i}$ are independent, using SMD we can identify 0 or 1 parameter based on the value of $E(X_i)$. This is the same case for a continuous or discrete $Z_i$.

\item Model (4\label{identi:model03}): $$y_i = \beta_1 W_{i,1}+ \beta_2 W_{i,1} * X_{i} +  \gamma_1 X_{i} + u_i$$
$$W_{i,1} = \pi_{1} Z_i+ v_{i,1}$$
$$W_{i,1} * Z_{i} = \pi_{1} Z_i * X_{i} + v_{i,1} * X_{i}$$

If $Z_i$ and $X_{i}$ are dependent, RSMD identifies both parameters using a continuous or discrete $Z_i$. If $Z_i$ and $X_{i}$ are independent, we can use RSMD to identify 0 or 1 parameter based on the value of $E(X_i)$ with continuous or discrete $Z_i$.

\item D-RSMD has extra terms in its formula and it is not symmetric. It will identify both parameters when RSMD cannot.
\end{enumerate}

\subsection{Proofs and Simulation Examples}

Following that, we move on to sections with more detailed explanations, proofs, and simulation studies to provide more information on the identification issue for methods incorporating SMD type estimators, such as SMD, RSMD, and D-RSMD.

\begin{enumerate}
  \item Prove the identification of the key parameters using the SMD method for the simplest model in the following (Model \ref{identi:model01}).

$$y_i = \beta_1 W_{i,1}+ \beta_2 W_{i,2} + u_i$$
$$W_{i,1} = \pi_1 Z_i+ v_{i,1}$$
$$W_{i,2} = \pi_2 Z_i+ v_{i,2}$$

  \begin{proof}

The SMD estimator is
\begin{eqnarray*}
 \begin{pmatrix} \beta_1 \\ \beta_2 \end{pmatrix}  &=& [E(\kappa_{j,l} W_j W_l')]^{-1} E(\kappa_{j,l} W_j y_l)  \\
   &=& \left[ E \left(\kappa_{j,l} \begin{pmatrix} W_{j,1} \\ W_{j,2} \end{pmatrix} \begin{pmatrix} W_{l,1} \\ W_{l,2} \end{pmatrix}' \right)\right]^{-1} E(\kappa_{j,l} W_j y_l)
\end{eqnarray*}

We need to prove that $E(\kappa_{j,l} W_j W_l')$ is nonsingular. To show that $E(\kappa_{j,l} W_j  W_l')$ is nonsingular, we consider the associated quadratic form, and show that it is positive definite. For any $a$ real vector of size $p$, we have:
\begin{align*}
E(a' W_j W_l'a \kappa_{j,l}) & = E(\kappa_{j,l} a 'E( W_j|Z_j) E(W_l'a|Z_l)) \\
& =  E(\int_{R^{q_w}} e^{it'(Z_j - Z_l)} d \mu(t) a' E(W_j|Z_j) E( W_l'|Z_l)a )\\
& = \int_{R^{q_w}} E[e^{it'(Z_j - Z_l)} a' E(W_j|Z_j) E(W_l'|Z_l)a]d \mu(t) \\
& = \int_{R^{q_w}} E[e^{it' Z_j}  a' E(W_j|Z_j) E(W_l'|Z_l)ae^{-it' Z_l} ]d \mu(t) \\
& = \int_{R^{q_w}} E[a'e^{it' Z_j} E(W_j|Z_j)]E[E(W_l'|Z_l)a e^{-it' Z_l}]  d \mu(t) \\
& = \int_{R^{q_w}} E[a'e^{it' Z_j} E(W_j|Z_j)]E[E(W_j'|Z_j)a e^{-it' Z_j}]  d \mu(t) \\
& = \int_{R^{q_w}} |\left(\int_{R^{q_w}}a'e^{it' Z_j} E(W_j|Z_j) f_Z(Z_j) d Z_j \right) |^2d \mu(t) \\
& = (2 \pi)^{2q_w}\int_{R^{q_w}} |\left( \mathcal{F}[a' E(W_j|Z_j) f_Z(Z_j)](t)  \right) |^2d \mu(t) \\
& \geq 0
\end{align*}
with $\mu$ strictly positive on $\mathbb{R}^p$ and $\mathcal{F}[g]$ the Fourier transform of a well-defined function $g(.)$ on $\mathbb{R}^{q_w}$ is formally defined as,
\begin{eqnarray} \label{eq:Fourier transform}
\mathcal{F}[g](t) = \frac{1}{(2\pi)^{q_w}} \int \exp^{it'u} g(u) du \, .
\end{eqnarray}
We then have:
\begin{eqnarray*}
a'E(\kappa_{j,l} W_j W_l')a = 0
&\Leftrightarrow& \exists \ a\neq 0 \ s.t. \ a'E(W_j|Z_j)f(W_j)=0 \ a.s. \\
&\Leftrightarrow& \exists \ a\neq 0 \ s.t. \ a'E(W_j|Z_j)=0 \ a.s.
\end{eqnarray*}

In the specific case, $$\Leftrightarrow \exists \ a \neq 0 \ s.t. \ a'E(W_j|Z_j) = a_1 \pi_1 Z_i + a_2 \pi_2 Z_i = (a_1 \pi_1+ a_2 \pi_2) Z_i=0 \ a.s.$$

$\beta_1$ and $\beta_2$ are not identified. To identify $\beta_1$ and $\beta_2$, the model must contain a nonlinear part of $Z_i$ or another control variable $X_i$ inside one of the equations for $W_i$.
\end{proof}

  \item Prove the identification of the key parameters using the SMD method for the model with an interaction term (Model (\ref{identi:model02})).
$$y_i = \beta_1 W_{i,1}+ \beta_2 W_{i,1} * X_{i} + u_i$$
$$W_{i,1} = \pi_1 Z_i+ v_{i,1}$$
$$W_{i,1} * X_{i} = \pi_1 Z_i * X_{i} + v_{i,1} * X_{i}$$

\begin{proof}
We need to prove that $E(\kappa_{j,l} W_j W_l')$ is nonsingular. Follow the same logic as in the previous proof.
\begin{equation*}
a'E(\kappa_{j,l} W_j W_l')a = 0
\Leftrightarrow \exists \ a\neq 0 \ s.t. \ a'E(W_j|Z_j)=0 \ a.s.
\end{equation*}
$$E(W_{i,1}|Z_i) = \pi_1 Z_i$$
\begin{eqnarray*}
E(W_{i,1} * X_{i}|Z_i) &=& E((\pi_1 Z_i+ v_{i,1}) * X_{i}|Z_i) \\
   &=& E(\pi_1 Z_i  * X_{i} + v_{i,1} * X_{i}|Z_i) \\
   &=& \pi_1 Z_i E(X_{i}|Z_i) + E(v_{i,1} * X_{i}|Z_i)
\end{eqnarray*}
$$a'E(W_i|Z_i) = a_1 \pi_1 Z_i + a_2 \pi_1 Z_i E(X_{i}|Z_i) + a_2E( v_{i,1} * X_{i}| Z_i)$$

If $v_{i,1} \perp (X_{i},  Z_i)$, $$E(v_{i,1} * X_{i}| Z_i) = 0$$
$$a'E(W_i|Z_i) = a_1 \pi_1 Z_i + a_2 \pi_1 Z_i E(X_{i}|Z_i)$$

When $Z_i$ and $X_{i}$ are dependent, and $Z_i$ is continuous, there are no $a_1\neq 0$ and $a_2 \neq 0$, $s.t. \ a'E(W_j|Z_j)=0 \ a.s.$. Both parameters are identified.

When $Z_i$ is binary, there will be a problem.

For instance,

$$a'E(W_i|Z_i = 0) = a_1 \pi_1 * 0 + a_2 \pi_1 *0 * E(X_{i}|Z_i=0) = 0$$
$$a'E(W_i|Z_i = 1) = a_1 \pi_1 * 1 + a_2 \pi_1 *1 * E(X_{i}|Z_i=1) = 0$$

When $E(X_{i}|Z_i=1) \neq 0$ we can find a $a_1\neq 0$ and $a_2 \neq 0$, $s.t. \ a'E(W_j|Z_j)=0 \ a.s.$

When $W_i$ is 3-value instrument, both parameters will be identified.

\end{proof}

Here are 2 examples for simulation illustration when using the SMD estimator. We consider the case that $W_i$ and $Z_{i}$ are independent for the second example.
\begin{enumerate}
\item
$$y_i = 2 W_{i,1} + 3 W_{i,2} + u_i$$
$$W_{i,1} = 4 Z_{i,1} + Z_{i,2} + v_{i,1}$$
$$W_{i,2} = Z_{i,1}  + 3 Z_{i,2} + v_{i,2}$$

  When $Z_{i,1}$ and $Z_{i,2}$ are independent, we are not able to identify both parameters inside the model for $y_i$ using SMD with only one instrument $Z_{i,1}$ or $Z_{i,2}$ when $E(Z_{i,2})$ or $E(Z_{i,1})$ are zero. Indeed, identification of SMD with $Z_{i,1}$ yields the following equation. $$a'E(W_i|Z_{i,1}) = a_1 E(4 Z_{i,1} + Z_{i,2} + v_{i,1}|Z_{i,1})+ a_2 E(Z_{i,1}  + 3 Z_{i,2} + v_{i,2}|Z_{i,1})$$ $$ = a_1 4 Z_{i,1}+a_1 E(Z_{i,2}|Z_{i,1})+ a_2 Z_{i,1}  + a_2E(3 Z_{i,2} |Z_{i,1})$$

  When $Z_{i,1}$ and $Z_{i,2}$ are independent, $$a'E(W_i|Z_{i,1})= a_1 4 Z_{i,1}+a_1 E(Z_{i,2})+ a_2 Z_{i,1}  + a_2E(3 Z_{i,2})$$ $$a'E(W_i|Z_{i,1})= (4a_1 + a_2) Z_{i,1}+(a_1 + 3 a_2 )E(Z_{i,2})$$
  \begin{enumerate}
    \item When $Z_{i,1}$ and $Z_{i,2}$ are continuous, we need $E(Z_{i,2}) \neq 0$ to identify both parameters.
    \item \label{sim:ident1} When $E(Z_{i,2}) = 0$, there are infinite possible combinations for $a_1$ and $a_2$ to have $$a'E(W_i|Z_{i,1}) =0$$. In this case, neither parameter is identified.
    \item \label{sim:ident2} When $Z_{i,1}$ and $Z_{i,2}$ are binary ${0,1}$, $E(Z_{i,2}) \neq 0$. $a_1=0$ and $a_2=0$ to have $$a'E(W_i|Z_{i,1}) =0$$. Thus, we will identify both parameters.
  \end{enumerate}

\item
$$y_i = 2 W_{i,1}+ 3 W_{i,1} * X_{i} + u_i$$
$$W_{i,1} = 2 Z_i+ v_{i,1}$$
$$W_{i,1} * X_{i} = 2 Z_i * X_{i} + v_{i,1} * X_{i}$$
When $Z_i$ and $X_{i}$ are independent and $E(X_{i})= 0$, we are able to identify the first parameter. When $Z_i$ and $X_{i}$ are independent and $E(X_{i}) \neq 0$, there is no identification for both parameters. Indeed,
$$a'E(W_i|Z_i) = a_1 E(2 Z_i+ v_{i,1}|Z_i) + a_2 E(2 Z_i * X_{i} + v_{i,1} * X_{i}|Z_i)$$
$$ = 2 a_1 Z_i + 2 a_2 Z_i * E(X_{i}|Z_i)$$
When $Z_i$ and $X_{i}$ are independent, $$a'E(W_i|Z_i) = 2 a_1 Z_i + 2 a_2 Z_i * E(X_{i})$$
\begin{enumerate}
  \item \label{sim:identint1} When $E(X_{i})= 0$, $a_1$ must be 0 for a continuous or discrete $Z_i$. $a_2$ can be any value. Thus, we can only identify the first parameter.
  \item \label{sim:identint2} When $E(X_{i})= 1$, there are infinite combinations for $(a_1, a_2)$ to make  $$a'E(W_i|Z_{i}) =0$$. Both parameters are not identified.
\end{enumerate}
All of these examples are verified used simulation studies. The simulation results are included in the Table \ref{sim:idenfi}.

\end{enumerate}

\newpage
  \item Prove the identification of the key parameters using the RSMD method for the following model (Model (\ref{identi:model03})).

$$y_i = \beta_1 W_{i,1}+ \beta_2 W_{i,1} * X_{i} +  X_{i} + u_i$$
$$W_{i,1} = \pi_1 Z_i+ v_{i,1}$$
$$W_{i,1} * X_{i} = \pi_1 Z_i * X_{i} + v_{i,1} * X_{i}$$

\begin{proof}
We need \begin{equation*}
a'E(\kappa_{j,l} \tilde W_j \tilde W_l')a = 0
\Leftrightarrow \exists \ a\neq 0 \ s.t. \ a'E(\tilde W_j|Z_j)=0 \ a.s.
\end{equation*} to show that the parameters are not identified.
\begin{eqnarray*}
E(\tilde W_{i,1}|Z_i) &=& E(W_{i,1} - E(W_{i,1}|X_i)|Z_i)\\
&=& E(\pi_1 Z_i + v_{i,1} - E(\pi_1 Z_i+ v_{i,1}|X_i)|Z_i)\\
&=& E(\pi_1 Z_i - E(\pi_1 Z_i|X_i)|Z_i)\\
&=& \pi_1 Z_i -  \pi_1 E(E(Z_i|X_i)|Z_i)
\end{eqnarray*}
\begin{eqnarray*}
E(W_{i,1}X_{i} - E(W_{i,1}X_{i}|X_i)|Z_i) &=&  E(W_{i,1}X_{i} - E(W_{i,1}|X_i) X_{i}|Z_i)\\
&=& E(\pi_1 Z_i X_{i} + v_{i,1}  X_{i} - E(W_{i,1}|X_i)  X_{i}|Z_i)\\
&=& E(\pi_1 Z_i X_{i}  - E(\pi_1 Z_i+ v_{i,1}|X_i) X_{i}|Z_i)\\
&=& E(\pi_1 Z_i X_{i}  - E(\pi_1 Z_i|X_i) * X_{i}|Z_i)\\
&=& \pi_1 Z_i E(X_{i}|Z_i)  - \pi_1E(E( Z_i|X_i) X_{i}|Z_i)
\end{eqnarray*}
\begin{eqnarray*}
a'E(\tilde{W}_i|Z_i)&=& a_1 (\pi_1 Z_i -  \pi_1 E(E(Z_i|X_i)|Z_i)) \\
&+& a_2 (\pi_1 Z_i E(X_{i}|Z_i)  - \pi_1E(E( Z_i|X_i) X_{i}|Z_i))\\
\end{eqnarray*}
Thus, when $a'E(\tilde{W}_i|Z_i) = 0$,
\begin{equation}\label{eq:identiillu1}
a_1 Z_i - a_1  E(E(Z_i|X_i)|Z_i) + a_2 Z_i E(X_{i}|Z_i)  - a_2 E(E(Z_i|X_i) X_{i}|Z_i)=0
\end{equation}
\begin{enumerate}
\item When the instrument $Z_i$ is binary, if $Z_i = 0$, Equation \ref{eq:identiillu1} becomes:
$$- a_1 E(E(Z_i|X_i)|Z_i=0) - a_2 E(E(Z_i|X_i)X_{i}|Z_i=0)=0$$
If $Z_i = 1$, Equation \ref{eq:identiillu1} becomes:
$$ a_1  - a_1 E(E(Z_i|X_i)|Z_i=1) + a_2  E( X_{i}|Z_i=1)- a_2 E(E( Z_i|X_i) X_{i}|Z_i=1)=0$$
We need to solve for $a_1$ and $a_2$. If both of them are 0, both parameters will be identified. If one of them is 0, only one parameter will be identified.
\small
$$\begin{pmatrix} E(E(Z_i|X_i)|Z_i=0) & E(E(Z_i|X_i)X_{i}|Z_i=0) \\ 1-E(E(Z_i|X_i)|Z_i=1) & (E(X_{i}|Z_i=1) - E(E(Z_i|X_i) X_{i}|Z_i=1)) \end{pmatrix}\begin{pmatrix} a_1 \\ a_2 \end{pmatrix} = \begin{pmatrix} 0 \\ 0 \end{pmatrix}$$
\normalsize
We will have a problem if $Z_i \perp X_i$. The determinant of the above matrix is 0. In that scenario, there are many possible solutions for $a_1$ and $a_2$. We cannot identify both parameters.
\end{enumerate}

Another straightforward way to illustrate that is in the following.

$$a'E(\tilde{W}_i|Z_i) = a_1 E(\tilde W_{i,1}|Z_i) + a_2 E(\tilde W_{i,1}* X_{i}|Z_i)$$
$$a'E(\tilde{W}_i|Z_i=0) = a_1 E(\tilde W_{i,1}|Z_i=0) + a_2 E(\tilde W_{i,1}* X_{i}|Z_i=0) = 0$$
$$a'E(\tilde{W}_i|Z_i=1) = a_1 E(\tilde W_{i,1}|Z_i=1) + a_2 E(\tilde W_{i,1}* X_{i}|Z_i=1) = 0$$
$$\begin{pmatrix}  E(\tilde W_{i,1}|Z_i=0) & E(\tilde W_{i,1}* X_{i}|Z_i=0)  \\  E(\tilde W_{i,1}|Z_i=1) & E(\tilde W_{i,1}* X_{i}|Z_i=1) \end{pmatrix} \begin{pmatrix} a_1 \\ a_2 \end{pmatrix} = \begin{pmatrix} 0 \\ 0 \end{pmatrix}$$
$$A \begin{pmatrix} a_1 \\ a_2 \end{pmatrix} = \begin{pmatrix} 0 \\ 0 \end{pmatrix}$$

If matrix A is invertible or $det(A)$ is not 0, then $a_1$ and $a_2$ are zeros. Both parameters are identified. If $det(A)$ is 0, we cannot identify both parameters. $det(A) = E(\tilde W_{i,1}|Z_i=0) E(\tilde W_{i,1}* X_{i}|Z_i=1) - E(\tilde W_{i,1}* X_{i}|Z_i=0) E(\tilde W_{i,1}|Z_i=1)$.
\end{proof}

Additionally, with a binary instrument, the RSMD identifies 2 or less than 2 parameters. For instance, if we want to identify 3 parameters with 2 parameters for 2 interaction terms, RSMD can only identify 2 parameters at most.

$$a'E(\tilde{W}_i|Z_i) = a_1 E(\tilde W_{i}|Z_i) + a_2 E(\tilde W_{i} X_{i,1}|Z_i) + a_3 E(\tilde W_{i}X_{i,2}|Z_i)$$
$$a'E(\tilde{W}_i|Z_i=0) = a_1 E(\tilde W_{i}|Z_i=0) + a_2 E(\tilde W_{i}X_{i,1}|Z_i=0) + a_3 E(\tilde W_{i}X_{i,2}|Z_i=0)$$
$$a'E(\tilde{W}_i|Z_i=1) = a_1 E(\tilde W_{i}|Z_i=1) + a_2 E(\tilde W_{i}X_{i,1}|Z_i=1) + a_3 E(\tilde W_{i}X_{i,2}|Z_i=1)$$

\item Prove the identification of the key parameters using the D-RSMD method for Model (\ref{identi:model03}).

The D-RSMD estimator relies on the invertibility of the following matrix. Because it is not symmetric, we cannot use the same method to prove the identification problem. There is an alternative proof.

$E\left[ \kappa_{j,l}\left( \tilde P_j -  \frac{g_{0,\tilde P_{m}}(X_l)}{g_{0,\kappa_{m,l}}(X_l)} \right)\tilde P_l' \right] = E\left[ \kappa_{j,l} \tilde P_j \tilde P_l' \right]-E\left[ \kappa_{j,l} \frac{g_{0,\tilde P_{m}}(X_l)}{g_{0,\kappa_{m,l}}(X_l)} \tilde P_l' \right]$

The invertibility of the matrix is illustrated by a simple example. $X_i$ and $W_i$ are random variables. $P_i = [W_i, W_i X_i]'$. $W$ is the treatment. $X$ is the control. $Z$ is the instrument.

\;

$\tilde P_j \equiv \begin{pmatrix}\tilde W_j \\ \tilde W_j X_j\end{pmatrix} = \begin{pmatrix} W_j - E(W_j|X_j)\\W_j X_j - E(W_j|X_j)X_j\end{pmatrix}$

$g_{0,\tilde P_{1,m}}(X_l) = E[(P_{1m} - g_{0,P_1}(X_m)) \kappa_{m,l} |X_l] = E[\tilde W_j \kappa_{j,l} |X_l] = a(X_l)$,

$g_{0,\tilde P_{2,j}}(X_l) = E[(P_{2j} - g_{0,P_2}(X_j)) \kappa_{j,l} |X_l] = E[\tilde W_j X_j\kappa_{j,l} |X_l] = b(X_l)$,

$g_{0,\kappa_{m,l}}(X_l) = E[\kappa_{m,l} |X_l]  = c(X_l)$.

Based on the definition of $\tilde P_j$, $E(\tilde P_j) = 0$.
$$E(\kappa_{j,l} \tilde P_j \tilde P_l') = E \begin{bmatrix} \kappa_{j,l} \tilde W_j\tilde W_l & \kappa_{j,l} \tilde W_j\tilde W_l X_l \\ \kappa_{j,l} \tilde W_j\tilde W_l X_j & \kappa_{j,l} \tilde W_j\tilde W_l X_l X_j \end{bmatrix}$$
$$E\left[ \kappa_{j,l} \frac{g_{0,\tilde P_{m}}(X_l)}{g_{0,\kappa_{m,l}}(X_l)} \tilde P_l' \right] = E \begin{bmatrix} \kappa_{j,l} \frac{a(X_l)}{c(X_l)} \tilde W_l & \kappa_{j,l} \frac{a(X_l)}{c(X_l)} \tilde W_l X_l \\ \kappa_{j,l} \frac{b(X_l)}{c(X_l)} \tilde W_l  & \kappa_{j,l} \frac{b(X_l)}{c(X_l)} \tilde W_l X_l \end{bmatrix}$$

\;

$E(\kappa_{j,l} \tilde W_j\tilde W_l)= E[\kappa_{j,l} E(\tilde W_j|Z_j) E(\tilde W_l|Z_l)]$

$=E[\kappa_{j,l} E(\tilde W_j|Z_j) E(\tilde W_l|Z_l)|Z_j=1,Z_l=1] P_r(Z_j=1,Z_l=1)$

$+E[\kappa_{j,l} E(\tilde W_j|Z_j) E(\tilde W_l|Z_l)|Z_j=1,Z_l=0] P_r(Z_j=1,Z_l=0)$

$+E[\kappa_{j,l} E(\tilde W_j|Z_j) E(\tilde W_l|Z_l)|Z_j=0,Z_l=1] P_r(Z_j=0,Z_l=1)$

$+E[\kappa_{j,l} E(\tilde W_j|Z_j) E(\tilde W_l|Z_l)|Z_j=0,Z_l=0] P_r(Z_j=0,Z_l=0)$

In the simulations, we use Gaussian CDF for $\mu(t)$ inside $\kappa_{j,l}$. $\kappa_{j,l}=1$ when instrument variables $Z_j=Z_l$ and $\kappa_{j,l}=a \sim 0.61$ when $Z_j \neq Z_l$.

$E(\kappa_{j,l} \tilde W_j\tilde W_l)=E(\tilde W_j|Z_j=1) E(\tilde W_l|Z_l=1) P_r(Z_j=1) P_r(Z_l=1)$

$+ a E(\tilde W_j|Z_j=1) E(\tilde W_l|Z_l=0) P_r(Z_j=1) P_r(Z_l=0)$

$+ a E(\tilde W_j|Z_j=0) E(\tilde W_l|Z_l=1) P_r(Z_j=0) P_r(Z_l=1)$

$+ E(\tilde W_j|Z_j=0) E(\tilde W_l|Z_l=0) P_r(Z_j=0) P_r(Z_l=0)$

\;

Because $E(\tilde W_j|Z_j=1) E(\tilde W_l|Z_l=1) P_r(Z_j=1) P_r(Z_l=1)$

$+ E(\tilde W_j|Z_j=1) E(\tilde W_l|Z_l=0) P_r(Z_j=1) P_r(Z_l=0)$

$= E(\tilde W_j|Z_j=1)P_r(Z_j=1) [E(\tilde W_l|Z_l=1) P_r(Z_l=1) + E(\tilde W_l|Z_l=0)  P_r(Z_l=0)] $

$= 0$,

$E(\kappa_{j,l} \tilde W_j\tilde W_l)=(1-a) E(\tilde W_j|Z_j=1) E(\tilde W_l|Z_l=1) P_r(Z_j=1) P_r(Z_l=1)$

$+(1-a) E(\tilde W_j|Z_j=0) E(\tilde W_l|Z_l=0) P_r(Z_j=0) P_r(Z_l=0)$

Denote $E(\tilde W_l|Z_l=1)$ as $\theta_1$, $E(\tilde W_l|Z_l=0)$ as $\theta_2$, and $P_r(Z_j=1)$ as $b$. We have
$$E(\kappa_{j,l} \tilde W_j\tilde W_l) = (1-a) (\theta_1^2 b^2 + \theta_2^2 (1-b)^2)$$

Follow the same steps,

$E[\kappa_{j,l} \tilde W_j\tilde W_l X_l]=(1-a) E(\tilde W_j|Z_j=1) E(X_l \tilde W_l|Z_l=1) P_r(Z_j=1) P_r(Z_l=1)$

$+(1-a) E(\tilde W_j|Z_j=0) E(X_l \tilde W_l|Z_l=0) P_r(Z_j=0) P_r(Z_l=0)$

Denote $E(X_l \tilde W_l|Z_l=1)$ as $\theta_3$, $E(X_l \tilde W_l|Z_l=0)$ as $\theta_4$.

$$E[\kappa_{j,l} \tilde W_j\tilde W_l X_l]=(1-a)(\theta_1 \theta_3 b^2 + \theta_2 \theta_4 (1-b)^2)$$

$E[\kappa_{j,l} \tilde W_j\tilde W_l X_l X_j]=(1-a) E(X_j \tilde W_j|Z_j=1) E(X_l \tilde W_l|Z_l=1) P_r(Z_j=1) P_r(Z_l=1)$

$+(1-a) E(X_j \tilde W_j|Z_j=0) E(X_l \tilde W_l|Z_l=0) P_r(Z_j=0) P_r(Z_l=0)$

$$E[\kappa_{j,l} \tilde W_j\tilde W_l X_l X_j]=(1-a) (\theta_3^2 b^2 + \theta_4^2(1-b)^2)$$

$$E(\kappa_{j,l} \tilde P_j \tilde P_l') =  \begin{bmatrix} (1-a) (\theta_1^2 b^2 + \theta_2^2 (1-b)^2) & (1-a)(\theta_1 \theta_3 b^2 + \theta_2 \theta_4 (1-b)^2) \\ (1-a)(\theta_1 \theta_3 b^2 + \theta_2 \theta_4 (1-b)^2) & (1-a) (\theta_3^2 b^2 + \theta_4^2(1-b)^2) \end{bmatrix}$$

\;

\begin{eqnarray*}
E \left[ \kappa_{j,l} \frac{a(X_l)}{c(X_l)} \tilde W_l \right] &=& E\left[\kappa_{j,l} E \left(\frac{a(X_l)}{c(X_l)} \tilde W_l |Z_l \right)  \right] \\
&=& E\left[\kappa_{j,l} E \left(\frac{a(X_l)}{c(X_l)} \tilde W_l |Z_l \right) |Z_j=1,Z_l=1 \right] P_r(Z_j=1,Z_l=1) \\
&+& E\left[\kappa_{j,l} E \left(\frac{a(X_l)}{c(X_l)} \tilde W_l |Z_l \right) |Z_j=1,Z_l=0\right] P_r(Z_j=1,Z_l=0)\\
&+& E\left[\kappa_{j,l} E \left(\frac{a(X_l)}{c(X_l)} \tilde W_l |Z_l \right) |Z_j=0,Z_l=1\right] P_r(Z_j=0,Z_l=1)\\
&+& E\left[\kappa_{j,l} E \left(\frac{a(X_l)}{c(X_l)} \tilde W_l |Z_l \right) |Z_j=0,Z_l=0\right] P_r(Z_j=0,Z_l=0)
\end{eqnarray*}
\begin{eqnarray*}
E \left[ \kappa_{j,l} \frac{a(X_l)}{c(X_l)} \tilde W_l \right]&=& E\left(\frac{a(X_l)}{c(X_l)} \tilde W_l|Z_l=1\right) P_r(Z_j=1) P_r(Z_l=1) \\
&+& a E\left(\frac{a(X_l)}{c(X_l)} \tilde W_l|Z_l=0\right) P_r(Z_j=1) P_r(Z_l=0)\\
&+& a E\left(\frac{a(X_l)}{c(X_l)} \tilde W_l|Z_l=1\right) P_r(Z_j=0) P_r(Z_l=1)\\
&+& E\left(\frac{a(X_l)}{c(X_l)} \tilde W_l|Z_l=0\right) P_r(Z_j=0) P_r(Z_l=0)\\
\end{eqnarray*}
$$E\left(\frac{a(X_l)}{c(X_l)} \tilde W_l\right) = E\left(\frac{a(X_l)}{c(X_l)} E(\tilde W_l | X_l)\right)=0$$ with $\tilde P_j \equiv [\tilde W_j, \tilde W_j X_j]' = [W_j - E(W_j|X_j), W_j X_j - E(W_j|X_j)X_j] $
\begin{eqnarray*}
E \left[ \kappa_{j,l} \frac{a(X_l)}{c(X_l)} \tilde W_l \right]&=&  P_r(Z_j=1) (1-a)  E\left(\frac{a(X_l)}{c(X_l)} \tilde W_l|Z_l=1\right) P_r(Z_l=1) \\
&+& a P_r(Z_j=1) E\left(\frac{a(X_l)}{c(X_l)} \tilde W_l|Z_l=1\right) P_r(Z_l=1) \\
&+& a P_r(Z_j=1) E\left(\frac{a(X_l)}{c(X_l)} \tilde W_l|Z_l=0\right) P_r(Z_l=0)\\
&+& a E\left(\frac{a(X_l)}{c(X_l)} \tilde W_l|Z_l=1\right) P_r(Z_j=0) P_r(Z_l=1)\\
&+& E\left(\frac{a(X_l)}{c(X_l)} \tilde W_l|Z_l=0\right) P_r(Z_j=0) P_r(Z_l=0)\\
&=& P_r(Z_j=1) (1-a) E\left(\frac{a(X_l)}{c(X_l)} \tilde W_l|Z_l=1\right) P_r(Z_l=1)\\
&+& P_r(Z_j=0) (1-a) E\left(\frac{a(X_l)}{c(X_l)} \tilde W_l|Z_l=0\right) P_r(Z_l=0)\\
\end{eqnarray*}

When we have $P_r(Z_j=1)= 0.5$, $d_1=0$.

Denote $E\left(\frac{a(X_l)}{c(X_l)} \tilde W_l|Z_l=1\right)$ as $\delta_1$ and $E\left(\frac{a(X_l)}{c(X_l)} \tilde W_l|Z_l=0\right)$ as $\delta_2$.

$$E \left[ \kappa_{j,l} \frac{a(X_l)}{c(X_l)} \tilde W_l \right] = (1-a) (b^2 \delta_1+ (1-b)^2 \delta_2)$$

\;

Follow the same steps,

\begin{eqnarray*}
E \left[\kappa_{j,l} \frac{a(X_l)}{c(X_l)} \tilde W_l X_l \right]&=&  P_r(Z_j=1) (1-a)  E\left(\frac{a(X_l)}{c(X_l)} \tilde W_l X_l|Z_l=1\right) P_r(Z_l=1) \\
&+& a P_r(Z_j=1) E\left(\frac{a(X_l)}{c(X_l)} \tilde W_l X_l|Z_l=1\right) P_r(Z_l=1) \\
&+& a P_r(Z_j=1) E\left(\frac{a(X_l)}{c(X_l)} \tilde W_l X_l|Z_l=0\right) P_r(Z_l=0)\\
&+& a E\left(\frac{a(X_l)}{c(X_l)} \tilde W_l X_l|Z_l=1\right) P_r(Z_j=0) P_r(Z_l=1)\\
&+& E\left(\frac{a(X_l)}{c(X_l)} \tilde W_l X_l|Z_l=0\right) P_r(Z_j=0) P_r(Z_l=0)\\
&=& P_r(Z_j=1) (1-a) E\left(\frac{a(X_l)}{c(X_l)} \tilde W_l X_l|Z_l=1\right) P_r(Z_l=1)\\
&+& P_r(Z_j=0) (1-a) E\left(\frac{a(X_l)}{c(X_l)} \tilde W_l X_l|Z_l=0\right) P_r(Z_l=0)\\
\end{eqnarray*}

Denote $E\left(\frac{a(X_l)}{c(X_l)} \tilde W_lX_l|Z_l=1\right)$ as $\delta_3$ and $E\left(\frac{a(X_l)}{c(X_l)} \tilde W_lX_l|Z_l=0\right)$ as $\delta_4$.

$$E \left[\kappa_{j,l} \frac{a(X_l)}{c(X_l)} \tilde W_l X_l \right] = (1-a)(b^2 \delta_3+ (1-b)^2 \delta_4)$$

\;

Denote $E\left(\frac{b(X_l)}{c(X_l)} \tilde W_l|Z_l=1\right)$ as $\delta_5$ and $E\left(\frac{b(X_l)}{c(X_l)} \tilde W_l|Z_l=0\right)$ as $\delta_6$.
\begin{eqnarray*}
E \left[ \kappa_{j,l} \frac{b(X_l)}{c(X_l)} \tilde W_l \right] &=& (1-a)(b^2 \delta_5+ (1-b)^2 \delta_6)
\end{eqnarray*}

Denote $E\left(\frac{b(X_l)}{c(X_l)} \tilde W_lX_l|Z_l=1\right)$ as $\delta_7$ and $E\left(\frac{b(X_l)}{c(X_l)} \tilde W_lX_l|Z_l=0\right)$ as $\delta_8$.

\begin{eqnarray*}
E \left[ \kappa_{j,l} \frac{b(X_l)}{c(X_l)} \tilde W_l X_l \right] &=& (1-a)(b^2 \delta_7+ (1-b)^2 \delta_8)
\end{eqnarray*}

$E\left[ \kappa_{j,l}\left( \tilde P_j -  \frac{g_{0,\tilde P_{m}}(X_l)}{g_{0,\kappa_{m,l}}(X_l)} \right)\tilde P_l' \right]$ becomes:

$$ \begin{bmatrix} (1-a) (\theta_1^2 b^2 + \theta_2^2 (1-b)^2) - (1-a) (b^2 \delta_1+ (1-b)^2 \delta_2) & e_1 \\ (1-a)(\theta_1 \theta_3 b^2 + \theta_2 \theta_4 (1-b)^2) -(1-a)(b^2 \delta_5+ (1-b)^2 \delta_6) & e_2 \end{bmatrix}$$

$e_1 = (1-a)(\theta_1 \theta_3 b^2 + \theta_2 \theta_4 (1-b)^2)-(1-a)(b^2 \delta_3+ (1-b)^2 \delta_4)$

$e_2 = (1-a) (\theta_3^2 b^2 + \theta_4^2(1-b)^2) -(1-a)(b^2 \delta_7+ (1-b)^2 \delta_8)$

Since a is not 1, the determinant of $E\left[ \kappa_{j,l}\left( \tilde P_j -  \frac{g_{0,\tilde P_{m}}(X_l)}{g_{0,\kappa_{m,l}}(X_l)} \right)\tilde P_l' \right]$ depends on the exact values for these numbers.

\item Simulation illustration for RSMD and D-RSMD

Here are one simulation example for illustration (Model (\ref{identi:model03})).

$$y_i = W_{i,1}+ W_{i,1} * X_{i} +  X_{i} + u_i$$
$$W_{i,1} = 2 Z_i+ v_{i,1}$$
$$W_{i,1} * X_{i} = 2 Z_i * X_{i} + v_{i,1} * X_{i}$$

\begin{enumerate}
  \item For RSMD,
  \begin{eqnarray*}
   a'E(\tilde{W}_i|Z_i) &=& 2 a_1 Z_i - 2 a_1 E(E(Z_i|X_i)|Z_i)\\
    &+& 2 a_2  Z_i E( X_{i}|Z_i)  - 2 a_2 E(E( Z_i|X_i) X_{i}|Z_i)\\
    &=& 0
  \end{eqnarray*}
When $Z_i$ and $X_{i}$ are independent and $E(Z_{i})$ is a number (for instance, $E(Z_{i})=1$), $$ a_1 Z_i - a_1 E(Z_i) +  a_2 Z_i E(X_{i})  - a_2 E(X_i) E(Z_i) =0$$ $$ (a_1 + a_2 E(X_{i})) Z_i - (a_1+ a_2 E(X_i)) E(Z_i) =0$$

\begin{enumerate}
\item \label{sim:identint3} When $E( Z_{i})= 0$, $a_1$ must be 0 when $Z_i$ is continuous or discrete. $a_2$ can be any value. Hence, the first parameter is identified.
\item \label{sim:identint4} When $E( Z_{i})= 1$, there are infinite combinations for $(a_1, a_2)$ to make $$a'E(X_i|Z_{i}) =0$$. Both parameters are not identified.
\end{enumerate}
  \item For D-RSMD, because we have extra terms, it should identify both parameters when RSMD cannot.

\end{enumerate}

All of the simulation results are included in Table \ref{sim:idenfi}.
\end{enumerate}

\begin{table}
  \centering
  \scalebox{.6}{
  \begin{tabular}{l c c c c c c c c}
    \hline \hline
  \textit{Estimator} & \textit{Instrument} & Med.Bias  & MAD & Med.SE & RR & Mean.Bias & Mean.SE & Mean.AHSE \\ \hline
  \multicolumn{9}{c}{\textbf{Example \ref{sim:ident1}: Continuous instrument}} \\ \hline
    SMD $\beta_1$ & $(Z_{i,1},Z_{i,2})$ & 0.000 & 0.005 & 0.007 & 0.055 & 0.000 & 0.007 & 0.007 \\
    SMD $\beta_2$ & $(Z_{i,1},Z_{i,2})$ & 0.000 & 0.006 & 0.009 & 0.052 & 0.000 & 0.009 & 0.009 \\
    SMD $\beta_1$ & $Z_{i,1}$ & -0.011 & 0.034 & 0.065 & 0.026 & -0.021 & 1.598 & 37.825 \\
    SMD $\beta_2$ & $Z_{i,1}$ & 0.042 & 0.132 & 0.256 & 0.041 & 0.087 & 6.394 & 149.586 \\
    SMD $\beta_1$ & $Z_{i,2}$ & 0.018 & 0.106 & 0.209 & 0.022 & -0.050 & 18.259 & 783.448 \\
    SMD $\beta_2$ & $Z_{i,2}$ & -0.007 & 0.036 & 0.070 & 0.014 & 0.008 & 5.521 & 238.702 \\
    TSLS $\beta_1$ & $(Z_{i,1},Z_{i,2})$ & 0.000 & 0.004 & 0.006 & 0.063 & 0.000 & 0.006 & 0.006 \\
    TSLS $\beta_2$ & $(Z_{i,1},Z_{i,2})$ & 0.000 & 0.006 & 0.008 & 0.055 & 0.000 & 0.008 & 0.008 \\
    \hline
   \multicolumn{9}{c}{\textbf{Example \ref{sim:ident2}: Binary instrument}} \\ \hline
    SMD $\beta_1$ & $(Z_{i,1},Z_{i,2})$ & 0.000 & 0.008 & 0.012 & 0.052 & 0.000 & 0.012 & 0.012 \\
    SMD $\beta_2$ & $(Z_{i,1},Z_{i,2})$ & 0.000 & 0.010 & 0.015 & 0.047 & 0.000 & 0.015 & 0.015 \\
    SMD $\beta_1$ & $Z_{i,1}$ & 0.000 & 0.011 & 0.017 & 0.053 & 0.000 & 0.017 & 0.017 \\
    SMD $\beta_2$ & $Z_{i,1}$ & -0.001 & 0.017 & 0.026 & 0.049 & 0.000 & 0.026 & 0.026 \\
    SMD $\beta_1$ & $Z_{i,2}$ & 0.000 & 0.013 & 0.020 & 0.046 & 0.000 & 0.020 & 0.021 \\
    SMD $\beta_2$ & $Z_{i,2}$ & 0.000 & 0.014 & 0.021 & 0.046 & 0.000 & 0.021 & 0.021 \\
    TSLS $\beta_1$ & $(Z_{i,1},Z_{i,2})$ & 0.000 & 0.008 & 0.013 & 0.054 & 0.000 & 0.013 & 0.013 \\
    TSLS $\beta_2$ & $(Z_{i,1},Z_{i,2})$ & 0.000 & 0.011 & 0.017 & 0.048 & 0.000 & 0.017 & 0.017 \\
    \hline
  \multicolumn{9}{c}{\textbf{Example \ref{sim:identint1}: Binary instrument}} \\
  \hline
    SMD $\beta_1$ & $Z_{i}$ & 0.000 & 0.013 & 0.019 & 0.036 & 0.000 & 0.024 & 0.027 \\
    SMD $\beta_2$ & $Z_{i}$ & 0.000 & 0.043 & 0.077 & 0.039 & 0.019 & 0.738 & 0.596 \\
  \hline
  \multicolumn{9}{c}{\textbf{Example \ref{sim:identint2}: Binary instrument}} \\
  \hline
    SMD $\beta_1$ & $Z_{i}$ & 0.000 & 0.046 & 0.077 & 0.039 & -0.020 & 0.742 & 0.602 \\
    SMD $\beta_2$ & $Z_{i}$ & 0.000 & 0.043 & 0.077 & 0.039 & 0.019 & 0.738 & 0.596 \\
  \hline
  \multicolumn{9}{c}{\textbf{Example \ref{sim:identint3}: Continuous instrument}} \\
  \hline
    RSMD $\beta_1$ & $Z_{i}$ & -0.001 & 0.009 & 0.013 & 0.032 & 0.000 & 0.171 & 1.937 \\
    RSMD $\beta_2$ & $Z_{i}$ & 0.000 & 0.056 & 0.138 & 0.006 & 0.020 & 4.528 & 77.709 \\
    D-RSMD $\beta_1$ & $Z_{i}$ & -0.001 & 0.009 & 0.013 & 0.041 & -0.001 & 0.021 &  0.114 \\
    D-RSMD $\beta_2$ & $Z_{i}$ & 0.000 & 0.041 & 0.066 & 0.029 & 0.026 &  1.093 &  7.282\\
  \hline
  \multicolumn{9}{c}{\textbf{Example \ref{sim:identint4}: Continuous instrument}} \\
  \hline
    RSMD $\beta_1$ & $Z_{i}$ & -0.002 & 0.057 & 0.137 & 0.008 & -0.039 &  7.188  & 120.469 \\
    RSMD $\beta_2$ & $Z_{i}$ &  0.001 & 0.055 & 0.137 & 0.005 & 0.038 & 7.331 & 122.732 \\
    D-RSMD $\beta_1$ & $Z_{i}$ & -0.001 & 0.043 & 0.067 & 0.031 & -0.002 & 0.970 & 2.922 \\
    D-RSMD $\beta_2$ & $Z_{i}$ & 0.002 & 0.041 & 0.066 & 0.029 & 0.001 & 0.970 & 2.919 \\
  \hline \hline
  \end{tabular}
  }
  \caption{Identification Illustration}\label{sim:idenfi}
  \par Notes: Mean.AHSE stands for Mean.Asympt.Heterosk.SE. There are 5000 rounds of simulation and 2000 observations for each simulation study.
\end{table}

\end{document}